\documentclass[twocolumn]{aastex631}

\usepackage{graphicx}
\usepackage{amsmath}
\usepackage{tabularx}
\let\splitbox\relax
\usepackage{adjustbox}

\tabletypesize{5pt}

\shorttitle{AS2COSMOS: X-ray- and SED-selected AGNs}
\begin{document}

\title{ALMA/SCUBA-2 COSMOS Survey: Properties of X-ray- and SED-selected AGNs in Bright Submillimeter Galaxies}

\author[0000-0001-6653-779X]{Ryosuke Uematsu}
\affiliation{Department of Astronomy, Kyoto University, Sakyo-ku, Kyoto, Japan}

\author[0000-0001-7821-6715]{Yoshihiro Ueda}
\affiliation{Department of Astronomy, Kyoto University, Sakyo-ku, Kyoto, Japan}

\author[0000-0002-5896-6313]{David M. Alexander}
\affiliation{Centre for Extragalactic Astronomy, Department of Physics, Durham University, South Road, Durham, DH1 3LE, UK}

\author[0000-0003-1192-5837]{A. M. Swinbank}
\affiliation{Centre for Extragalactic Astronomy, Department of Physics, Durham University, South Road, Durham, DH1 3LE, UK}

\author[0000-0003-3037-257X]{Ian Smail}
\affiliation{Centre for Extragalactic Astronomy, Department of Physics, Durham University, South Road, Durham, DH1 3LE, UK}

\author[0000-0002-5580-4298]{Carolina Andonie}
\affiliation{Centre for Extragalactic Astronomy, Department of Physics, Durham University, South Road, Durham, DH1 3LE, UK}

\author[0000-0002-3805-0789]{Chian-Chou Chen}
\affiliation{Academia Sinica Institute of Astronomy and Astrophysics (ASIAA), No. 1, Sec. 4, Roosevelt Road, Taipei 106216, Taiwan}

\author[0000-0003-4748-0681]{Ugne \,Dudzevi{\v{c}}i{\={u}}t{\.{e}}}
\affiliation{Max-Planck-Institut f\"{u}r Astronomie, K'onigstuhl 17, D-69117, Heidelberg, Germany}

\author{Soh Ikarashi}
\affiliation{National Astronomical Observatory of Japan, Osawa 2-21-1, Mitaka, Tokyo 181-8588, Japan}
\affiliation{Junior College, Fukuoka Institute of Technology, 3-30-1 Wajiro-higashi, Higashi-ku, Fukuoka, 811-0295 Japan}
\affiliation{Department of Physics, General Studies, College of Engineering, Nihon University, 1 Nakagawara, Tokusada, Tamuramachi, Koriyama, Fukushima, 963-8642, Japan}

\author[0000-0002-4052-2394]{Kotaro Kohno}
\affiliation{Institute of Astronomy, Graduate School of Science, The University of Tokyo, 2-21-1 Osawa, Mitaka, Tokyo 181-0015, Japan}
\affiliation{Research Center for the Early Universe, Graduate School of Science, The University of Tokyo, 7-3-1 Hongo, Bunkyo-ku, Tokyo 113-0033, Japan}

\author[0000-0003-1747-2891]{Yuichi Matsuda}
\affiliation{National Astronomical Observatory of Japan, Osawa 2-21-1, Mitaka, Tokyo 181-8588, Japan}
\affiliation{Graduate University for Advanced Studies (SOKENDAI), Osawa 2-21-1, Mitaka, Tokyo 181-8588, Japan}

\author[0000-0001-9369-1805]{Annagrazia Puglisi}
\affiliation{Centre for Extragalactic Astronomy, Department of Physics, Durham University, South Road, Durham, DH1 3LE, UK}
\affiliation{School of Physics and Astronomy, University of Southampton, Highfield SO17 1BJ, UK}

\author[0000-0003-1937-0573]{Hideki Umehata}
\affiliation{Institute for Advanced Research, Nagoya University, Furocho, Chikusa, Nagoya 464-8602, Japan}
\affiliation{Department of Physics, Graduate School of Science, Nagoya University, Furocho, Chikusa, Nagoya 464-8602, Japan}

\author[0000-0003-2588-1265]{Wei-Hao Wang}
\affiliation{Institute of Astronomy and Astrophysics, Academia Sinica, Taipei 10617, Taiwan}

%% Note that the \and command from previous versions of AASTeX is now
%% depreciated in this version as it is no longer necessary. AASTeX
%% automatically takes care of all commas and "and"s between authors names.

%% AASTeX 6.31 has the new \collaboration and \nocollaboration commands to
%% provide the collaboration status of a group of authors. These commands
%% can be used either before or after the list of corresponding authors. The
%% argument for \collaboration is the collaboration identifier. Authors are
%% encouraged to surround collaboration identifiers with ()s. The
%% \nocollaboration command takes no argument and exists to indicate that
%% the nearby authors are not part of surrounding collaborations.

%% Mark off the abstract in the ``abstract'' environment.
\begin{abstract}

We investigate the properties of active galactic nuclei (AGNs) in the brightest submillimeter galaxies (SMGs) in the COSMOS field. We utilize the bright sample of ALMA/SCUBA-2 COSMOS Survey (AS2COSMOS), which consists of 260 SMGs with $S_{\mathrm{870}\, \mu \mathrm{m}}=0.7\text{--}19.2\,\mathrm{mJy}$ at $z=0\text{--}6$. We perform optical to millimeter spectral energy distribution (SED) modeling for the whole sample. We identify 24 AGN-host galaxies from the SEDs. Supplemented by 23 X-ray detected AGNs (X-ray AGNs), we construct an overall sample of 40 AGN-host galaxies. The X-ray luminosity upper bounds indicate that the X-ray undetected SED-identified AGNs are likely to be nearly Compton thick or have unusually suppressed X-ray emission. From visual classification, we identify $25^{+6}_{-5}$\% of the SMGs without AGNs as major merger candidates. This fraction is almost consistent with the general galaxy population at $z\sim2$, suggesting that major mergers are not necessarily required for the enhanced star formation in SMGs. We also identify $47^{+16}_{-15}$\% of the AGN hosts as major merger candidates, which is about twice as high as that in the SMGs without AGNs. This suggests that major mergers play a key role in triggering AGN activity in bright SMGs.

\end{abstract}

%% Keywords should appear after the \end{abstract} command.
%% The AAS Journals now uses Unified Astronomy Thesaurus concepts:
%% https://astrothesaurus.org
%% You will be asked to selected these concepts during the submission process
%% but this old "keyword" functionality is maintained in case authors want
%% to include these concepts in their preprints.

%% From the front matter, we move on to the body of the paper.
%% Sections are demarcated by \section and \subsection, respectively.
%% Observe the use of the LaTeX \label
%% command after the \subsection to give a symbolic KEY to the
%% subsection for cross-referencing in a \ref command.
%% You can use LaTeX's \ref and \label commands to keep track of
%% cross-references to sections, equations, tables, and figures.
%% That way, if you change the order of any elements, LaTeX will
%% automatically renumber them.
%%
%% We recommend that authors also use the natbib \citep
%% and \citet commands to identify citations.  The citations are
%% tied to the reference list via symbolic KEYs. The KEY corresponds
%% to the KEY in the \bibitem in the reference list below.

\section{Introduction}

The supermassive black holes (SMBHs) at the centers of galaxies may play a critical role in regulating star formation within the interstellar matter (ISM) of massive galaxies (e.g., \citealt{2006MNRAS.370..645B}). Previous studies have suggested a close co-evolution of galaxies and SMBHs inspired by the tight bulge-mass-to-SMBH-mass correlation found in the local universe (see \citealt{2013ARA&A..51..511K} for a review). This idea is also supported by the similarity of the cosmological evolution of star-formation rate density and the SMBH accretion rate density across $z=0\text{--}5$ \citep{2014ARA&A..52..415M,2003ApJ...598..886U,2014ApJ...786..104U}. However, the main physical mechanism driving the co-evolution is still unclear. Active galactic nuclei (AGNs) are key targets to solve this problem, as they are the observed manifestation of growing SMBHs \citep{2012NewAR..56...93A}. Some studies suggest that feedback from AGNs can affect the star-formation activity of their host galaxies through outflowing material via winds and/or relativistic jets (see \citealt{2012ARA&A..50..455F} for a review). Therefore, studying the properties of AGNs and their host galaxies is crucial to investigate the nature of galaxy-SMBH coevolution.

Dusty star-forming galaxies (DSFGs) are an important population in this context. DSFGs are characterized by their luminous far-infrared emission from dust heated by stars indicating their intense star formation. Theoretical studies have indicated that intense star-formation activity can trigger AGN activity by injecting turbulence in the gas disks and making the gas fall into the nuclear regions (e.g., \citealt{2011MNRAS.413.2633H}). Some studies also suggested that starburst activity can be triggered by galaxy mergers, which subsequently trigger AGN activity (e.g., \citealt{2008ApJS..175..356H,2019MNRAS.488.2440M}). Hence, DSFGs may be useful targets for studying the triggering mechanism of AGNs. Sub/millimeter observations are powerful tools to detect high-redshift DSFGs. At higher redshift, the peak of dust emission is redshifted to the sub/millimeter bands. Thus, high-redshift DSFGs are commonly referred to as submillimeter galaxies (SMGs) due to their selection wavelength. The Atacama Large Millimeter/submillimeter Array (ALMA) is a key observational facility to study high-redshift sources in the sub/millimeter bands. Due to its high angular resolution and sensitivity, it can precisely identify sub/millimeter sources (e.g., \citealt{2019MNRAS.487.4648S}).

The properties of AGNs in SMGs have been intensely studied over the last two decades. For instance, \citet{2005ApJ...632..736A} investigated the X-ray properties of SMGs in the 2 Ms Chandra Deep Field North (CDF-N), using a spectroscopically identified SCUBA 850-$\mu$m sample with radio counterparts from \citet{2005ApJ...622..772C} ($S_{\mathrm{850\mu m}}=2.4\text{--}17.4\,\mathrm{mJy}$). They showed that the majority (15/20) of those radio-detected SCUBA galaxies host X-ray AGNs, from which the large fraction (12/15) are moderately/highly absorbed ($N_{\mathrm{H}}\geq10^{23}\,\mathrm{cm}^{-2}$). On the basis of the ALMA 26 arcmin$^2$ Survey of GOODS-S at One-millimeter (ASAGAO; \citealt{2018PASJ...70..105H}) combined with the deepest 7 Ms Chandra X-ray survey, \citet{2018ApJ...853...24U} found $90^{+8}_{-19}$ per cent and $57^{+23}_{-25}$ per cent of millimeter-selected galaxies with $L_{\mathrm{IR}} > 10^{12}\,L_{\odot}$ and $10^{11}\,L_{\odot} <L_{\mathrm{IR}} < 10^{12}\,L_{\odot}$ at $z=1.5\text{--}3$ contains X-ray detected AGNs. The GOODS fields were also studied by an ultradeep SCUBA-2 survey (a submillimeter perspective on the GOODS fields [SUPER GOODS]; \citealt{2017ApJ...837..139C,2018ApJ...865..106C,2019ApJ...887...23B,2022ApJ...934...56B}). As a study in a wider field with relatively shallow X-ray observations, \citet{2013ApJ...778..179W} studied SMGs in the Extended Chandra Deep Field South (E-CDF-S), utilizing the ALMA LABOCA-E-CDF-S Submillimeter Survey (ALESS; \citealt{2013ApJ...768...91H}; $S_{\mathrm{870\mu m}}=1.3\text{--}9.0\,\mathrm{mJy}$). They identified eight X-ray AGNs from 91 SMGs in the shallower Chandra footprints, from which six sources are moderately/highly absorbed. Similarly, \citet{2019MNRAS.487.4648S} examined the properties of SMGs in the UKIDSS Ultra Deep Survey (UDS) field, based on the ALMA SCUBA-2 UDS survey (AS2UDS; \citealt{2018ApJ...860..161S}; $S_{\mathrm{870\mu m}}=0.6\text{--}13.6\,\mathrm{mJy}$). They found 23 X-ray AGNs from 274 SMGs in the shallow Chandra coverage. They also found 37 potential AGNs that were not detected in X-ray from 162 IRAC-selected SMGs at $z<3$, utilizing the IRAC-color selection of \citet{2012ApJ...748..142D}. Moreover, studies of high-density regions by \citet{2010ApJ...724.1270T} and \citet{2015ApJ...815L...8U,2019Sci...366...97U} have investigated the properties of AGNs in a protocluster at $z=3.09$ (SSA22). Although these studies constrained the physical properties of AGNs in SMGs, the derived values vary from study to study. For a more comprehensive understanding of the nature of AGN in SMGs, it is necessary to study a larger sample and wider parameter ranges of SMGs.

The detection of AGNs in galaxy samples is always challenging. Historically, many studies used X-ray observations and/or optical spectroscopy to study the nature of AGNs. However, these methods have a bias against the most heavily obscured sources (see \citealt{2018ARA&A..56..625H} for a review). In heavily obscured AGNs, X-ray and ultra-violet (UV) emission from the inner regions of the accretion disk around the SMBH is absorbed by surrounding material and is re-emitted in the mid- to far-infrared bands. Color-color diagrams in the mid-infrared wavebands are able to identify such heavily obscured systems, but are less reliable for high-redshift sources ($z>3$; \citealt{2012ApJ...748..142D}) and lower-luminosity AGN where the mid-IR emission can be strongly contaminated by the host galaxy (see also \citealt{2008ApJ...675.1171P}). Spectral energy distribution (SED) modeling of broad-band (rest-frame UV to far infrared) photometries is a promising method to robustly identify hidden AGNs missed at X-ray and optical energies. Although this method is dependent on the quality of the multi-wavelength photometry and model assumptions, it has successfully identified AGN candidates that are not detected in X-ray observations (e.g., \citealt{2022MNRAS.517.2577A,2023ApJ...950L...5Y,2024ApJ...965..108U,2024MNRAS.532..719C,2024ApJ...961..226L}).

In this study, we investigate the properties of AGNs hosted by bright SMGs in the COSMOS field, utilizing the ALMA/SCUBA-2 COSMOS Survey (AS2COSMOS; \citealt{2020MNRAS.495.3409S}). This survey is a follow-up of the brightest ${\sim}180$ submillimeter sources that were detected in the SCUBA-2 COSMOS Survey (S2COSMOS; \citealt{2019ApJ...880...43S}), and is effectively complete for S2COSMOS sources brighter than $S_{\mathrm{850}\, \mu \mathrm{m}}=6.2\,\mathrm{mJy}$. We perform rest-frame UV to far-infrared SED modeling for the AS2COSMOS SMG sample and identify AGNs. We also perform an X-ray spectral analysis of the bright X-ray detected AGNs. On the basis of these results, we discuss the properties of the AGNs and their host galaxies. The structure of this paper is as follows: In Section~\ref{section:data}, we describe the observations and data reduction. In Section~\ref{section:analysis}, we describe the SED modeling and the X-ray spectral analysis. Section~\ref{section:results} presents our results and discussions. A summary is given in Section~\ref{section:summary}. Throughout the paper, we assume a flat universe with $H_0=70.4\,\mathrm{km\,s^{-1}\,Mpc^{-1}},\,\Omega_{\mathrm{M}}=0.272$ \citep{2011ApJS..192...18K}. The Chabrier initial mass function (IMF) is adopted \citep{2003PASP..115..763C}. We use the AB magnitude system \citep{1983ApJ...266..713O}. If not specially mentioned otherwise, the quoted errors correspond to 1$\sigma$ confidence.

\section{Observations and Data Reduction}\label{section:data}

\subsection{Submillimeter Photometry}\label{subsection:submilli}

\begin{figure*}
  \centering
  \includegraphics[width=\linewidth]{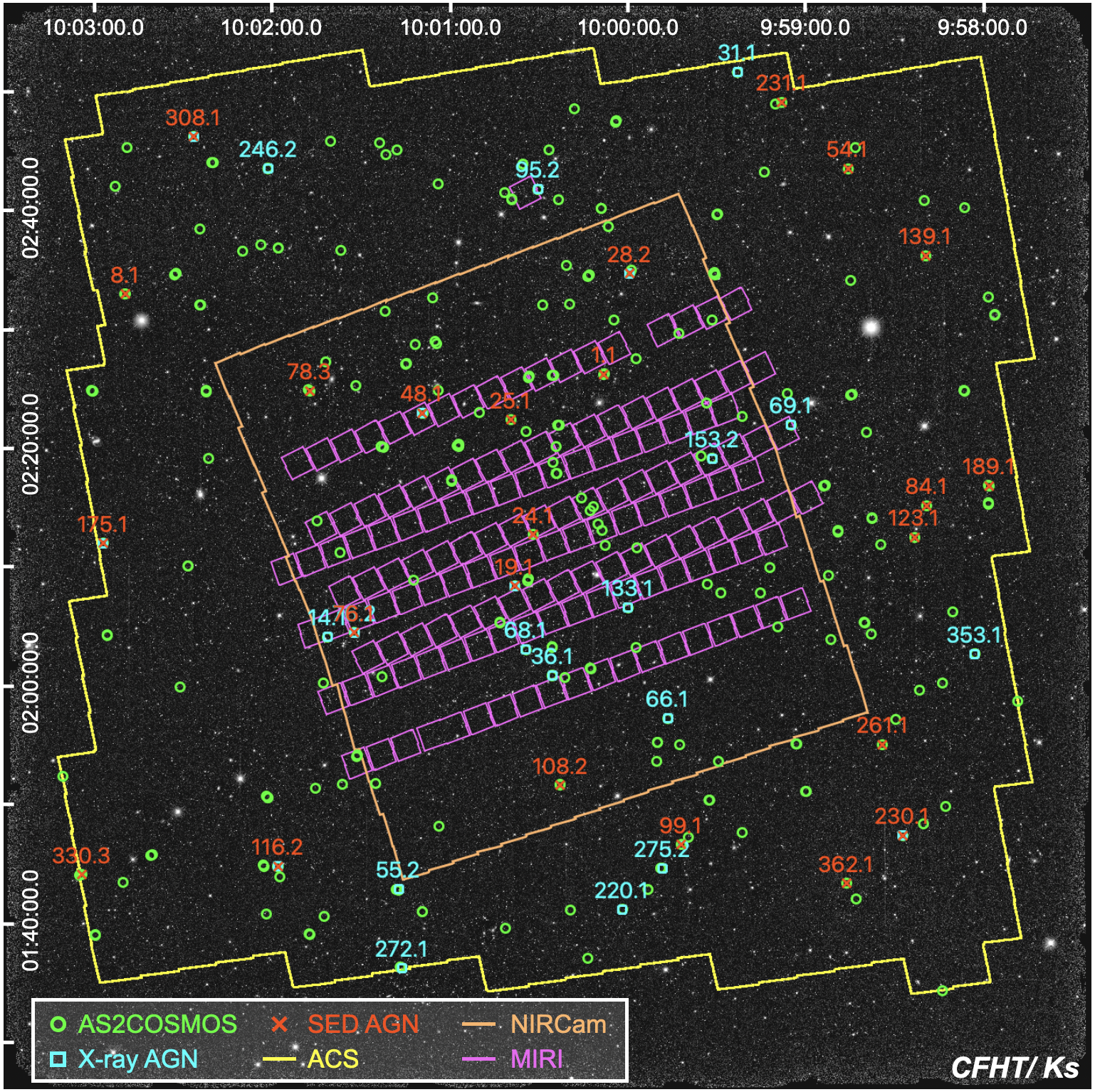}
  \caption{
  Positions of the AS2COSMOS sources. The cyan and red points correspond to the X-ray AGNs and SED AGNs, respectively (see Section~\ref{subsubsection:agn_identification} and Section~\ref{subsubsection:x-ray_LFIR}). The background is the CFHT $K_{\mathrm{s}}$-band image. The coverage of Hubble Space Telescope (HST)/ACS, JWST/NIRCam, and JWST/MIRI imaging are shown in yellow, orange, and magenta, respectively.}
  \label{figure:position}
\end{figure*} 

The ALMA submillimeter source catalog that is the basis of our analysis was presented in \citet{2020MNRAS.495.3409S}. The pilot study of AS2COSMOS was performed in ALMA Cycle 4. This survey followed up 160 of the brightest SCUBA-2 850-$\mu$m sources in the COSMOS field from the S2COSMOS survey \citep{2019ApJ...880...43S}, of which 158 had deboosted/deblended 850 $\mu$m flux densities of $S_{\mathrm{850}\, \mu \mathrm{m}}>6.2\,\mathrm{mJy}$. Observations were conducted between 2018 May 15 and 21 covering the 7.5-GHz bandwidth centered at 343 GHz (870 $\mu$m; band 7). Supplemented by the archival ALMA band 7 imaging, \citet{2020MNRAS.495.3409S} constructed a sample of 182 brightest S2COSMOS sources (deboosted/deblended flux densities of $S_{\mathrm{850}\, \mu \mathrm{m}}>6.2\,\mathrm{mJy}$) with ALMA band-7 imaging. All the ALMA data were calibrated and reduced with the Common Astronomy Software Applications ({\sc casa}) package (v5.1.1). The final images have a median synthesized beam size of $0.80\times0.79$ arcsec full width at half maximum (FWHM) and a median sensitivity of $\sigma_{\mathrm{870}\, \mu \mathrm{m}}=0.19\,\mathrm{mJy}\,\mathrm{beam}^{-1}$.

Source extraction was performed with {\sc sextractor}. First, \citet{2020MNRAS.495.3409S} extracted sources within the ALMA primary beams with a significance of ${>}4.8\sigma$ at a peak or ${>}4.9\sigma$ in an aperture with a $1.5\times$ FWHM of the synthesized beam. The number of false detections fell to zero with these criteria. In this step, \citet{2020MNRAS.495.3409S} constructed a robust sample of 254 SMGs in 182 ALMA maps. Then, they extracted sources from the outer region of the ALMA primary beam with a slightly higher peak significance of ${>}5.1\sigma$, where the false detection rate again fell to zero. As part of this step, six additional SMGs were detected at 8.9--11.5 arcsec offsets from the phase center. The final sample contains 260 SMGs out of the 182 brightest S2COSMOS sources. This sample effectively comprises all S2COSMOS sources brighter than $S_{\mathrm{850}\, \mu \mathrm{m}}=6.2\,\mathrm{mJy}$. However due to the influence of flux boosting on the SCUBA-2 catalog, this corresponds to an overall completeness of 50 per cent for SMGs at $S_{\mathrm{850}\, \mu \mathrm{m}}=7.2$ mJy in the 1.6 deg$^2$ survey area. We note that \citet{2020MNRAS.495.3409S} confirmed a small astrometric offset of $\Delta \mathrm{R.A.}=0.08\pm0.01$ arcsec between the AS2COSMOS and COSMOS2015 catalog \citep{2016ApJS..224...24L}. In this study, we always correct the source astrometry for this offset by referring to the COSMOS2015 catalog. Figure~\ref{figure:position} shows the positions of the AS2COSMOS sources on a Canada France Hawaii Telescope (CFHT) $K_{\mathrm{s}}$-band image.

\subsection{Spectroscopic Redshifts}

Spectroscopic redshifts of AS2COSMOS sources were obtained from several studies. We utilized the following spectroscopic surveys: the Sloan Digital Sky Surveys (SDSS DR16; \citealt{2020ApJS..249....3A}), zCOSMOS survey (DR3; \citealt{2009ApJS..184..218L}), the VIMOS VLT Deep Survey (VVDS final data release; \citealt{2013A&A...559A..14L}), the PRIsm MUlti-object Survey (PRIMUS DR1; \citealt{2011ApJ...741....8C,2013ApJ...767..118C}), the FMOS-COSMOS survey (version 2; \citealt{2015ApJS..220...12S,2019ApJS..241...10K}), the 3D-HST survey \citep{2014ApJS..214...24S,2016ApJS..225...27M}, the Complete Calibration of the Color-Redshift Relation (C3R2) survey (DR3; \citealt{2019ApJ...877...81M}), hCOSMOS survey \citep{2018ApJS..234...21D}, and the DEIMOS 10K spectroscopic survey \citep{2018ApJ...858...77H}. In addition to these catalogs, we used the results of the submillimeter line scan of the brightest AS2COSMOS sources performed by \citet{2022ApJ...929..159C} and \citet{2024ApJ...961..226L} (see also \citealt{2021MNRAS.501.3926B}). Moreover, we utilized the spectroscopic study of a galaxy cluster performed by \citet{2016ApJ...828...56W} (AS2COS0003.1, AS2COS0003.2, AS2COS0003.3, and AS2COS0003.4) and the spectroscopic study of [C\,{\sc ii}] emitters performed by \citet{2021ApJ...907..122M} (AS2COS0001.1, AS2COS0001.2, AS2COS0006.1, AS2COS0034.1, and AS2COS0034.2).

We cross-matched the AS2COSMOS sources with these catalogs within 1.0 arcsec radii and only used ``secure'' redshifts. With priority given to the submillimeter redshifts, 80 out of 260 sources had spectroscopic redshifts ranging from 0.033 to 5.3 (hereafter ``spec-$z$ sample''). In Section~\ref{subsubsection:sed_procedure} and Appendix~\ref{subsection:misidentified}, we show that the original optical counterpart of AS2COS0159.1 ($z_{\mathrm{spec}}=0.033$) was misidentified. Therefore, the final spectroscopic redshift sample contains 79 SMGs within $z_{\mathrm{spec}}=0.3\text{--}5.3$. For the other sources, we calculate their photometric redshifts by utilizing {\sc cigale} to simultaneously treat redshift uncertainties (hereafter ``photo-$z$ sample''; Section~\ref{subsubsection:redshift}).

\subsection{Size Measurement with ALMA}\label{subsection:size}

The typical synthesized beam of the ALMA 870-$\mu$m observations in AS2COSMOS from \citet{2020MNRAS.495.3409S} was $\sim$\,0.8$''$ FWHM. Given that the characteristic sizes of the bright dust continuum emission SMGs are believed to be $\sim$\,0.2$''$--0.3$''$ FWHM \citep[e.g.,][]{2015ApJ...799...81S,2015ApJ...810..133I,2017ApJ...849L..36I,2016ApJ...833..103H,2018ApJ...861....7F,2019MNRAS.490.4956G}, the coarse resolution of our observations would have made it challenging to measure reliable sizes for such compact sources in the image domain. However, the high S/N of our ALMA detections (median integrated S/N\,$\sim$\,30, with a range of 17--52) meant that it would be possible to obtain constraints on the size of the dust continuum emission in the brighter sources in our sample using a visibility-domain analysis, as has been demonstrated in the past \citep[e.g.,][]{2006ApJ...640L...1I,2008ApJ...688...59Y,2015ApJ...810..133I,2017ApJ...849L..36I,2019MNRAS.490.4956G}.

The ALMA dust continuum size measurements were undertaken by Ikarashi et al.\ (in prep.) and followed the methodology described in \citet{2015ApJ...810..133I}, with the exception that they adopted a Sersic $n$\,$=$\,1 exponential disk profile (rather than a Gaussian profile used in that earlier work) as suggested by high-angular resolution submillimetre imaging of SMGs in the literature \citep[e.g.,][]{2016ApJ...833..103H,2018ApJ...861....7F,2019ApJ...876..130H,2019MNRAS.490.4956G}. The first step in their analysis was to remove any other significant sources (S/N\,$\geq$\,5) from the ALMA map of the primary source whose size was being measured. This is necessary as the visibility-domain analysis assumes there is only a single source in the map. To remove the secondary sources these were modeled using the {\sc clean} task in {\sc casa} employing its scratch option. This involved masking the other sources and cleaning down to 1-$\sigma$ noise level, before subtracting the modeled source visibilities from the ALMA visibility data.

Then $uv$--amplitude plots were constructed for each source, by shifting the phase center to the position of the source and measuring the circularized, average flux in each $uv$-distance bin. A single exponential-disk light profile ($n$\,$=$\,1) model was then fitted to the $uv$--amplitude plot to obtain the best-fit size as a circularised half-light radius, $R_{\rm circ, 0.5}$. Ikarashi et al.\ (in prep.) also tested the reliability of their size measurements by performing simulations which involved injecting mock sources with known light profiles and sizes into the real ALMA visibility data and measuring their sizes in the same manner as for the real sources. This provided robust estimates of the errors in our size measurements. To determine the final size for each source, Ikarashi et al.\ (in prep.) assessed whether the reported size is measurable (or only provides a limit) given the claimed S/N and resulting size using these Monte Carlo simulations, which also provided appropriate uncertainties on the source size measurements.

\subsection{Optical to Far-infrared Photometry}\label{subsection:opt-fir}

The optical-to-radio photometric catalog for the SMG counterparts was also constructed by \citet{2020MNRAS.495.3409S}, by updating the COSMOS2015 photometry catalog \citep{2016ApJS..224...24L} to include deeper observations in the critical optical and near-infrared bands. By cross-matching the AS2COSMOS sources with the COSMOS2015 catalog, 179/260 optical/near-infrared counterparts were identified within search radii of 0.85 arcsec, where the possibility of a false match is estimated at ${\sim}6.6$ per cent. After that, the $YJHK_{s}$ band photometry was substituted with those in DR4 of the Ultravista survey \citep{2012A&A...544A.156M}, which is up to ${\sim}0.5$ mag deeper than DR2 used in the COSMOS2015 catalog. Photometry was measured using 2-arcsec diameter apertures centered at the SMG positions. In addition, the $BVriz$ photometry of COSMOS2015 catalog was replaced with the $grizY$ photometry in the second data release of the Hyper Suprime-Cam Subaru Strategic Program (HSC-SSP; \citealt{2019PASJ...71..114A}) if available. HSC-SSP DR2 reached ${\sim}1$ mag deeper than the optical imaging used in the COSMOS2015 catalog, and 158/260 counterparts were identified within 0.85 arcsec. In total, 199/260 optical/near-infrared counterparts are identified in this step.

To obtain mid-infrared photometry, the 3.6--8.0 $\mu$m images of the Spitzer Large Area Survey combined with Hyper Supreme-Cam (SPLASH; \citealt{2014ApJ...791L..25S}) were utilized. The images were reduced with {\sc iracclean} \citep{2012ApJS..203...23H}, where sources are deblended by referring to the stacked $zYJHK_{s}$ detection images and ALMA 870 $\mu$m source positions. The far-infrared imaging of the COSMOS field was provided by three large programs; the COSMOS-Spitzer program (24 $\mu$m; \citealt{2007ApJS..172...86S}), the PACS Evolutionary Probe (PEP) survey (100 and 160 $\mu$m; \citealt{2011A&A...532A..90L}), and the Herschel Multi-tiered Extragalactic Survey (HerMES; 250, 350, and 500 $\mu$m; \citealt{2012MNRAS.424.1614O}). The ``superdeblended'' catalog \citep{2018ApJ...864...56J}, which contains deblended 24--160-$\mu$m photometry for $K_{s}$- and 3-GHz-selected sources, was cross-matched to the SMGs to improve completeness at 24--100-$\mu$m band and photometry was also obtained by cross-matching the AS2COSMOS sources with the PACS/PEP survey catalog \citep{2011A&A...532A..90L}. Finally, 250--500-$\mu$m photometry was derived following the method described in \citet{2014MNRAS.438.1267S}, where the Spitzer 24 $\mu$m, VLA 3 GHz and ALMA 870 $\mu$m source positions were used to deblend the Herschel images.

The radio photometry was taken from the source catalog of the VLA-COSMOS 3 GHz Large Project \citep{2017A&A...602A...1S}. Cross-matching the AS2COSMOS sources with the VLA 3G-Hz catalog, 191 counterparts were identified within 1 arcsec. Since two pairs of SMGs were confused in the 3-GHz map (AS2COS0051.1/.2 and AS2COS0228.1/.2), the {\sc casa/imfit} routine was applied to deblend the fluxes of these sources.

\begin{figure*}
  \centering
  \includegraphics[width=\linewidth]{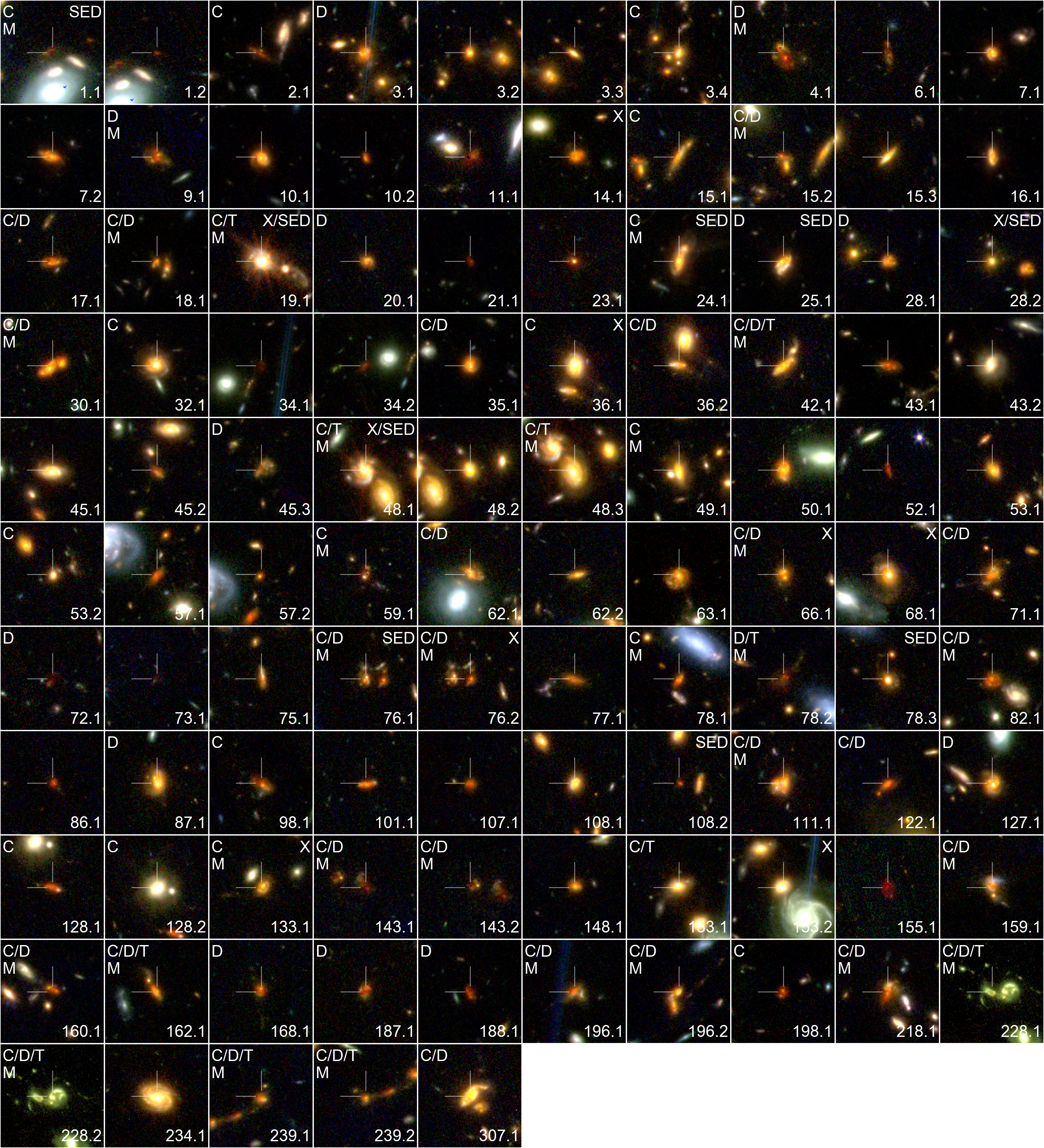}
  \caption{
  8 arcsec $\times$ 8 arcsec JWST images of the 105 AS2COSMOS sources in the coverage of NIRCam (AS2COS0005.1 and AS2COS0005.2 are excluded). The blue, green, and red colors correspond to the F115W+F150W, F277W, and F444W filters, respectively. We label the merger candidates, which have tidal features (T), disturbed morphology (D) or possible companions (C) (see Section~\ref{subsection:morphology}). The major merger candidates are indicated by ``M''. The X-ray AGNs and the SED AGNs are indicated by ``X'' and ``SED'' (see Section~\ref{subsubsection:agn_identification} and Section~\ref{subsubsection:x-ray_LFIR}). Note that the images of AS2COS0107.1 and AS2COS0122.1 are constructed from F115W (blue), F277W (green), and F444W (red) filters as they are not covered by F150W. For the same reason, the image of AS2COS0155.1 is constructed from F115W (blue), F2150W (green), and F277W (red), and those of AS2COS0228.1 and AS2COS0228.2 are constructed from F115W (blue), F115W+F150W (green), and F150W (red), respectively.}
  \label{figure:JWST_nircam}
\end{figure*}

\subsection{High-resolution Imaging with JWST}\label{subsection:nir}

The COSMOS field was observed with the James Webb Space Telescope (JWST) through two Cycle 1 JWST treasury programs. The first was PRIMER \citep{2021jwst.prop.1837D}, which provides a contiguous 144 arcmin$^2$ with NIRCam in eight filters (F090W, F115W, F150W, F200W, F277W, F356W, F444W and F410M) and a non-contiguous 112 arcmin$^2$ with MIRI in two filters (F770W and F1800W). The second was COSMOS-Web \citep{2023ApJ...954...31C}, which provides a contiguous 0.54 deg$^2$ with NIRCam in four filters (F115W, F150W, F277W, and F444W) and a non-contiguous 0.18 deg$^2$ with MIRI in one filter (F770W). 107/260 AS2COSMOS sources fall within the coverage of the NIRCam imaging, whereas 43/260 sources are in the MIRI coverage (Figure~\ref{figure:position}). We utilized the reprocessed images provided by the DAWN JWST Archive (DJA) and constructed the color-composite images of those sources. Figure~\ref{figure:JWST_nircam} shows the color-composite images of the 107 AS2COSMOS sources in the NIRCam coverage. The color-composite images of the 40 AS2COSMOS sources in the coverage of both NIRCam and MIRI are shown in Figure~\ref{figure:JWST_miri} in Appendix~\ref{appendix:miri}. In Section~\ref{subsection:morphology}, we perform a morphological analysis using these images. Note that the near-infrared counterpart of AS2COSMOS0005.1 has distorted morphology, indicating it is a strong gravitational lens \citep{2021ApJ...913....6H,2024MNRAS.52712044P,2024A&A...690L..16J}. This is due to gravitational lensing by a foreground galaxy. Thus, we exclude AS2COSMOS0005.1 and AS2COSMOS0005.2 in the following analysis.

\subsection{X-ray Observations}\label{subsection:x-ray}

\subsubsection{Observation Overview and Cross Matching}\label{subsubsection:x-ray_obs}

The COSMOS field was observed with the Chandra X-ray observatory \citep{2002PASP..114....1W} through two large programs. The first was the Chandra COSMOS survey (C-COSMOS; \citealt{2009ApJS..184..158E}), which was conducted from 2006 November to 2007 June. This survey covered the central 0.5 deg$^2$ of the COSMOS field with an exposure time of ${\sim}160$ ks, and an outer 0.4 deg$^2$ area of ${\sim}80$ ks. The second program was the Chandra COSMOS-Legacy survey \citep{2016ApJ...819...62C}, which was carried out from 2012 November to 2014 March. During this survey, the outer region of C-COSMOS was observed to a depth of ${\sim}160$ ks, achieving an effective exposure of ${\simeq}160$ ks over the central 1.5 deg$^2$ and of ${\simeq}80$ ks in the outer 0.7 deg$^2$. The whole survey detected 4016 point sources to limiting depths of $2.2\times10^{-16}\,\mathrm{erg\,s^{-1}\,cm^{-2}}$, $1.5\times10^{-15}\,\mathrm{erg\,s^{-1}\,cm^{-2}}$, and $8.9\times10^{-16}\,\mathrm{erg\,s^{-1}\,cm^{-2}}$ in the 0.5--2.0, 2.0--10.0, and 0.5--10.0 keV bands, respectively.

Cross-matching the AS2COSMOS sources with the latest catalog by \citet{2016ApJ...819...62C}, we identified 23 X-ray counterparts within a radius of 1.4 arcsec. Since \citet{2016ApJ...819...62C} reported that 95 per cent of the X-ray sources used for the astrometry correction have their optical counterparts within 1.4 arcsec, we employed this value for the cross-matching radius (the total false matching rate is ${\sim}0.3$ across the entire field).

In this study, we performed an X-ray spectral analysis for the X-ray detected sources to derive their X-ray luminosities and line-of-sight absorption. We also estimated the X-ray luminosity upper bounds for X-ray undetected SMGs assuming typical X-ray spectra of AGNs.\footnote{Here we use the term ``upper bound'' to describe the upper boundary of confidence intervals (see \citealt{2010ApJ...719..900K}).} Details of the X-ray spectral analysis are described in Section~\ref{subsection:x-ray_spec}. Here we reanalyzed all the X-ray data, consisting of 117 pointings, using the Chandra interactive analysis of observations software ({\sc ciao v4.15}; \citealt{2006SPIE.6270E..1VF}) and the latest calibration database ({\sc CALDB v4.10.7}). Note that AS2COS0353.2 was initially identified as an X-ray source that was later confirmed as the counterpart of AS2COS0353.1 by visual inspection (Appendix~\ref{appendix:x-misid}).

\subsubsection{Source Extraction}\label{subsubsection:x-ray_extraction}

Before extracting X-ray spectra of the SMGs, we performed X-ray astrometric corrections for all the observations, following \citet{2016ApJ...819...62C}. First, we reprocessed the data with \texttt{chandra\_repro} script, employing the \texttt{vfaint} mode for ACIS background cleaning. Next, we performed source detection in each 0.5--7.0 keV image with \texttt{wavdetect} tool. The point spread function (PSF) maps were created at the 39.3\% encircled count fraction (ECF) radius, assuming a single power-law spectrum with a photon index of 1.4 and Galactic absorption of $N_{\mathrm{H}}=2.6\times 10^{20}\,\mathrm{cm}^{-2}$ \citep{2005A&A...440..775K}. We employed a ``2 sequence'' of wavelet scales (i.e., 1, 1.414, 2, 2.828, 4, 5.656, 8, 11.314, and 16 pixels) and a false-positive probability threshold of $10^{-6}$. Then, we corrected the astrometry using the HSC $i$-band-selected subsample (17--24 AB mag) of the COSMOS2020 catalog \citep{2022ApJS..258...11W}. In this step, only secure sources whose significance is higher than $5\sigma$ and the PSF size is smaller than 2 pixels (${\sim}1$ arcsec) were cross-matched with the reference within 1.5 arcsec. On average, 15 sources were used to perform reprojection in each image. Finally, we combined all the images using \texttt{merge\_obs} script. To confirm the astrometric accuracy, we again performed source detection in the combined image with \texttt{wavdetect} tool. In this step, the length scale was varied from 0.5$''$ to 16$''$ in steps of $\sqrt{2}$. The detailed analysis of the Chandra astrometry is summarized in Appendix~\ref{appendix:x-misid}.

We then extracted the X-ray spectra using circular apertures centered at the X-ray source positions. Here we determined the X-ray source positions by the frame that has the sharpest PSF at the detected position. The aperture radii were set to the 80 ECF radius in each frame, but an upper limit was set to 5.0 arcsec for sources that are detected over 3$\sigma$ in the 0.5--7.0 keV band\footnote{Here we used the signal-to-noise ratios that were reported in \citet{2016ApJ...819...62C}.} and 4.0 arcsec for the others.\footnote{AS2COS95.2 was detected over 3$\sigma$ in the 0.5--7.0 keV band, but the upper limit on the extraction aperture was set to 4.0 arcsec to avoid contamination from a nearby source.} Background spectra were taken from annuli with inner and outer radii of 8.0 and 25.0 arcsec, respectively, by masking any X-ray sources with circular regions with radii of 8.0 arcsec. For X-ray undetected SMGs, we estimated the X-ray upper bound count rates at 99.73 per cent confidence intervals (3$\sigma$) in the 0.5--7.0 keV band, utilizing {\sc srcflux} script. For the source regions, we employed circular apertures with radii of 3.0 arcsec centered at the 870-$\mu$m source positions. The background count rates were estimated by annular regions with inner and outer radii of 8.0 and 15.0 arcsec, respectively, by masking any X-ray sources with circular regions with radii of 8.0 arcsec.

\section{Analysis}\label{section:analysis}

\subsection{Optical to Submillimeter SED Analysis}\label{subsection:sed}

\subsubsection{SED Fitting}\label{subsubsection:sed_procedure}

\begin{figure*}
  \centering
  \includegraphics[width=\linewidth]{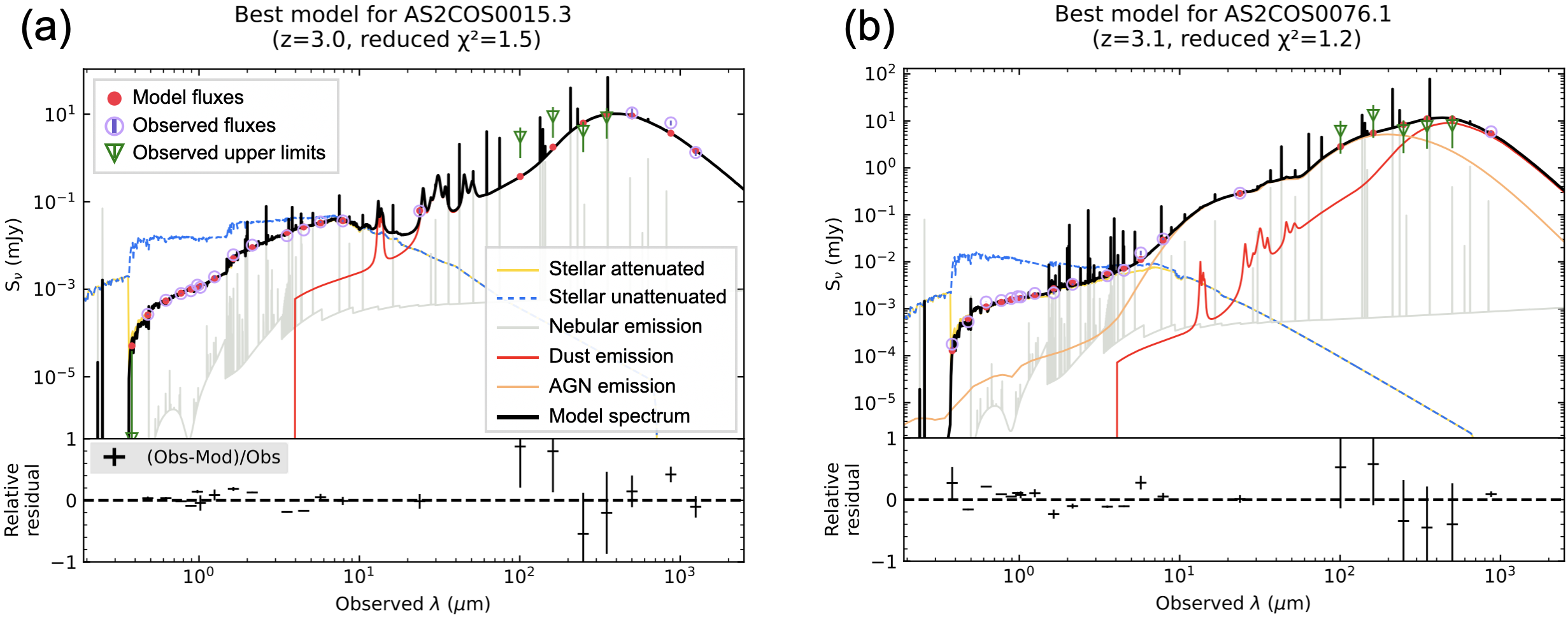}
  \caption{
  Example SEDs of a galaxy fitted \textit{(a)} without and \textit{(b)} with AGN templates. The black solid line represents the best-fit template SED solution. The yellow solid line illustrates the stellar emission attenuated by interstellar dust. The blue dashed line depicts the unattenuated stellar emission for reference. The orange line corresponds to the emission from the AGN. The red line shows the infrared emission from interstellar dust. The gray line denotes the nebulae emission from the host galaxy. The observed data points are represented by purple circles, accompanied by 1$\sigma$ error bars. The bottom panel displays the relative residuals between the best-fitting template solution and the photometry.}
  \label{figure:example}
\end{figure*}

\begin{figure}
  \centering
  \includegraphics[width=\linewidth]{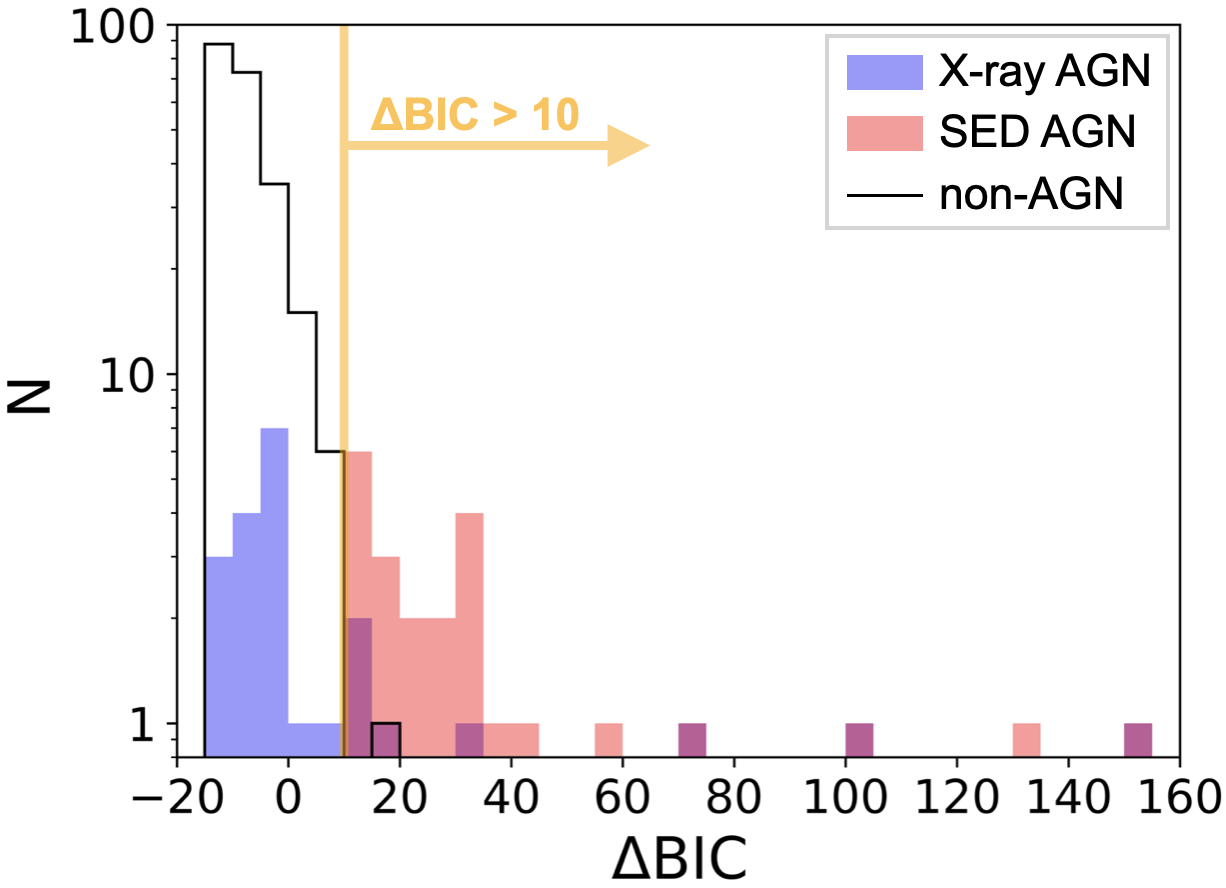}
  \caption{
  Distribution of $\Delta \mathrm{BIC}$ of the X-ray detected AS2COSMOS sources (X-ray AGN; see Section~\ref{subsubsection:x-ray_LFIR}), SED AGNs, and all the AS2COSMOS sources. The vertical orange solid lines show the adopted threshold to identify an SED AGN ($\Delta \mathrm{BIC}=10$). Most of the X-ray detected sources (16 out of 23) have BICs lower than 10, suggesting that the SED AGNs are the complementary sample to X-ray detected AGNs.}
  \label{figure:deltaBIC}
\end{figure}

We perform optical to submillimeter SED modeling for the whole AS2COSMOS sample. We use {\sc cigale v2022.0} \citep{2019A&A...622A.103B,2020MNRAS.491..740Y,2022ApJ...927..192Y}, but incorporating modifications for the dust emission model as described below. The {\sc cigale} code is designed to calculate the likelihoods of all the models on a user-defined grid and return the likelihood-weighted mean of the marginalized probability distribution function (PDF) as a Bayesian estimation. In {\sc cigale}, users can choose several options for each SED component. We employ a delayed star-formation history (SFH), accompanied by a recent starburst with a constant star-formation rate. The composite stellar populations are calculated based on the stellar template of \citet{2003MNRAS.344.1000B} and the Chabrier IMF \citep{2003PASP..115..763C}. The dust extinction of the stellar component is modeled with the modified Calzetti starburst attenuation law \citep{2000ApJ...533..682C,2002ApJS..140..303L,2009A&A...507.1793N}. For the dust emission model, we use the {\sc ethemis} model \citep{2024ApJ...965..108U}, which is a high-temperature extension of the {\sc themis} model \citep{2017A&A...602A..46J}. The optical to far-infrared emission from an AGN is modeled with the {\sc skirtor} model \citep{2012MNRAS.420.2756S,2016MNRAS.458.2288S}, where a polar-dust component is implemented with a single optically-thin graybody. The UV-to-optical emission from the accretion disk is modeled by a broken power law \citep{2005A&A...437..861S}. In the {\sc skirtor} model, an AGN whose inclination angle is smaller (larger) than the half-opening angle of the torus is classified as a type-1 (type-2) AGN. We employ this classification in later sections.
 
In {\sc cigale}, the number of parameter grid points is often limited by computational cost. Hence, we need to optimize the parameter set to adequately reproduce the SEDs given this limitation. Moreover, there remain some possible errors in extracting photometry such as remaining blending effects or calibration issues. For these reasons, we separate the SED modeling into two steps. In the first step, we analyze all the SEDs with a robust parameter set, which is summarized in Table~\ref{table:modelbasic} in Appendix~\ref{appendix:sed_detail}. In this step, we adequately reproduce most of the SEDs with reduced $\chi^2$ values typically lower than five. Then, in the second step, we reanalyze the remaining sources by optimizing the photometry or adjusting the parameter settings. Details of this iterations are summarized in Appendix~\ref{appendix:sed_detail}. With this approach, we adequately reproduce the SEDs of the AS2COSMOS sources ($\chi^2/\mathrm{d.o.f.}<7$; see Figure~\ref{figure:chi2} in Appendix~\ref{subsection:chi2}). Figure~\ref{figure:example} illustrates two example SEDs of AS2COSMOS sources fitted with and without AGN templates. Note that the radio component is not simultaneously treated in the SED modeling to save the computational cost.

When calculating the physical properties, we vary the following parameters logarithmically: stellar mass ($M_*$), star-formation rate (SFR), dust mass ($M_{\mathrm{dust}}$), and infrared luminosity ($L_{\mathrm{IR}}$)\footnote{In {\sc cigale}, the infrared luminosity is calculated by the sum of the dust luminosity from the AGN and the dust luminosity from the host galaxy. This is almost the same as the integrated luminosity between 8--1000 $\mu$m.}, while the following are varied linearly: dust temperature ($T_{\mathrm{dust}}$), color excess of stellar continuum attenuation ($E(B-V)$), power-law index to modify the Calzetti attenuation slope ($\delta$), AGN luminosity fraction in the total dust luminosity ($f_{\mathrm{AGN}}$), bolometric AGN luminosity ($L_{\mathrm{AGN,\,bol}}^{\mathrm{SED}}$), and redshift ($z$). For SFR, we employ 10 Myr averaged values. The specific SFRs (sSFR) are calculated by dividing the SFRs by the stellar masses. The radio-IR correlation parameters ($q_{\mathrm{IR}}$) are calculated by the following equation:
\begin{equation}
  q_{\mathrm{IR}}=\log\left(\frac{L_{\mathrm{IR}}}{L_{\nu,\mathrm{21\,cm}}\times3.75\times10^{12}\,\mathrm{Hz}}\right)
\end{equation}
The radio luminosity density at 21 cm is converted from the radio flux density at 3 GHz assuming a spectral index of 0.75 (e.g., \citealt{1992ARA&A..30..575C,2009MNRAS.397..281I,2010MNRAS.401L..53I}). Note that the spectral index was reported to have large uncertainty (a standard deviation of 0.29; \citealt{2010MNRAS.401L..53I}) and care should be taken. The SFR is directly calculated from the stellar templates, and this value corresponds to the sum of UV (unobscured) and FIR (obscured) SFRs.

\subsubsection{Treatment of the Redshift Uncertainty}\label{subsubsection:redshift}

In the SED modeling, the redshifts of the SMGs without spectroscopic redshifts are allowed to vary within $0.1<z<6.0$ (photo-$z$ sample). Hence, the physical quantities derived by {\sc cigale} include the redshift uncertainties. However, in the X-ray spectral analysis, redshifts are always fixed at the spectroscopic redshifts or the Bayesian-estimated photometric redshifts derived by SED modeling (Section~\ref{subsection:x-ray_spec}). This inconsistency can cause a problem in comparing the physical properties. Therefore, we repeat the SED modeling by fixing the redshifts at either the spectroscopic ones or the Bayesian-estimated photometric ones. The values estimated in this way are used for comparison with the X-ray properties.

\subsubsection{Identification of AGN}\label{subsubsection:agn_identification}

To identify whether a galaxy hosts an AGN or not, we utilize the Bayesian Information Criterion (BIC). The BIC is calculated as $\mathrm{BIC}=\chi^2+k\times \ln(n)$, where $\chi^2$ is the non-reduced $\chi^2$ value, $k$ is the number of degrees of freedom, and $n$ is the number of photometric points. In this study, we regard the sources with $\Delta \mathrm{BIC} = \mathrm{BIC}_{\mathrm{w/o\,AGN}} - \mathrm{BIC}_{\mathrm{w/\,AGN}} >10$ as hosting AGNs (SED AGNs; see also \citealt{2020ApJ...899...35T}). The median number of photometric points in the SED fitting is 21, and the number of degrees of freedom for the AGN module is 4. Thus, this criterion roughly corresponds to the improvement in reduced $\chi^2$ of 1. However, in some cases, this method gives unreasonable results, due to the limited parameter range of the SED model. Thus, we also visually examine the SEDs and conclude that there is no evidence of AGNs in the SED of AS2COS0001.2, where both mid-infrared excess and a flat far-infrared SED are not confirmed. Finally, we identify 24 SED AGNs in the AS2COSMOS sample, where seven sources are also detected in X-ray. 

Figure~\ref{figure:deltaBIC} shows the histogram of $\Delta \mathrm{BIC}$. Notably, most of the X-ray detected sources (16 out of 23) exhibit BICs lower than 10. This shows that the SED AGNs provide a complementary AGN sample to the X-ray AGNs in our sample (see Section~\ref{subsubsection:x-ray_LFIR}). In addition, all three type-1 SED AGNs (AS2COS0019.1, AS2COS0175.1, and AS2COS0230.1) are detected in X-ray, which is consistent with the unobscured nature of those systems. We emphasize that this method strongly depends on the quality of the multi-wavelength photometry and model assumptions. In particular, at high redshift ($z\geq3$) or in optical(/near-infrared)-dark systems, the indicators of AGNs are sometimes limited to a slight excess at observed-frame 24-$\mu$m flux density or relatively flat SED at observed-frame 100--1000 $\mu$m, which are difficult to distinguish from the PAH emission from the host-galaxy dust or hot dust emission associated with a starburst. In our SED AGN sample, the evidence for AGNs in AS2COS0025.1, AS2COS0084.1, AS2COS0099.1, AS2COS0108.2, and AS2COS0330.3 appear to be fairly limited and we should be careful of these sources. For sources other than the SED AGNs, we estimate the 99.73 per cent (3$\sigma $) upper bound of the bolometric AGN luminosity by integrating the marginalized PDFs with a uniform prior. Except for the SED AGNs, the physical properties of the galaxies are calculated without the AGN component.

\subsubsection{SED Modeling Check}\label{subsubsection:validation}
 
For a consistency check, we compare some observational properties with the physical properties derived by the SED modeling that they are expected to most strongly correlate with. Figure~\ref{figure:consistency_check} in Appendix~\ref{appendix:sed_consistency} presents these comparisons. We confirm a strong correlation between dust mass and 870-$\mu$m flux density as reported in \citet{2020MNRAS.494.3828D}. Additionally, we confirm that the 5.8 $\mu$m flux density, which corresponds to the rest-frame H-band flux density at the median redshift of $z=2.47$, shows a positive correlation with the stellar mass. Moreover, the photometric redshifts derived by {\sc cigale} are reasonably correlated with the spectroscopic redshifts; the median difference and the standard deviation are $\Delta z/(1+z) = (z_{\mathrm{photo}}-z_{\mathrm{spec}})/z_{\mathrm{spec}} = -0.04\pm 0.19$. Note that the photometric redshifts derived by {\sc cigale} are also reasonably aligned with those derived by {\sc eazy} and {\sc lephare}, which is listed in the COSMOS2020 catalog; the median differences and the standard deviations are $\Delta z/(1+z) = (z_{\text{\sc eazy/lephare}}-z_{\text{\sc cigale}})/z_{\text{\sc cigale}} = 0.14\pm 0.62$ for {\sc eazy} and $0.09\pm0.41$ for {\sc lephare}, respectively. Determining the physical properties of galaxies with SED analysis are sometimes challenging or unreliable, particulary for a type-1 SED AGN where the AGN emission can mask the host-galaxy emission. To address this issue, we perform a mock analysis, which is a procedure provided by {\sc cigale} (see Section 4.3 in \citealt{2019A&A...622A.103B}), for all the AS2COSMOS sources (Appendix~\ref{appendix:sed_degeneracy}). The median differences and the standard deviations are $\Delta \log \mathrm{SFR}=\log \mathrm{SFR(mock)} - \log \mathrm{SFR} = -0.01\pm0.19$ and $\Delta \log M_*=\log M_*\mathrm{(mock)} - \log M_* = 0.02\pm0.19$, respectively.

\begin{figure*}
  \centering
  \includegraphics[width=\linewidth]{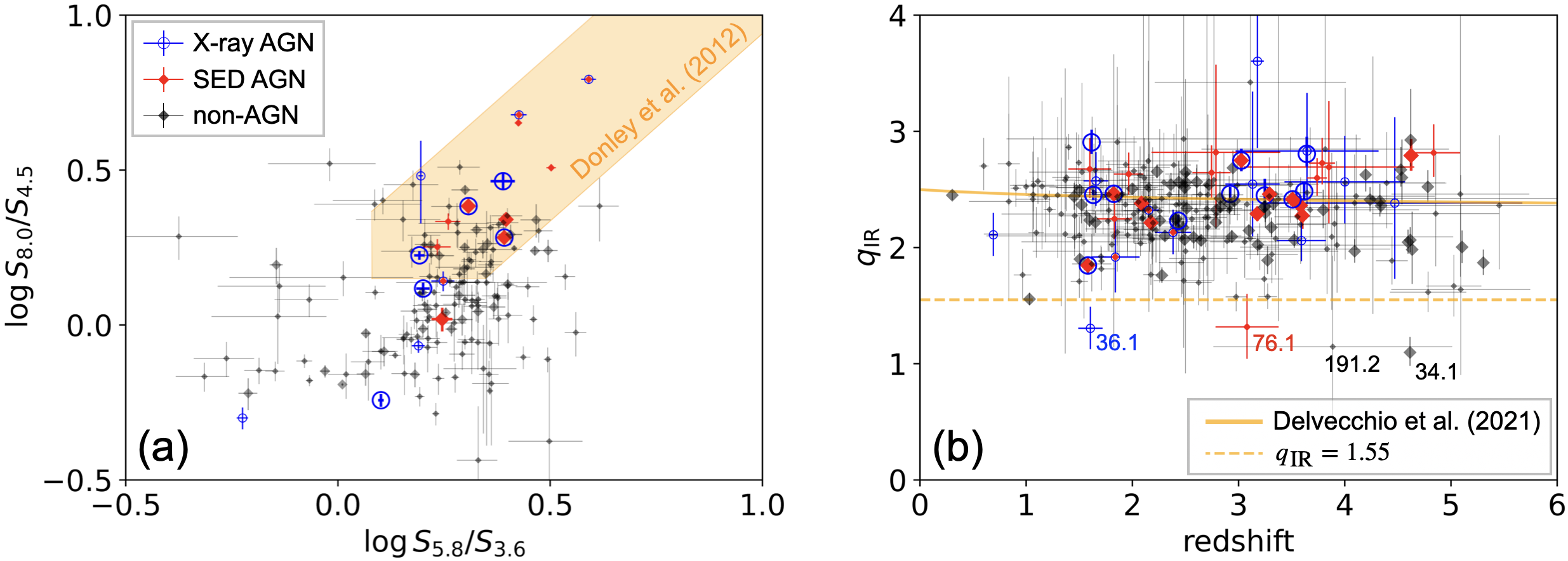}
  \caption{\textit{(a)} 
  Mid-infrared color-color diagram of the AS2COSMOS sources. The shaded area shows the AGN selection criteria by \citet{2012ApJ...748..142D}. Note that only the sources that are detected in all the mid-infrared bands (Spitzer 3.6 $\mu$m, 4.5 $\mu$m, 5.8 $\mu$m, and 8.0 $\mu$m) are plotted in this figure. Moreover, the sample is limited to $z<3$, where the selection criteria by \citet{2012ApJ...748..142D} are applicable. We confirm that most of the SED AGNs meet the AGN criteria by \citet{2012ApJ...748..142D}. \textit{(b)} Distribution of $q_{\mathrm{IR}}$ as a function of redshift. The solid line illustrates the empirical relation of star-formation galaxies at $M_*=10^{11}M_{\odot}$ \citep{2021A&A...647A.123D}, while the dashed line indicates $q_{\mathrm{IR}}=1.55$, which is used as the threshold of radio-excess AGNs in \citet{2020ApJ...903..138A}. The large symbols show the spec-$z$ sample, while the smaller ones show the photo-$z$ sample. We confirm that most of the AS2COSMOS sources follow the empirical relation of star-forming galaxies, while two AGNs and two non-AGNs show radio-loud characteristics.}
  \label{figure:agn_selection}
\end{figure*}

To validate our AGN selection, we compare the properties of our sources to the conventional AGN selection using a color-color diagram in the mid-infrared bands. Figure~\ref{figure:agn_selection} (a) displays the mid-infrared color-color diagram of the AS2COSMOS sources. The orange area denotes the AGN criteria by \citet{2012ApJ...748..142D}. Note that only the sources detected in all the mid-infrared bands (Spitzer 3.6 $\mu$m, 4.5 $\mu$m, 5.8 $\mu$m, and 8.0 $\mu$m) are plotted in this figure. We also limit the sample to $z<3$ because the stellar emission can produce a similar shape as the AGN emission at high redshift \citep{2012ApJ...748..142D}. We confirm that 9/11 of the SED AGNs are located in the AGN region, while two SED AGNs (AS2COS0025.1 and AS2COS0230.1) are not. AS2COS0025.1 shows a possible excess in the Spitzer 24-$\mu$m band, which appears to drive the AGN classification. AS2COS0230.1 has a flat rest-frame optical SED, which may favor the presence of an AGN.

A significant fraction of galaxies in the AGN region (40/49) are not classified as hosting AGNs by the BIC selection. 24 out of these 40 AGN-unclassified galaxies in the AGN region have relatively high redshifts ($z\gtrsim2.4$), which may make it difficult to distinguish AGN-host galaxies from dusty star-forming galaxies in the mid-infrared color-color diagram \citep{2019MNRAS.487.4648S}. The {\sc cigale} fits for the other 13 sources with lower redshift ($1.0\lesssim z\lesssim2.0$) are suggested to have strong 3.3 $\mu$m PAH emission, which may caused the misidentification in the mid-infrared color-color diagram. However, we need to consider the possibility that the near-infrared AGN emission is misidentified as the 3.3 $\mu$m PAH emission in the SED fitting. The remaining three sources lie near the edge of the selection criteria. In summary, we conclude that the AGN selection by BIC provides a more conservative sample of SED AGNs than the conventional selection using the mid-infrared color-color diagram at $z<3$.

We also check the radio loudness of the SMGs. Figure~\ref{figure:agn_selection} (b) plots the radio-IR correlation factors as a function of the redshifts. The solid and dashed lines denote the empirical relation of star-formation galaxies at $z=0\text{--}4$ \citep{2021A&A...647A.123D} and a criterion for radio-excess AGNs ($q_{\mathrm{IR}}\leq1.55$; \citealt{2020ApJ...903..138A}), respectively. We confirm that our sample mostly aligns with the empirical relation of star-forming galaxies. One X-ray detected AS2COSMOS source and an X-ray undetected SED AGN have especially low radio-IR correlation factors ($q_{\mathrm{IR}}=1.3\pm0.2$ and $1.3\pm0.3$ for AS2COS0036.1 and AS2COS0076.1, respectively), suggesting that these sources are radio-excess AGNs. Two galaxies that are not detected in X-ray nor identified as SED AGNs also have low radio-IR correlation factors ($q_{\mathrm{IR}}=1.1\pm0.2$ and $1.1^{+0.8}_{-0.7}$ for AS2COS0034.1 and AS2COS0191.2, respectively). This might suggest that they are candidates for radio-excess AGNs. These sources are located at high redshift, which may have made it difficult to identify AGNs by SED modeling. In total, the AS2COSMOS sample contains 4/258 ($1.5^{+1.2}_{-0.7}$ per cent) radio-excess sources. This fraction is consistent with the previous studies of the AS2UDS sample (12/659; $1.8\pm0.5$ per cent; \citealt{2020ApJ...903..138A}).

The 18 brightest and not strongly lensed AS2COSMOS sources ($S_{\mathrm{870}\, \mu \mathrm{m}}=12.4\text{--}19.2$) were also analyzed in \citet{2024ApJ...961..226L}. They performed X-ray to radio SED modeling with {\sc cigale}, and found that the luminosity fractions of AGNs in the total infrared luminosities ($f_{\mathrm{AGN}}$) are always lower than ${\sim}10$\% in their sample. However, in our study, some of these sources have moderately high AGN luminosity fraction ($f_{\mathrm{AGN}} = 0.56\pm0.13,\,0.38\pm0.05,\,0.55\pm0.13,\,\mathrm{and}\,0.49\pm0.03$ for AS2COS0001.1, AS2COS0008.1, AS2COS0054.1, and AS2COS0139.1, respectively), indicating significant contributions of AGNs in those systems. These inconsistencies can be attributed to the different parameter settings used for {\sc cigale}. In particular, since the radio photometry of some of the AS2COSMOS sources has a high signal-to-noise ratio, the radio loudness parameter can sometimes behave as the normalization for the AGN component. In \citet{2024ApJ...961..226L}, the radio loudness parameter was sampled at one dex intervals in the fitting. Thus $f_{\mathrm{AGN}}$ might not be estimated with an accuracy of less than one dex, which might cause potential trouble in the identification of AGNs. In addition, care needs to be taken about the possible degeneracies between the parameters in SED modules. For example, the host-galaxy dust emission module have a potential degeneracy with the AGN module. Thus, if we limit parameter range of the host-galaxy dust emission module, the contribution of the AGN module can be overestimated. Currently, we are not able to calculate all the possible parameter grids and test the degeneracies between the SED modules due to the computational cost. Future studies on the SED modules themselves are needed to solve this problem.

\subsubsection{Selection Bias of SED AGNs}\label{subsection:bias}

Detection of type-2 AGNs using SED modeling mostly occurs due to the dominance of AGN emission in the infrared bands. Consequently, it is more difficult to identify an AGN in a strongly star-forming galaxy because the host-galaxy emission will mask the emission from the AGN (see also \citealt{2022MNRAS.517.2577A}). This selection can be characterized by the AGN luminosity fraction in the total infrared luminosity ($f_{\mathrm{AGN}}$). In our sample, the type-2 SED AGNs have $f_{\mathrm{AGN}}$ larger than ${\sim}0.3$, which is considered to be the detection limit of the SED AGNs in our study. Assuming a typical SED of a type-2 AGN and the tight SFR-$L_{\mathrm{IR}}$ correlation, this detection limit can be expressed as follows:
\begin{eqnarray}
  \mathrm{SFR}\,[M_{\odot}\,\mathrm{yr}^{-1}]
  &\gtrsim& \frac{1 - f_{\mathrm{AGN}}}{f_{\mathrm{AGN}}} \times \kappa_{\mathrm{IR}}^{-1} \times L_{\mathrm{AGN,\,bol}}\,[\mathrm{erg\,s^{-1}}]\nonumber\\
  &\times& \kappa \prime_{\mathrm{IR}}\nonumber\\
  &=& 4.2 \times 10^{-44} \times L_{\mathrm{AGN,\,bol}}\,[\mathrm{erg\,s^{-1}}]
\end{eqnarray}
where $\kappa_{\mathrm{IR}}$ is the infrared to bolometric correction factor of an AGN, and $\kappa\prime_{\mathrm{IR}}^{-1}$ is the conversion factor of infrared luminosity to SFR. Here we assume $\kappa_{\mathrm{IR}}=1.6$, which is calculated by the {\sc skirtor} model with $\tau_{9.7}=7$, $p=1$, $q=1$, $\Delta=40\degr$, $R=20$, $\theta=70\degr$, $E(B-V)=0.1$, and $T_{\mathrm{pol}}=100\,\mathrm{K}$. For $\kappa \prime_{\mathrm{IR}}$, we employ the calibration by \citet{1998ApJ...498..541K}, where a factor of 0.63 is applied to convert Salpeter IMF to Chabrier IMF. Note that this detection limit is not as applicable to type-1 AGNs, because type-1 AGNs can also be selected by their flat UV to optical spectra. Moreover, the detection of AGNs by SED modeling is expected to have a potential bias against high-redshift ($z>3$) AGNs unless 24-$\mu$m flux densities are available, since the hot dust emission from an AGN is redshifted out of the mid-infrared bands ($3.6\text{--}8.0$ $\mu$m) and can only be identified by the 24-$\mu$m flux density.

\subsection{X-ray Analysis}\label{subsection:x-ray_spec}

\subsubsection{X-ray Spectral Analysis for X-ray Detected Sources}\label{subsubsection:x-ray_spec_bright}

We perform an X-ray spectral analysis of the X-ray detected AS2COSMOS sources. To appropriately treat the low-count statistics, we use the C statistic \citep{1979ApJ...228..939C}. We first construct a background model utilizing the combined background spectrum. We employ an analytical model referring to \citet{2021A&A...655A.116S}, but including a simplification to avoid unreasonable fits. The background model used in this study is expressed as follows in the {\sc xspec} terminology \citep{1996ASPC..101...17A}:
\small
\begin{eqnarray}
  \mathrm{model\_back} &=& \texttt{gaussian} + \texttt{gaussian} + \texttt{gaussian}\\
  &+& \texttt{gaussian} + \texttt{gaussian} + \texttt{gaussian}\nonumber\\
  &+& \texttt{gaussian} + \texttt{gaussian} + \texttt{gaussian}\nonumber\\
  &+& \texttt{gaussian}\nonumber\\
  &+& \texttt{powerlaw} + \texttt{gabs} * \texttt{expdec}\nonumber
\end{eqnarray}
\normalsize
The nine Gaussian lines represent the fluorescence lines from the instrument: Al K$\alpha$ (1.487 keV), Al K$\beta$ (1.557 keV), Ni K$\alpha$ (7.478 keV), Ni K$\beta$ (8.265 keV), Au M$\alpha$ (2.123 keV), Au M$\beta$ (2.205 keV), Au M$\gamma$ (2.410 keV), Au L$\alpha_1$ (9.713 keV), and Au L$\alpha_2$ (9.628 keV). The widths of these lines are fixed at 1 eV. The remaining Gaussian line corresponds to the broad line component at 2.7 keV produced by the inappropriate correction of charge transfer inefficiency. The continuum background spectrum is approximated by a power-law and an exponential decay with Gaussian-profile absorption. We note that this model is not multiplicated by the ancillary response file, assuming that the background spectrum is dominated by particle-induced events. We confirm that this model adequately reproduces the combined background spectrum ($\mathrm{C\text{-}statistic}=826$ for 634 degrees of freedom). The best-fit parameters are summarized in Table~\ref{table:modelback} in Appendix~\ref{appendix:background}. In the later sections, we fix the parameters except for the normalizations of additive components, which are left as free parameters to treat the slight differences among the observations.

\begin{figure*}
  \centering
  \includegraphics[width=\linewidth]{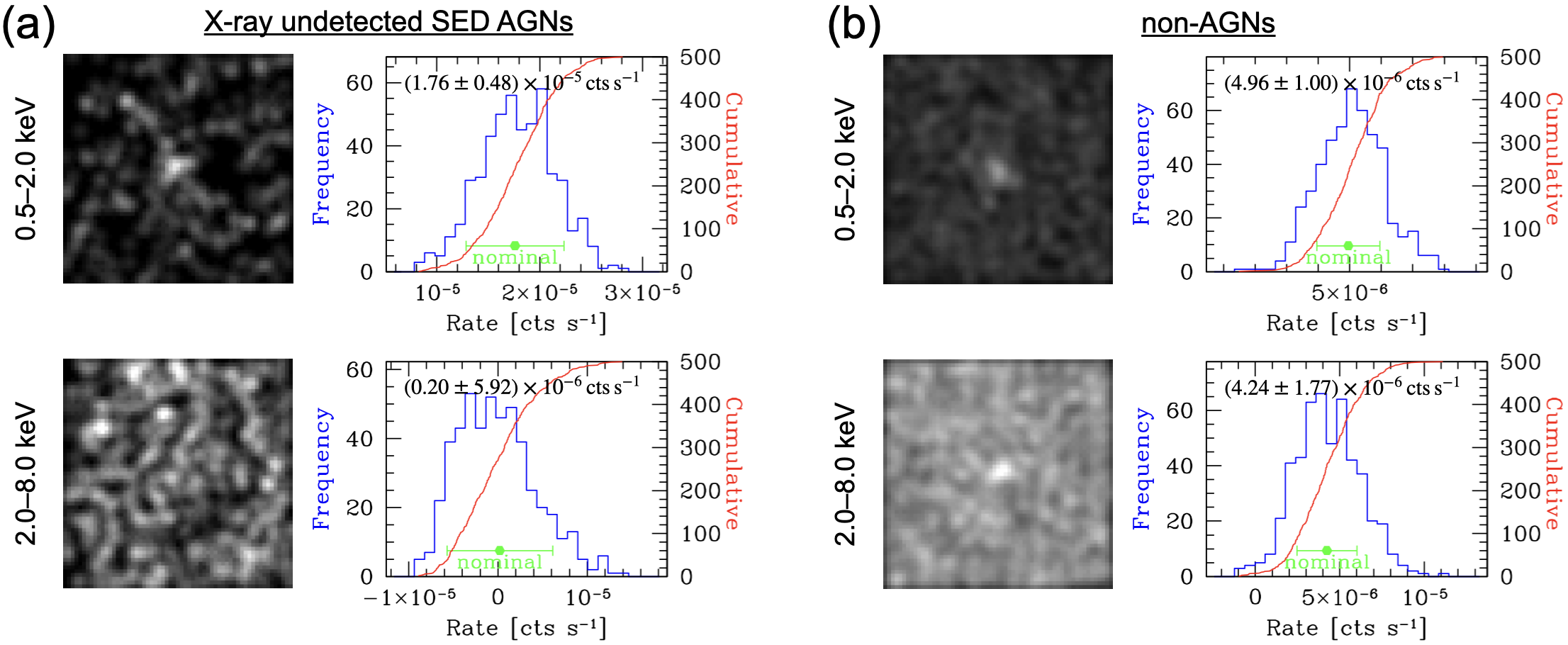}
  \caption{
  Results of X-ray stacking analysis for \textit{(a)} the X-ray undetected SED AGNs and \textit{(b)} the other X-ray undetected AS2COSMOS sources. The left panels show the stacked X-ray images, while the right panels show the bootstrap histograms of the net count rates. The mean count rates and the confidence intervals are shown in green and also reported in the histograms. Both samples are detected in the soft band (0.5--2.0 keV) over 3$\sigma$.}
  \label{figure:xstack}
\end{figure*}

Adopting this approach, we simultaneously fit the source and the background spectra. Here, we assume that the source X-ray spectra are dominated by AGN emission. The X-ray spectrum of an AGN is approximated to three components: (1) a direct component from the nucleus, (2) reflection components from the torus and/or the accretion disk, and (3) a scattered component with emission from the photoionized plasma. The photoionized emission is usually much fainter than the direct or the reflection components at rest-frame energies above 2.0 keV. Thus we do not consider this component in the spectral analysis. The reflection from an accretion disk is also excluded because its contribution to the X-ray spectrum is still under debate \citep{2019ApJ...875..115O,2021ApJ...906...84O}. The combined spectrum is expressed as follows in the {\sc xspec} terminology:
\small
\begin{eqnarray}
  \mathrm{model\_source} &=& \texttt{phabs} * (\texttt{zphabs}*\texttt{cabs}*\texttt{zcutoffpl}\\
  &+& \texttt{const}*\texttt{zcutoffpl}\nonumber\\
  &+& \texttt{atable\{xclumpy\_v01\_RC.fits\}}\nonumber\\
  &+& \texttt{atable\{xclumpy\_v01\_RL.fits\}})\nonumber
\end{eqnarray}
\normalsize
The first factor represents the galactic absorption, which is calculated following the method of \citet{2013MNRAS.431..394W}. The first term in the parenthesis represents the direct component. The second term corresponds to the unabsorbed scattered component. The third and fourth terms are the reflection component from the torus. In this study, we employ the {\sc xclumpy} model, which reproduces the X-ray reflection from a clumpy torus \citep{2019ApJ...877...95T}. The photon index, cutoff energy, and normalization of the reflection and scattered components are linked to those of the direct component. The line-of-sight hydrogen column density ($N_{\mathrm{H,X}}^{\mathrm{LOS}}$) of the direct component is set to be consistent with the torus geometry (see Equation~3 in \citealt{2019ApJ...877...95T}). Redshifts are fixed at the spectroscopic ones or the Bayesian estimated values derived by the SED modeling. Because of the modest data quality, scattered fraction, photon index, and cutoff energy are fixed at 0.01, 1.9 and 370 keV, respectively. Note that the photon index is confirmed to be unconstrained except for AS2COS0019.1, where AS2COS0019.1 has fits in which a photon index of 1.9 lies within the 90 per cent confidence interval. Moreover, the torus angular width ($\sigma$) and the equatorial hydrogen column density ($N_{\mathrm{H,X}}^{\mathrm{Equ}}$) are fixed at 20 degrees and $10^{25}\,\mathrm{cm}^{-2}$, respectively. Finally, the free parameters in the X-ray spectral fitting are the inclination angle ($i$; 18--87 degrees), which is linked to the line-of-sight hydrogen column density, and the normalization of the direct component.

With this approach, we successfully reproduce the X-ray spectra of the 23 X-ray detected sources with an AGN model (C-statistic $< 1140$ for 1285 degrees of freedom). The X-ray spectra and the plots of the goodness of fit as a function of the line-of-sight hydrogen column densities are summarized in Appendix~\ref{appendix:figure_detail}. We confirm that all the sources have solutions in the Compton-thin regime ($N_{\mathrm{H,X}}^{\mathrm{LOS}}<10^{24.2}\,\mathrm{cm}^{-2}$), with 12/23 sources showing significant absorption ($N_{\mathrm{H,X}}^{\mathrm{LOS}}>10^{22}\,\mathrm{cm}^{-2}$). Note that seven SED AGNs are detected in X-rays, from which three are classified as type-1 and four are classified as type-2 from the SED analysis. All the X-ray detected type-2 SED AGNs show significant absorption ($N_{\mathrm{H,X}}^{\mathrm{LOS}}>10^{22}\,\mathrm{cm}^{-2}$), and two of the X-ray detected type-1 SED AGNs do not. This shows the consistency of the X-ray analysis and the SED analysis of these sources. However, AS2COS0019.1 shows significant absorption in X-rays ($\log N_{\mathrm{H,X}}^{\mathrm{LOS}}/\mathrm{cm}^{-2}=22.4^{+0.1}_{-0.2}$), while it is classified as a type-1 SED AGN. This might be attributed to the time variability of the AGN or the uncertainties associated with the model assumptions. AS2COS0036.1, AS2COS0055.2, AS2COS0066.1, AS2COS0076.2, AS2COS0095.2, AS2COS0153.2, AS2COS0220.1, and AS2COS0275.2 present alternative solutions in the Compton-thick regime which are not distinguishable from the Compton-thin solutions at a 90 per cent confidence level. In later sections, we investigate these solutions separately to consider both possibilities. For the Compton-thick case, we calculate the X-ray luminosities by fixing the line-of-sight hydrogen column density\footnote{This is the hard limit of the {\sc xclumpy} model (v1).} to $N_{\mathrm{H,X}}^{\mathrm{LOS}}=9.8\times10^{24}\,\mathrm{cm}^{-2}$. The X-ray luminosities are calculated by correcting the line-of-sight absorption and the reflection from the torus. The median X-ray luminosity is estimated as $L_{\mathrm{2\text{--}10\,\mathrm{keV}}}=2.6\times 10^{44}\,\mathrm{erg\,s^{-1}\,cm^{-2}}$ for the Compton-thin cases, while $L_{\mathrm{2\text{--}10\,\mathrm{keV}}}=3.1\times 10^{45}\,\mathrm{erg\,s^{-1}\,cm^{-2}}$ for the Compton-thick cases. Note that the X-ray spectral analysis of the sources in the COSMOS field is also performed by \citet{2023MNRAS.518.2546L}. We confirm that our estimations of the line-of-sight hydrogen column densities and the X-ray luminosities are almost consistent with those in \citet{2023MNRAS.518.2546L} within the errors.

\subsubsection{Upper Bound Estimation for X-ray-Undetected Sources}\label{subsubsection:x-ray_spec_undetected}

For X-ray undetected SED AGNs (Section~\ref{subsection:sed}), we convert the 3$\sigma$ upper bound count rates to the intrinsic X-ray luminosities by assuming a typical AGN spectrum. We employ the same model as in Section~\ref{subsubsection:x-ray_spec_bright}, but the inclination angle is fixed to either $i=47$ degrees ($N_{\mathrm{H,X}}^{\mathrm{LOS}}=10^{23}\,\mathrm{cm}^{-2}$) or $i=60$ degrees ($N_{\mathrm{H,X}}^{\mathrm{LOS}}=10^{24}\,\mathrm{cm}^{-2}$) assuming Compton-thin absorptions. We also calculate the Compton-thick case by fixing the inclination angle to $i=87$ degrees ($N_{\mathrm{H,X}}^{\mathrm{LOS}}=9.8\times10^{24}\,\mathrm{cm}^{-2}$). Note that all the type-1 SED AGNs are detected with X-rays, so the X-ray upper bounds are only calculated for type-2 AGNs. The X-ray luminosities are calculated by correcting the line-of-sight absorption and the reflection from the torus. The median X-ray luminosity upper bound is estimated as $L_{\mathrm{2\text{--}10\,\mathrm{keV}}}=1.5\times 10^{44}\,\mathrm{erg\,s^{-1}\,cm^{-2}}$ for $N_{\mathrm{H,X}}^{\mathrm{LOS}}=10^{23}\,\mathrm{cm}^{-2}$, $L_{\mathrm{2\text{--}10\,\mathrm{keV}}}=6.5\times 10^{44}\,\mathrm{erg\,s^{-1}\,cm^{-2}}$ for $N_{\mathrm{H,X}}^{\mathrm{LOS}}=10^{24}\,\mathrm{cm}^{-2}$, and $L_{\mathrm{2\text{--}10\,\mathrm{keV}}}=4.7\times 10^{45}\,\mathrm{erg\,s^{-1}\,cm^{-2}}$ for $N_{\mathrm{H,X}}^{\mathrm{LOS}}=10^{25}\,\mathrm{cm}^{-2}$.

\subsubsection{Stacking Analysis of X-ray Undetected Sources}\label{subsection:x-ray_stack}

We conduct an X-ray stacking analysis for the X-ray undetected AS2COSMOS sources, using {\sc cstack v4.5} \citep{2008HEAD...10.0401M} with default settings for the maximum off-axis angle, the source region radius, and the exclusion of obvious X-ray sources. We separately stack the 17 X-ray undetected SED AGNs and the 220 AS2COSMOS sources that are neither detected in X-rays nor identified as SED AGNs (non-AGNs). Figure~\ref{figure:xstack} shows the results. Both samples are detected over 3$\sigma$ in the soft band (0.5--2.0 keV), but not in the hard band (2.0--8.0 keV). The mean count rate of the X-ray undetected SED AGN sample is about three times higher than that of the non-AGN sample. This might show the weak X-ray emission from the X-ray undetected SED AGNs. However, given that the median far-infrared luminosity of the X-ray undetected SED AGN sample is also about three times higher than that of the non-AGN sample, the difference does not necessarily trace the X-ray emission associated with the AGN activity. Moreover, the non-AGN sample is marginally detected in the 2.0--8.0 keV band (2.4$\sigma$), although the significance is not high. The inferred spectral shape is quite hard (photon index of ${\sim}1$), implying a significant absorption of these sources ($N_{\mathrm{H,X}}^{\mathrm{LOS}}\sim 7\times 10^{22}\,\mathrm{cm}^{-2}$ for the AGN model in Section~\ref{subsubsection:x-ray_spec_bright} at the median redshift of $z=2.45$). This provides marginal evidence for the existence of obscured AGNs in the non-AGN sample.

Utilizing the integrated count rate derived from our stacking analysis, we calculate the typical X-ray luminosity of the X-ray undetected SED AGN sample. Here, we assume the same X-ray spectral model as in Section~\ref{subsubsection:x-ray_spec_undetected} placed at the median redshift of the X-ray undetected SED AGN sample ($z=3.18$). Since all these AGNs are classified as type-2 (see Section~\ref{subsubsection:sed_procedure}), we set the inclination angle at either $i=47$ degrees ($N_{\mathrm{H,X}}^{\mathrm{LOS}}=10^{23}\,\mathrm{cm}^{-2}$) or $i=60$ degrees ($N_{\mathrm{H,X}}^{\mathrm{LOS}}=10^{24}\,\mathrm{cm}^{-2}$) assuming Compton-thin absorption. However, considering that the stacked X-ray undetected SED AGN sample is detected only in the soft band, we explore an unobscured AGN scenario by setting the inclination angle to $i=15$ degrees ($N_{\mathrm{H,X}}^{\mathrm{LOS}}=10^{20}\,\mathrm{cm}^{-2}$). In addition, we examine a Compton-thick scenario by fixing the inclination angle to $i=87$ degrees ($N_{\mathrm{H,X}}^{\mathrm{LOS}}=9.8\times10^{24}\,\mathrm{cm}^{-2}$). With this recipe, we estimate the stacked X-ray luminosity of the X-ray undetected SED AGNs as $L_{\mathrm{2\text{--}10\,keV}}=(1.5\pm0.4)\times 10^{43}\,\mathrm{erg\,s^{-1}}$ for $N_{\mathrm{H,X}}^{\mathrm{LOS}}=10^{20}\,\mathrm{cm}^{-2}$, $L_{\mathrm{2\text{--}10\,keV}}=(2.4\pm0.7)\times 10^{43}\,\mathrm{erg\,s^{-1}}$ for $N_{\mathrm{H,X}}^{\mathrm{LOS}}=10^{23}\,\mathrm{cm}^{-2}$, $L_{\mathrm{2\text{--}10\,keV}}=(2.4\pm0.6)\times 10^{44}\,\mathrm{erg\,s^{-1}}$ for $N_{\mathrm{H,X}}^{\mathrm{LOS}}=10^{24}\,\mathrm{cm}^{-2}$, and $L_{\mathrm{2\text{--}10\,keV}}=(1.1\pm0.3)\times 10^{45}\,\mathrm{erg\,s^{-1}}$ for $N_{\mathrm{H,X}}^{\mathrm{LOS}}=9.8\times10^{24}\,\mathrm{cm}^{-2}$, respectively. Moreover, we evaluate the X-ray luminosity due to star-formation activity from the integrated count rate of the non-AGN sample. In this calculation, we employ a power-law spectrum with a photon index of 2.1 \citep{2017MNRAS.468.2249S} placed at the median redshift of $z=2.45$. The stacked X-ray luminosity of the non-AGNs is estimated as $L_{\mathrm{2\text{--}10\,keV}}=(2.3\pm0.5)\times 10^{42}\,\mathrm{erg\,s^{-1}}$.

\begin{table*}
  \caption{
    Physical properties of X-ray AGNs and SED AGNs in AS2COSMOS SMG sample. (1), (2)\&(3) ALMA source ID and positions (same as \citealt{2020MNRAS.495.3409S}); (4) redshift (three decimal digits are for spectroscopic redshifts and two are for photometric redshifts); (5) stellar mass; (6) star-formation rate; (7) luminosity fraction of AGN in total infrared luminosity ($f_{\mathrm{AGN}}$); (8) change in BIC by the addition of an AGN component in the SED modeling; (9) bolometric AGN luminosity estimated by SED modeling (the uncertainties listed in this table are converted from those calculated in the linear space); (10) line-of-sight hydrogen column density (parameters that reach the model limits are indicated by ``lim''); (11) X-ray luminosity (the values in the parenthesis show the X-ray luminosities assuming Compton-thick absorption; $N_{\mathrm{H,X}}^{\mathrm{LOS}}=9.8\times 10^{24}\,\mathrm{cm}^{-2}$).
    The redshift uncertainties are not considered for the calculation of $L_{\mathrm{AGN,\,bol}}^{\mathrm{SED}}$, $N_{\mathrm{H,X}}^{\mathrm{LOS}}$, and $L_{\mathrm{2\text{--}10\,keV}}$. The upper bounds correspond to the 99.73 per cent confidence (3$\sigma$). Note that the reduced $\chi^2$ values of these AGNs are lower than 5 (well fitted).}
    \fontsize{5.9pt}{5.9pt}\selectfont
    \begin{tabularx}{\linewidth}{@{}ccccccccccc}
    \hline\hline
    ID & R.A. & Dec. & $z$ & Log$_{10}M_*$ & Log$_{10}\mathrm{SFR}$ & $f_{\mathrm{AGN}}$ & $\Delta \mathrm{BIC}$ & Log$_{10}L_{\mathrm{AGN,\,bol}}^{\mathrm{SED}}$ & Log$_{10}N_{\mathrm{H,X}}^{\mathrm{LOS}}$ & Log$_{10}L_{\mathrm{2\text{--}10\,keV}}$\\
    & [degrees] & [degrees] &  & $[M_{\odot}]$ & $[M_{\odot}\,\mathrm{yr}^{-1}]$ &  &  & $[\mathrm{ergs\,s^{-1}}]$ & $[\mathrm{cm^{-2}}]$ & $[\mathrm{ergs\,s^{-1}}]$\\
    (1) & (2) & (3) & (4) & (5) & (6) & (7) & (8) & (9) & (10) & (11)\\
    \hline
    \multicolumn{11}{c}{Only X-ray AGN}\\
    \hline
    AS2COS0014.1 & $150.42100$ & $2.06802$ & $2.921$ & $11.50\pm0.13$ & $2.89\pm0.05$ & \nodata & $-3.6$ & ${<}46.37$ & $23.4_{-0.4}^{+0.3}$ & $44.06_{-0.26}^{+0.25}$\\ 
    AS2COS0031.1 & $149.84586$ & $2.86042$ & $3.643$ & $11.24\pm0.12$ & $3.20\pm0.05$ & \nodata & $-3.0$ & ${<}46.24$ & $22.9_{\mathrm{lim}}^{+0.5}$ & $44.21_{-0.23}^{+0.22}$\\ 
    AS2COS0036.1 & $150.10616$ & $2.01437$ & $1.61\pm0.11$ & $11.69\pm0.07$ & $2.43\pm0.24$ & \nodata & $-0.7$ & ${<}46.39$ & $22.5_{-0.4}^{+0.3}$ & $43.63_{-0.14}^{+0.14}$\\ 
    AS2COS0055.2 & $150.32147$ & $1.71421$ & $0.69\pm0.03$ & $11.16\pm0.05$ & $0.54\pm0.31$ & \nodata & $-12.2$ & ${<}44.45$ & $21.9_{-0.4}^{+0.2}$ & $42.95_{-0.10}^{+0.10}$\\ 
    AS2COS0066.1 & $149.94410$ & $1.95418$ & $3.247$ & $11.07\pm0.15$ & $2.84\pm0.08$ & \nodata & $-10.6$ & ${<}46.08$ & $22.8_{\mathrm{lim}}^{+0.5}$ & $43.87_{-0.26}^{+0.25}$\\ 
    AS2COS0068.1 & $150.14291$ & $2.05061$ & $2.433$ & $11.39\pm0.02$ & $2.51\pm0.03$ & \nodata & $-1.1$ & ${<}46.37$ & $23.5_{-0.2}^{+0.1}$ & $44.61_{-0.13}^{+0.14}$\\ 
    AS2COS0069.1 & $149.77102$ & $2.36576$ & $3.618$ & $10.28\pm0.05$ & $3.27\pm0.02$ & \nodata & $-2.0$ & ${<}46.40$ & $23.8_{-0.3}^{+0.3}$ & $44.54_{-0.30}^{+0.30}$\\ 
    AS2COS0076.2 & $150.38353$ & $2.07446$ & $4.47\pm1.20$ & $10.30\pm0.37$ & $3.26\pm0.37$ & \nodata & $-4.1$ & ${<}45.42$ & $23.8_{-0.4}^{+0.4}$ & $44.74_{-0.32}^{+0.37}$\\ 
    AS2COS0095.2 & $150.12596$ & $2.69636$ & $1.66\pm0.12$ & $11.37\pm0.07$ & $2.12\pm0.20$ & \nodata & $-6.2$ & ${<}46.30$ & $22.2_{-1.3}^{+0.4}$ & $43.69_{-0.11}^{+0.11}$\\ 
    AS2COS0133.1 & $149.99979$ & $2.10924$ & $3.59\pm0.23$ & $10.46\pm0.31$ & $3.07\pm0.16$ & \nodata & $6.1$ & ${<}45.72$ & $23.7_{-0.2}^{+0.2}$ & $44.65_{-0.20}^{+0.19}$\\ 
    AS2COS0153.2 & $149.88157$ & $2.31819$ & $1.616$ & $10.84\pm0.18$ & $2.77\pm0.08$ & \nodata & $-5.1$ & ${<}45.96$ & $22.4_{-0.4}^{+0.3}$ & $43.74_{-0.11}^{+0.11}$\\ 
    AS2COS0220.1 & $150.00780$ & $1.68614$ & $4.00\pm0.57$ & $10.94\pm0.18$ & $3.03\pm0.18$ & \nodata & $-12.4$ & ${<}46.34$ & $23.8_{-0.3}^{+0.3}$ & $44.70_{-0.25}^{+0.27}$\\ 
    AS2COS0246.2 & $150.50454$ & $2.72503$ & $3.18\pm0.06$ & $10.35\pm0.04$ & $2.66\pm0.03$ & \nodata & $3.2$ & ${<}45.77$ & $22.7_{\mathrm{lim}}^{+0.4}$ & $44.28_{-0.13}^{+0.13}$\\ 
    AS2COS0272.1 & $150.31664$ & $1.60374$ & $3.65\pm0.67$ & $11.29\pm0.21$ & $3.22\pm0.23$ & \nodata & $-9.0$ & ${<}48.01$ & $22.8_{\mathrm{lim}}^{+0.4}$ & $44.42_{-0.17}^{+0.17}$\\ 
    AS2COS0275.2 & $149.95169$ & $1.74405$ & $1.632$ & $10.85\pm0.12$ & $2.92\pm0.04$ & \nodata & $-1.3$ & ${<}46.25$ & $22.6_{-0.7}^{+0.3}$ & $43.43_{-0.20}^{+0.19}$\\ 
    AS2COS0353.1 & $149.51390$ & $2.04452$ & $3.13\pm0.26$ & $9.54\pm0.20$ & $2.45\pm0.14$ & \nodata & $-8.3$ & ${<}46.41$ & $23.9_{-0.3}^{+0.3}$ & $44.51_{-0.35}^{+0.32}$\\ 
    \hline
    \multicolumn{11}{c}{X-ray AGN $\cap$ SED AGN}\\
    \hline
    AS2COS0019.1 & $150.15840$ & $2.13955$ & $1.825$ & $10.37\pm0.28$ & $3.15\pm0.08$ & $0.22\pm0.04$ & $74.9$ & $45.99\pm0.04$ & $22.4_{-0.2}^{+0.1}$ & $44.42_{-0.05}^{+0.05}$\\ 
    AS2COS0028.2 & $149.99795$ & $2.57821$ & $2.15\pm0.10$ & $10.14\pm0.25$ & $2.87\pm0.29$ & $0.29\pm0.17$ & $19.0$ & $46.42\pm0.03$ & $23.9_{-0.2}^{+0.2}$ & $44.48_{-0.27}^{+0.29}$\\ 
    AS2COS0048.1 & $150.28853$ & $2.38193$ & $1.581$ & $11.62\pm0.06$ & $1.91\pm0.06$ & $0.57\pm0.05$ & $11.0$ & $46.25\pm0.08$ & $22.6_{-0.2}^{+0.2}$ & $43.76_{-0.09}^{+0.10}$\\ 
    AS2COS0116.2 & $150.49037$ & $1.74637$ & $2.38\pm0.17$ & $10.85\pm0.23$ & $2.49\pm0.14$ & $0.61\pm0.04$ & $100.3$ & $46.75\pm0.05$ & $23.5_{-0.1}^{+0.1}$ & $44.73_{-0.10}^{+0.11}$\\ 
    AS2COS0175.1 & $150.73578$ & $2.19958$ & $3.509$ & $12.40\pm0.84$ & $1.78\pm1.34$ & $0.30\pm0.02$ & $151.3$ & $46.56\pm0.02$ & $20.0_{\mathrm{lim}}^{+2.6}$ & $44.59_{-0.07}^{+0.09}$\\ 
    AS2COS0230.1 & $149.61511$ & $1.78955$ & $1.84\pm0.23$ & $10.56\pm0.30$ & $2.71\pm0.17$ & $0.10\pm0.01$ & $14.8$ & $45.22\pm0.08$ & $22.2_{\mathrm{lim}}^{+0.6}$ & $43.72_{-0.15}^{+0.15}$\\ 
    AS2COS0308.1 & $150.60879$ & $2.76970$ & $3.026$ & $11.01\pm0.06$ & $3.22\pm0.02$ & $0.41\pm0.04$ & $33.3$ & $46.77\pm0.11$ & $22.6_{-0.6}^{+0.3}$ & $44.93_{-0.06}^{+0.06}$\\ 
    \hline
    \multicolumn{11}{c}{Only SED AGN}\\
    \hline
    AS2COS0001.1 & $150.03350$ & $2.43675$ & $4.625$ & $10.33\pm0.22$ & $3.19\pm0.12$ & $0.56\pm0.13$ & $57.2$ & $46.96\pm0.16$ & \nodata & ${<}44.89$\\ 
    AS2COS0008.1 & $150.70497$ & $2.54874$ & $3.581$ & $11.96\pm0.07$ & $2.27\pm0.09$ & $0.38\pm0.05$ & $15.8$ & $46.38\pm0.09$ & \nodata & ${<}44.93$\\ 
    AS2COS0024.1 & $150.13265$ & $2.21184$ & $2.176$ & $11.25\pm0.09$ & $2.54\pm0.14$ & $0.55\pm0.09$ & $37.9$ & $46.46\pm0.07$ & \nodata & ${<}44.47$\\ 
    AS2COS0025.1 & $150.16352$ & $2.37252$ & $2.086$ & $10.51\pm0.12$ & $2.70\pm0.05$ & $0.49\pm0.03$ & $13.3$ & $46.42\pm0.04$ & \nodata & ${<}44.41$\\ 
    AS2COS0054.1 & $149.69140$ & $2.72481$ & $3.176$ & $10.86\pm0.23$ & $2.88\pm0.15$ & $0.55\pm0.13$ & $24.3$ & $46.62\pm0.08$ & \nodata & ${<}44.66$\\ 
    AS2COS0076.1 & $150.38388$ & $2.07448$ & $3.08\pm0.30$ & $10.04\pm0.38$ & $2.49\pm0.17$ & $0.65\pm0.09$ & $31.3$ & $46.39\pm0.03$ & \nodata & ${<}45.07$\\ 
    AS2COS0078.3 & $150.44575$ & $2.41303$ & $2.75\pm0.18$ & $10.93\pm0.13$ & $2.35\pm0.16$ & $0.46\pm0.14$ & $13.8$ & $46.19\pm0.10$ & \nodata & ${<}45.04$\\ 
    AS2COS0084.1 & $149.58108$ & $2.25228$ & $3.600$ & $9.69\pm0.12$ & $2.64\pm0.06$ & $0.40\pm0.08$ & $21.6$ & $46.08\pm0.10$ & \nodata & ${<}44.90$\\ 
    AS2COS0099.1 & $149.92558$ & $1.77732$ & $1.60\pm0.20$ & $10.18\pm0.88$ & $2.04\pm1.62$ & $0.88\pm0.04$ & $27.2$ & $46.89\pm0.03$ & \nodata & ${<}44.30$\\ 
    AS2COS0108.2 & $150.09505$ & $1.86019$ & $3.85\pm0.80$ & $9.23\pm0.23$ & $2.16\pm0.19$ & $0.89\pm0.03$ & $14.9$ & $46.71\pm0.07$ & \nodata & ${<}44.81$\\ 
    AS2COS0123.1 & $149.59827$ & $2.20781$ & $4.84\pm0.25$ & $9.58\pm0.14$ & $2.53\pm0.08$ & $0.90\pm0.02$ & $16.4$ & $47.03\pm0.06$ & \nodata & ${<}45.00$\\ 
    AS2COS0139.1 & $149.58247$ & $2.60280$ & $3.292$ & $10.43\pm0.05$ & $3.42\pm0.02$ & $0.49\pm0.03$ & $131.2$ & $47.31\pm0.03$ & \nodata & ${<}44.75$\\ 
    AS2COS0189.1 & $149.49363$ & $2.28020$ & $2.79\pm0.61$ & $9.78\pm0.33$ & $2.62\pm0.28$ & $0.35\pm0.22$ & $30.1$ & $46.15\pm0.73$ & \nodata & ${<}44.76$\\ 
    AS2COS0231.1 & $149.78372$ & $2.81792$ & $1.97\pm0.14$ & $11.21\pm0.14$ & $2.31\pm0.14$ & $0.69\pm0.08$ & $44.8$ & $46.47\pm0.07$ & \nodata & ${<}44.48$\\ 
    AS2COS0261.1 & $149.64308$ & $1.91751$ & $1.83\pm0.16$ & $10.93\pm0.10$ & $2.17\pm0.11$ & $0.63\pm0.05$ & $12.6$ & $46.16\pm0.06$ & \nodata & ${<}44.59$\\ 
    AS2COS0330.3 & $150.76551$ & $1.73478$ & $3.74\pm0.09$ & $10.34\pm0.22$ & $2.74\pm0.15$ & $0.70\pm0.07$ & $27.4$ & $46.66\pm0.06$ & \nodata & ${<}45.05$\\ 
    AS2COS0362.1 & $149.69350$ & $1.72286$ & $3.79\pm0.12$ & $10.81\pm0.09$ & $2.50\pm0.12$ & $0.69\pm0.07$ & $31.6$ & $46.57\pm0.03$ & \nodata & ${<}44.89$\\
    \hline
  \end{tabularx}\label{table:results}
\end{table*}
\normalsize

\subsection{Morphological Study}\label{subsection:morphology}

Utilizing high-resolution imaging with JWST, we investigate the visual morphology of the 105/258\footnote{The strongly lensed system (AS2COS0005.1 and AS2COS0005.2) is excluded.} AS2COSMOS sources in the coverage of NIRCam imaging (see also \citealt{2024arXiv240808346M} for a similar morphological study in COSMOS). We follow the same approach as the previous study of bright SMGs in the PRIMER region by \citet{2024arXiv240603544G}. We assess the presence of distorted morphology, asymmetric structures, or tidal features by the color-composite images of the F115W, F150W, F277W, and F444W bands. Additionally, we check for the presence of possible companions in the F444W images. We classify sources with strongly disturbed morphology, tidal features, or potential bright companions (within a factor of four brightness of the target galaxy) as ``major'' merger candidates. Sources with less disturbed morphology or fainter companions are classified as ``minor'' merger candidates. The color-composite images of the 105 AS2COSMOS sources in the coverage of NIRCam are summarized in Figure~\ref{figure:JWST_nircam}. We identify 62/105 ($59^{+5}_{-6}$\%) as potential merger candidates, from which 33/105 ($31\pm5$\%) are classified as major mergers. These fractions are broadly consistent with the previous work by \citet{2024arXiv240603544G}. Note that we find six systems where both of the merging galaxies are included in the AS2COSMOS sample (AS2COS0036.1/36.2, AS2COS0048.1/48.3, AS2COS0076.1/76.2, AS2COS0143.2/143.2, AS2COS0228.1/228.2, and AS2COS0239.1/239.2). If these systems are counted once, the merging fractions are 56/99 ($57^{+5}_{-6}$\%) and 28/99 ($28\pm5$\%) for all merger candidates and major mergers, respectively. We stress that the possible companions are not spectroscopically confirmed and some of them may be foreground or background galaxies.

We also study the F770W-band morphology of the 38/258 AS2COSMOS sources in the coverage of both NIRCam and MIRI imaging (Figure~\ref{figure:JWST_miri} in Appendix~\ref{appendix:miri}), which is considered to better trace obscured AGN activitiy. We find that AS2COS0019.1 and AS2COS0048.1 have point-like morphology in the MIRI/F770W band. These sources are confirmed to host AGNs by SED analysis, and their mid-infrared emission are expected to be dominated by the AGN component. Thus, the point-like morphology in the MIRI/F770W band is consistent with AGN emission, as predicted by the SED analysis.

\section{Results and Discussion}\label{section:results}

In this section, we discuss the physical properties of the X-ray detected and SED AGN SMGs including a comparison of their far-infrared and X-ray luminosities, their SED-derived host galaxy properties, X-ray absorption and X-ray versus bolometric luminosities of the AGNs and their SFRs (Sections~\ref{subsubsection:x-ray_LFIR}--\ref{subsection:sfr_lbol}), before ending on a discussion of the population statistics of AGN in SMGs and their connection to dynamical interactions as indicated by JWST imaging. The sample of the X-ray and SED AGNs are summarized in Table~\ref{table:results}. The complete table including infrared luminosity, dust luminosity, dust mass, dust temperature, and color excess of dust attenuation for stellar emission is provided as supplemental material. The SEDs and the best-fit models of the SED AGNs are summarized in Appendix~\ref{appendix:figure_detail}. We also show the X-ray spectra and the line-of-sight hydrogen column densities of the 23 X-ray detected SMGs in Appendix~\ref{appendix:figure_detail}.

\begin{figure*}
  \centering
  \includegraphics[width=0.5\linewidth]{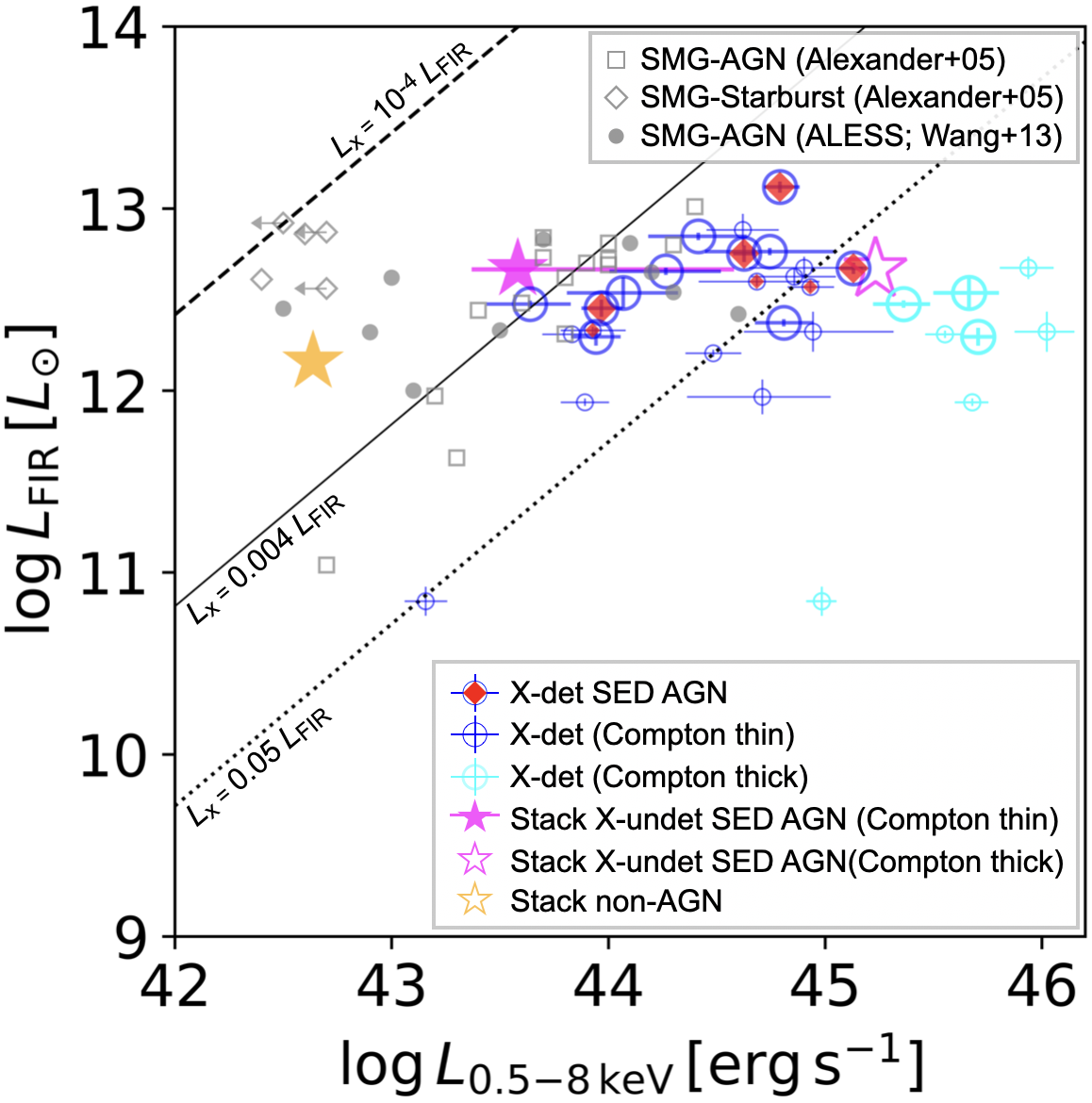}
  \caption{
  Comparison between far-infrared luminosity (rest-frame 40--120 $\mu$m) and X-ray (rest-frame 0.5--8 keV) luminosity for X-ray detected (X-det) SMGs. For the AS2COSMOS sources, we apply a factor of 1/1.91 to convert total infrared luminosity to far-infrared luminosity \citep{2012A&A...539A.155M}. In addition, we convert the rest-frame 2--10 keV luminosity to rest-frame 0.5--8 keV luminosity assuming an intrinsic power-law spectrum with a photon index of 1.9. The stacked values of X-ray undetected (X-undet) SED AGNs and non-AGNs are also plotted. The solid and dashed lines show the empirical relations of AGN-classified SMGs and starburst-classified SMGs, respectively \citep{2005ApJ...632..736A}. The dotted line shows the empirical relation of local quasars \citep{1994ApJS...95....1E}. The large symbols show the spec-$z$ sample, while the smaller ones show the photo-$z$ sample. The redshift uncertainties are not considered in these plots. The X-ray detected AS2COSMOS sources have more than one dex higher X-ray luminosities than those expected from the empirical relation of starburst-classified SMGs, suggesting that the X-ray emission is dominated by AGNs.}
  \label{figure:Lx_LFIR}
\end{figure*}

\subsection{Comparison of Far-infrared Luminosity and X-ray Luminosity}\label{subsubsection:x-ray_LFIR}

Figure~\ref{figure:Lx_LFIR} compares the rest-frame far-infrared (40--120 $\mu$m) luminosities and the X-ray luminosities for the X-ray detected AS2COSMOS sources. We also plot the empirical relation for local quasars \citep{1994ApJS...95....1E}, AGN-classified SMGs, and starburst-classified SMGs \citep{2005ApJ...632..736A}. We confirm that the X-ray detected AS2COSMOS sources have more than one dex higher X-ray luminosities than those expected from their star-formation activity ($L_{\mathrm{X}}=10^{-4}L_{\mathrm{FIR}}$). This suggests that the X-ray emission is dominated by AGNs, which verifies the assumption that the X-ray detected AS2COSMOS sources host AGNs. Thus, we regard these sources as ``X-ray AGNs'' in the following sections. Note that the X-ray to far-infrared luminosity correlation in galaxies without AGNs has also been studied by a recent Chandra study in the Great Observatories All-Sky LIRG survey \citep{2018A&A...620A.140T}. According to their relation, even galaxies with $L_{\mathrm{FIR}}=10^{46}\,\mathrm{erg\, s^{-1}}$ (${\sim}10^{12.4} L_{\odot}$) cannot be brighter than $L_{\mathrm{X}}=10^{42}\,\mathrm{erg\,s^{-1}}$. Thus, we may conservatively say that the X-ray emission in our X-ray AGN sample is dominated by the AGN activities. The X-ray AGN sample exhibits about a median of 6 times higher X-ray luminosities than the empirical relation of AGN-classified SMGs ($L_{\mathrm{X}}=0.004L_{\mathrm{FIR}}$), assuming Compton-thin absorption. This can be attributed to the selection using the relatively shallow X-ray observations in the COSMOS field, as compared with the deeper Chandra observations employed in \citet{2005ApJ...632..736A}.

\begin{figure*}
  \centering
  \includegraphics[width=\linewidth]{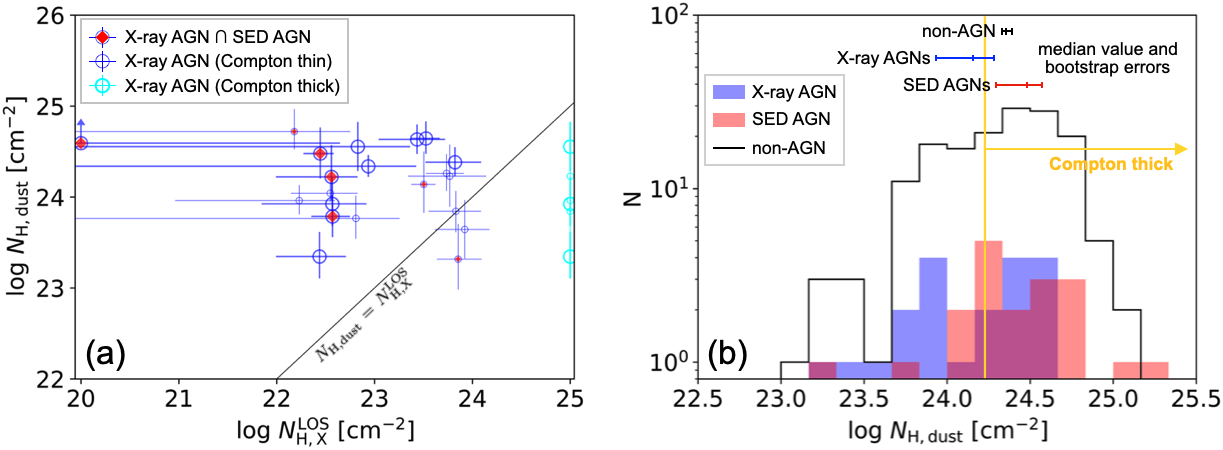}
  \caption{
  \textit{(a)}
  Comparison of line-of-sight hydrogen column density measured by X-ray spectral analysis ($N_{\mathrm{H,X}}^{\mathrm{LOS}}$) with that estimated from the dust properties ($N_{\mathrm{H,dust}}$). AS2COS0055.2 and AS2COS246.2 are not plotted in this panel because they are too faint to derive reliable sizes. The large symbols show the spec-$z$ sample, while the smaller ones show the photo-$z$ sample. \textit{(b)} Histogram of the line-of-sight hydrogen column density calculated from the dust properties for the X-ray AGN, the SED AGNs, and the rest of the AS2COSMOS sources (non-AGNs). Only the sources whose sizes are well-constrained are included in these histograms. The vertical solid line shows the threshold of Compton-thick absorption ($N_{\mathrm{H}}=1.7\times 10^{24}\,\mathrm{cm}^{-2}$). The redshift uncertainties are not considered when calculating the plotted values in these panels.}
  \label{figure:Nh_dust}
\end{figure*}

Regarding the X-ray stacking analysis, we plot in Figure~\ref{figure:Lx_LFIR} the median far-infrared luminosities and the stacked X-ray luminosities for the X-ray undetected SED AGNs. We find that the stacked values are consistent with the empirical relation of AGN-classified SMGs assuming Compton-thin absorption. This implies that the X-ray undetected SED AGNs are probably the same population as the AGN-classified SMGs reported in the previous studies using deeper X-ray observations \citep{2005ApJ...632..736A,2013ApJ...778..179W}. On the other hand, if we consider the Compton-thick case, the stacked values align with the empirical relation of local quasars ($L_{\mathrm{X}}=0.05L_{\mathrm{FIR}}$). To distinguish between these two possibilities, deeper hard X-ray observations above rest-frame 10 keV are needed. Moreover, we plot the median far-infrared luminosities and the stacked X-ray luminosities of the non-AGNs. We find that the stacked X-ray luminosity is ${\sim}0.7$ dex higher than that expected from their star-formation activity. This supports the prediction in Section~\ref{subsection:x-ray_stack} that there remain obscured AGNs that are not detected in X-ray nor identified as SED AGNs. Thus, we also calculate the upper bound of the AGN luminosity in the non-AGN sample from the integrated count rate derived from our stacking analysis, which is used in Section~\ref{subsection:sfr_lbol}. Here we assume the same X-ray spectral model as in Section~\ref{subsubsection:x-ray_spec_undetected} placed at the median redshift of $z=2.45$, and the line-of-sight hydrogen column density is fixed to $N_{\mathrm{H,X}}^{\mathrm{LOS}}\sim 10^{23.2}\,\mathrm{cm}^{-2}$ as predicted by the stacking analysis (Section~\ref{subsection:x-ray_stack}).

\subsection{X-ray Absorption by Host Galaxy Gas}\label{subsection:absorption}

If we make assumptions about the distribution of dust and gas-to-dust ratio, we can estimate the line-of-sight hydrogen column densities of the host-galaxy dust based on the dust masses and the sizes of the dust-emitting regions. We assume that the half mass of galaxy dust is uniformly distributed within a sphere with a half-light radius centered at the SMBH. The H$_2$-to-dust mass ratio is fixed to 90 \citep{2014MNRAS.438.1267S}. Based on these assumptions across the whole sample, the median hydrogen column density in the molecular phase is estimated as $N_{\mathrm{H,dust}}^{\mathrm{mol}}=8.7^{+0.9}_{-0.3}\times 10^{23}\,\mathrm{cm}^{-2}$, which is consistent with the previous study of the bright subsample of AS2UDS ($N_{\mathrm{H,dust}}^{\mathrm{mol}}=9.8^{+1.4}_{-0.7}\times 10^{23}\,\mathrm{cm}^{-2}$; \citealt{2017ApJ...839...58S}). Then, we convert the hydrogen column density in the molecular phase to total hydrogen column density by applying a factor of 2.5 following \citet{2024MNRAS.527L.144A}, which is based on theoretical models by \citet{2011MNRAS.418.1649L} and \citet{2012MNRAS.424.2701F}. Finally, we obtain the median hydrogen column density of $N_{\mathrm{H,dust}}=1.4^{+0.5}_{-0.6}\times 10^{24}\,\mathrm{cm}^{-2}$, $N_{\mathrm{H,dust}}=3.0^{+0.7}_{-1.1}\times 10^{24}\,\mathrm{cm}^{-2}$, and $N_{\mathrm{H,dust}}=2.2^{+0.2}_{-0.1}\times 10^{24}\,\mathrm{cm}^{-2}$ for the X-ray AGNs, SED AGNs, and the rest of the AS2COSMOS sources. These values are slightly higher than the previous report on the average ISM column density of 24 submillimeter-infrared quasars at $z\sim1\text{--}3$ in the COSMOS field: $\left\langle N_{\mathrm{H}, \text { ISM }}\right\rangle_{\text {submm-IRquasars }}=(0.8\pm0.1) \times 10^{24} \mathrm{~cm}^{-2}$ \citep{2024MNRAS.527L.144A}. Given the tight correlation between dust mass and 870-$\mu$m flux density \citep{2020MNRAS.494.3828D}, this may be attributed the bright submillimeter selection of our sample.

\begin{figure*}
  \centering
  \includegraphics[width=\linewidth]{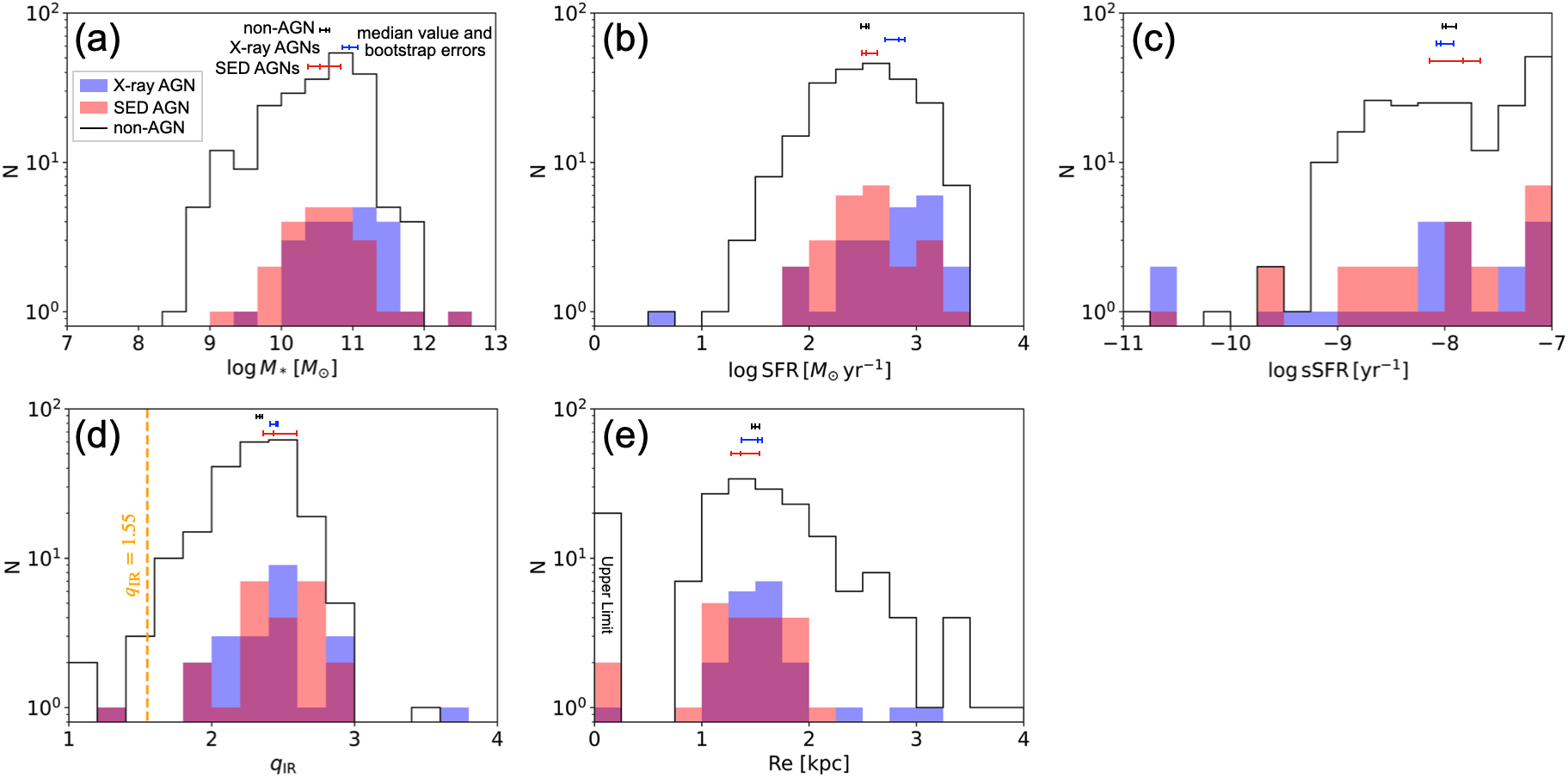}
  \caption{
  Histograms of \textit{(a)} stellar masses, \textit{(b)} SFRs, \textit{(c)} specific SFRs, \textit{(d)} radio-IR correlation factors ($q_{\mathrm{IR}}$), and \textit{(e)} the dust continuum sizes ($R_{\mathrm{e}}$) for the X-ray AGNs, SED AGNs, and the rest of the AS2COSMOS sources (non-AGNs). In the histogram of $R_{\mathrm{e}}$, the bins at $R_{\mathrm{e}}=0\,\mathrm{kpc}$ show the number of sources whose dust sizes are only given as the upper confidence intervals. We confirm no significant difference between the samples, suggesting the similarity of their host galaxies, regardless of the presence of an X-ray or SED AGN.}
  \label{figure:histograms}
\end{figure*}

Figure~\ref{figure:Nh_dust} (a) compares the line-of-sight hydrogen column density measured by X-ray spectral analysis ($N_{\mathrm{H,X}}^{\mathrm{LOS}}$) with that estimated by the dust properties ($N_{\mathrm{H,dust}}^{\mathrm{mol}}$). In the majority of cases (18/21), the line-of-sight hydrogen column density estimated from the dust properties is significantly larger than the X-ray measurements in the Compton-thin cases. This suggests that the host-galaxy clouds have non-spherical geometry and/or have complex (e.g., clumpy) distributions. We may also consider the possibility that the X-ray absorption of those sources is dominated by the opacity in the host galaxy rather than the torus. Moreover, this results might suggest that many of the X-ray AGNs are Compton thick and the line-of-sight hydrogen column densities are not properly constrained by the X-ray spectral analysis. Figure~\ref{figure:Nh_dust} (b) shows the histogram of the line-of-sight hydrogen column density estimated from the dust properties. We find that the SED AGNs have tentatively ${\sim}0.3$ dexes higher column densities than the X-ray AGNs. The median value of the SED AGNs is $N_{\mathrm{H,dust}}^{\mathrm{mol}}=1.2^{+0.3}_{-0.4}\times 10^{24}\,\mathrm{cm}^{-2}$, which is broadly equivalent to the threshold of Compton-thick absorption ($N_{\mathrm{H}}=1.7\times 10^{24}\,\mathrm{cm}^{-2}$). Thus, this might suggest that in bright SMGs AGNs can be heavily obscured by the host galaxies, which makes them difficult to detect with X-ray observations.

\subsection{Comparison of Physical Properties between Samples}\label{subsection:comparison}

Figure~\ref{figure:histograms} compares the stellar masses, SFRs, specific SFRs, radio-IR correlation factors, and the sizes of dust-emitting regions of the X-ray AGNs, the SED AGNs, and the rest of the AS2COSMOS sources (non-AGNs). We find that the median SFR and stellar mass of the X-ray AGN sample ($\log \mathrm{SFR}[M_{\odot}\,\mathrm{yr}^{-1}]=2.84^{+0.05}_{-0.13}$ and $\log M_*[M_{\odot}]=10.94^{+0.12}_{-0.09}$) are about two times higher than those of the non-AGN sample ($\log \mathrm{SFR}[M_{\odot}\,\mathrm{yr}^{-1}]=2.53^{+0.02}_{-0.06}$ and $\log M_*[M_{\odot}]=10.62^{+0.04}_{-0.09}$), while the median SFR and stellar mass of the SED AGN sample ($\log \mathrm{SFR}[M_{\odot}\,\mathrm{yr}^{-1}]=2.53^{+0.10}_{-0.04}$ and $\log M_*[M_{\odot}]=10.53^{+0.30}_{-0.16}$) agree with the non-AGNs. This might be attributed to the X-ray selection bias of these sources. Since the X-ray observations in the COSMOS field are relatively shallow, the X-ray AGNs are biased to bright AGNs, which likely host higher-mass SMBHs. Hence, if we assume the positive correlation between stellar mass and SMBH mass \citep{2013ARA&A..51..511K}, the X-ray AGNs are expected to be hosted by massive galaxies, which may have high SFR according to the scaling relation of SFR and stellar mass. This argument is supported by the similarity in the distributions of the specific SFRs between the X-ray AGNs, SED AGNs, and non-AGNs (the median specific SFRs of the X-ray AGN, SED AGN, and non-AGN samples are $-8.04^{+0.12}_{-0.04}$ yr$^{-1}$, $-7.83^{+0.23}_{-0.31}$ yr$^{-1}$, and $-7.99^{+0.10}_{-0.03}$ yr$^{-1}$, respectively). We also find that the radio-IR correlation factors and the dust-continuum sizes do not significantly vary between the samples. This implies that the vast majority of the X-ray AGNs and the SED AGNs are radio quiet and the properties of their host galaxy are similar to those of the non-AGNs.

\subsection{X-ray Luminosity versus Bolometric AGN Luminosity}\label{subsection:lx_lbol}

\begin{figure*}
  \centering
  \includegraphics[width=\linewidth]{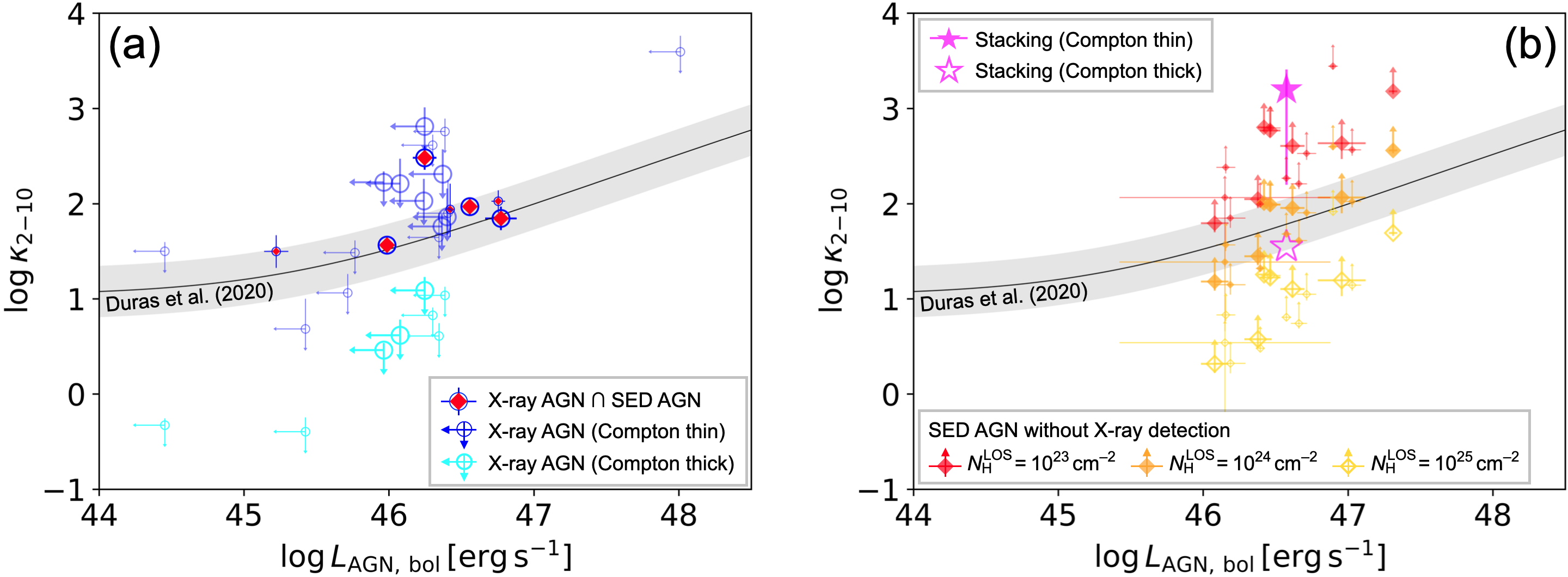}
  \caption{
  Comparison of X-ray to bolometric correction factor ($\kappa_{2\text{--}10}$) and bolometric AGN luminosity for \textit{(a)} the X-ray AGNs and \textit{(b)} the SED AGNs (the overlapped sources are plotted in panel (a)). The black solid line denotes the empirical relation at $z=0\text{--}4$ \citep{2020A&A...636A..73D}. Symbols with arrows show the 3$\sigma$ upper bound for each parameter. The redshift uncertainties are not considered when calculating the overall uncertainties of the plotted values. The large symbols show the spec-$z$ sample, while the smaller ones show the photo-$z$ sample. We find that the X-ray AGNs are predicted to be Compton-thin AGNs, while the majority of X-ray undetected SED AGNs are nearly Compton-thick AGNs, if the empirical relation holds.}
  \label{figure:lx_lagn}
\end{figure*}

Figure~\ref{figure:lx_lagn} compares the X-ray to bolometric correction factor ($\kappa_{2\text{--}10} = L_{\mathrm{AGN,\,bol}}/L_{\mathrm{2\text{--}10\,keV}}$) with the bolometric AGN luminosity for our AGN samples. We also plot the empirical relation of AGNs at $z<3.5$ from \citet{2020A&A...636A..73D}. We find that six out of the seven X-ray detected SED AGNs show good agreement with the empirical relation, while one source has about one dex higher X-ray luminosity than the empirical relation (AS2COS0048.1; $\kappa_{2\text{--}10}=326$). We also find that all the X-ray AGNs are predicted to be Compton-thin AGNs, while the majority of the X-ray undetected SED AGNs (14/17) appear to be nearly Compton thick ($N_{\mathrm{H,X}}^{\mathrm{LOS}}\geq 10^{24}\,\mathrm{cm}^{-2}$), assuming the empirical relation holds. This is consistent with the recent observation that a significant fraction of AGNs in local ultra/luminous infrared galaxies (U/LIRGs; $L_{\mathrm{IR}}/L_{\odot}>10^{11}$) show Compton-thick absorption (16/35; \citealt{2021MNRAS.506.5935R}). However, some studies have reported that AGNs in the local U/LIRGs are intrinsically X-ray weak ($\kappa_{2\text{--}10}=100\text{--}1000$; \citealt{2014ApJ...785...19T,2021ApJS..257...61Y}). If this phenomenon is applicable to our samples, the X-ray undetected SED AGNs do not have to be Compton thick. Note that the X-ray AGNs are not predicted to be intrinsically X-ray weak, based on the upper limits of the X-ray to bolometric correction factors.

Regarding the X-ray stacking analysis, we plot the average X-ray to bolometric correction factors and the median bolometric AGN luminosities for the X-ray undetected SED AGNs. We find that assuming the Compton-thin case, the inferred average X-ray to bolometric corrections for the X-ray undetected SED AGNs is more than one dex higher than the empirical relation and higher than any X-ray detected SED AGN in our sample. This may indicate that the X-ray undetected SED AGNs are either Compton thin and intrinsically X-ray weak or they are Compton thick.

\subsection{SFR versus Bolometric AGN Luminosity}\label{subsection:sfr_lbol}

\begin{figure*}
  \centering
  \includegraphics[width=0.5\linewidth]{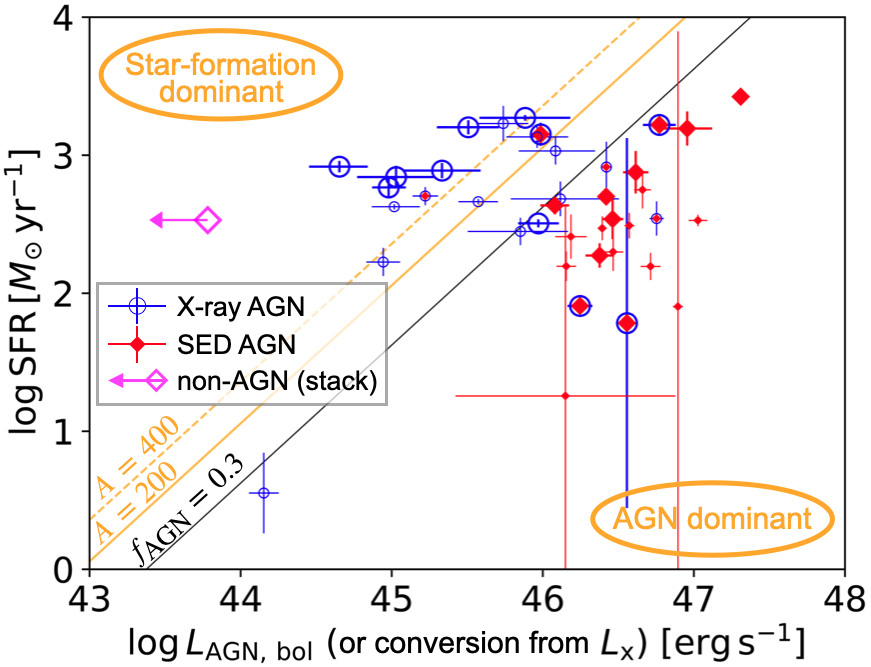}
  \caption{
  Comparison of SFR and bolometric AGN luminosity for the X-ray AGNs and the SED AGNs. The orange solid and dashed lines show the ``simultaneous evolution'' with $A=200$ (bulge only) and $A=400$ (bulge$+$disk), respectively. The black solid line shows the detection limit of the type-2 SED AGNs (see Section~\ref{subsection:bias}). The large symbols show the spec-$z$ sample, while the smaller ones show the photo-$z$ sample. We find that the SED AGNs are distributed in the AGN-dominant region, suggesting that in such a phase AGNs can be nearly Compton-thick and missed in X-ray.}
  \label{figure:sfr_lagn}
\end{figure*}

Figure~\ref{figure:sfr_lagn} compares the SFR with the bolometric AGN luminosity for the X-ray AGNs and SED AGNs. Here we use the bolometric AGN luminosity derived from the SED modeling for the SED AGNs. For X-ray AGNs, excluding X-ray AGNs that are also identified as SED AGNs, the bolometric AGN luminosities are converted from the X-ray luminosities using the calibration by \citet{2020A&A...636A..73D}. Note that we use the X-ray luminosities calculated with the Compton-thin assumption based on the discussion in Section~\ref{subsection:lx_lbol}. We also plot the median SFR and the upper bound of the bolometric AGN luminosity for the non-AGN sample calculated from the X-ray stacking (Section~\ref{subsubsection:x-ray_LFIR}), where the calibration by \citet{2020A&A...636A..73D} is adopted to convert the X-ray luminosity upper bound to the bolometric AGN luminosity upper bound. For comparison, we plot the ``simultaneous galaxy-SMBH evolution'' line for the co-evolution of SMBHs and the host galaxies. Assuming that SMBHs and the host galaxies co-evolve simultaneously across cosmic time, the relative ratio of star formation and AGN activity is expected to track this relationship. 
\begin{eqnarray}
  \mathrm{SFR}\times (1-R) &=& A\times \dot{M}_{\mathrm{BH}}\\
  &=& L_{\mathrm{AGN,\,bol}}\times (1-\eta)/\eta c^2\nonumber
\end{eqnarray}
where $R$ is the return function (the fraction of stellar masses that are ejected back to the interstellar medium), $A$ is the mass ratio of stars to SMBHs, $\eta$ is the accretion efficiency, and $c$ is the speed of light. We assume $R=0.41$ (Chabrier IMF) and $\eta=0.05$. For $A$, we calculate in two settings: $A=200$ and $A=400$. The former value is the typical mass ratio of stars in bulge components to SMBHs in the local universe \citep{2013ARA&A..51..511K}, while the latter value includes the disk components (see Section 4.4 in \citealt{2018ApJ...853...24U}). We find that the SED AGNs are distributed in the AGN dominant phase, where SMBHs grow faster than the host galaxies, while the X-ray AGNs are almost distributed around the simultaneous evolution lines. The bias of SED AGNs to higher AGN luminosity can be explained by the selection effect discussed in Section~\ref{subsection:bias}. We can potentially interpret these results within the merger-driven evolutionary scenario: on the basis of this scenario, galaxy mergers first trigger star formation activity and later evoke AGN activity after the gas has had time to reach the nuclear region (e.g., via stellar winds, supernovae, or through gravitational torques). Within this scenario, the SED AGNs would correspond to the transition phase, where merging has reached the nuclear regions and the AGN activity has been enhanced, but the star formation is not yet quenched. Hence, we might say that in such phases, AGNs are likely to be heavily obscured by dust or have unusually suppressed X-ray emissions, which makes them difficult to detect with X-ray observations. We note that the selection of the AGNs by SED modeling strongly depends on the quality of the multi-wavelength photometry and the assumptions of models. Thus, these results should be verified by other methods like future sensitive observations with Advanced Telescope for High Energy Astrophysics (Athena; \citealt{2013arXiv1306.2307N}).

\subsection{Population Statistics}\label{subsection:statistical}

\subsubsection{AGN Number Fraction}\label{subsubsection:agn_fraction}

\begin{figure*}
  \centering
  \includegraphics[width=\linewidth]{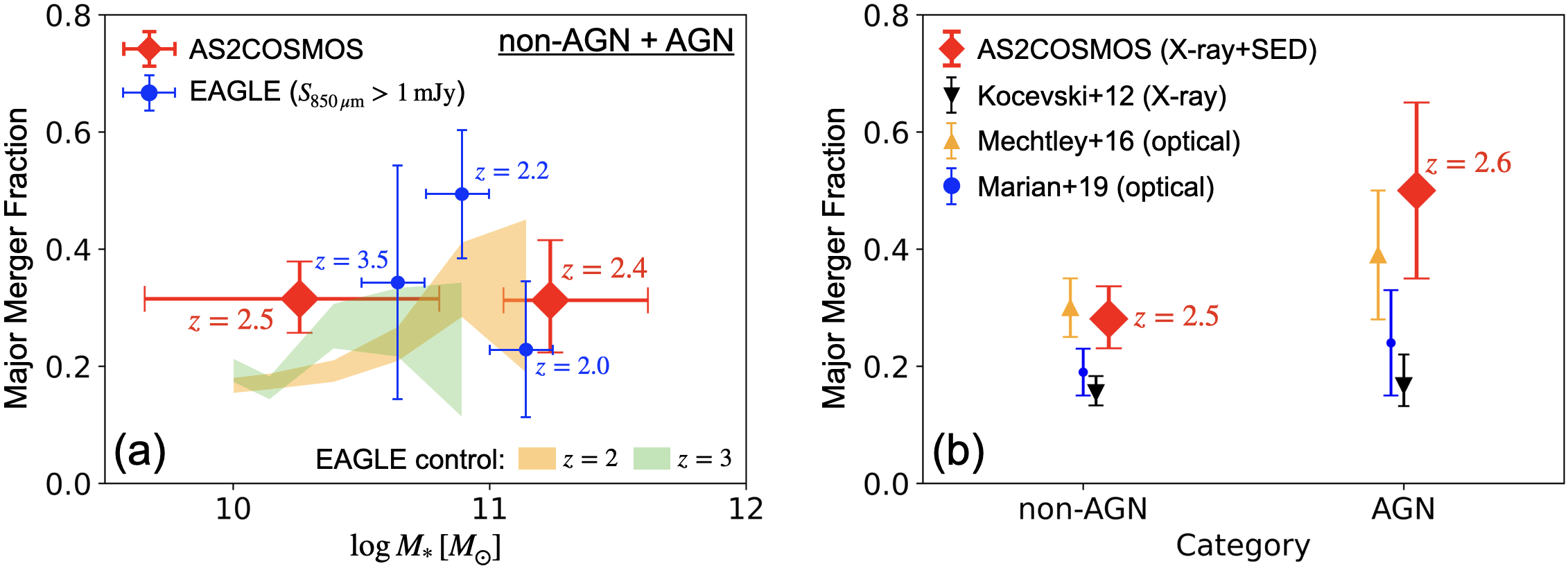}
  \caption{\textit{(a)} 
  Comparison of the major merger fraction in the AS2COSMOS sample with the theoretical prediction from EAGLE \citep{2019MNRAS.488.2440M}. The errors on $M_*$ of AS2COSMOS sample indicate the 16--84\% intervals of each stellar mass bin. \textit{(b)} Dependencies of the major merger fraction as a function of AGN activity. The X-ray and optical selected AGNs at $z\sim2$ are also plotted \citep{2012ApJ...744..148K,2016ApJ...830..156M,2019ApJ...882..141M}. Each data point is annotated with the median redshift of the galaxies in that bin. We find that major mergers may play a key role in triggering AGN activity in bright SMGs.}
  \label{figure:merger_fraction}
\end{figure*}

The depths of the X-ray observations in the COSMOS field are only moderate (${\sim}160$ ks per pixel). Thus, the X-ray AGNs in the AS2COSMOS sample are not complete even for X-ray bright sources above luminosities of $L_{\mathrm{X}}=10^{43}\,\mathrm{erg\,s^{-1}\,cm^{-2}}$. The SED AGNs are also incomplete due to the selection shown in Section~\ref{subsection:bias}. Hence, the AGN number fraction in our sample provides a lower limit for bright SMGs. In this study, the AGN number fraction in the AS2COSMOS sample is estimated as $16^{+3}_{-2}$ per cent for all AGNs (SED AGNs and X-ray AGNs) and $9\pm2$ per cent for only X-ray AGNs. The number fraction of X-ray AGNs in our sample is therefore consistent with that in the AS2UDS sample ($8\pm2$ per cent; \citealt{2019MNRAS.487.4648S}), where the depth of the X-ray observations used in that work are more comparable to those in the COSMOS field (X-UDS; 200--600 ks with Chandra; \citealt{2018ApJS..236...48K}).

To identify trends in AGN fraction with the SMG population, we separate the sample into four bins based on the redshifts and 870-$\mu$m flux densities. Since our sample is only complete for SMGs brighter than $S_{\mathrm{850}\, \mu \mathrm{m}}=6.2\,\mathrm{mJy}$, we separate the sample at $S_{\mathrm{870}\, \mu \mathrm{m}}=6.2\,\mathrm{mJy}$. We also separate the sample at $z=3$, because the selection of AGNs by SED modeling is more reliable at low redshift ($z<3$; Section~\ref{subsubsection:agn_identification}).

In the brighter SMG samples, the AGN number fraction (SED AGN and X-ray AGN) is about two times higher at high redshift ($24^{+8}_{-6}$\%) than at low redshift ($13^{+5}_{-4}$\%), which is also confirmed for X-ray AGNs ($10^{+6}_{-4}$\%  and $6^{+4}_{-2}$\%, respectively). This might suggest that the SMBH growth is more intense in high-redshift SMGs. However, we stress that the evidence for high-redshift SED AGNs ($z>3$) is fairly limited and care should be taken with this issue. For low-redshift samples, the SED AGN number fraction is higher in the brighter SMGs ($7^{+4}_{-3}$\%) than in the fainter SMGs ($1^{+2}_{-1}$\%), while the X-ray AGN number fractions are consistent within their confidence intervals ($6^{+4}_{-2}$\% and $8^{+4}_{-3}$\%, respectively). This might suggests that, in bright SMGs, AGNs tend to be heavily obscured or have unusually weak X-ray emission. As the 870-$\mu$m flux density is tightly correlated with the dust mass of a galaxy \citet{2020MNRAS.494.3828D}, this trend might be attributed to the dust obscuration by the host galaxies (Section~\ref{subsection:absorption}). Nevertheless, we need to be cautious about the selection bias due to the quality of multi-wavelength photometry. Brighter SMGs are more likely have higher S/N in the mid- to far-infrared photometry, which can highlight the imperfection of the host-galaxy models and cause the misidentification of SED AGNs. On the other hand, in fainter SMGs, the S/N in the mid- to far-infrared photometry is expected to be lower, which can miss the presence of SED AGNs.

\subsubsection{Merger Fraction}\label{subsubsection:merger_fraction}

The connection between galaxy mergers and AGN/host-galaxy properties is an important, but still unresolved issue. In the low-redshift universe ($z\leq1$), several observational studies have shown that galaxies in mergers are more likely to host AGNs than isolated galaxies (e.g., \citealt{2018PASJ...70S..37G,2020A&A...637A..94G}). While theoretical investigations predict that the fraction of merger-induced starburst in SMGs increases with submillimeter flux density \citep{2013MNRAS.428.2529H}, although other theoretical work investigating the dependence of merger fraction on the stellar mass and found that the merger fraction in SMGs is similar to the general galaxy population, suggesting that mergers are not the sole driver of the enhanced star formation in SMGs \citep{2019MNRAS.488.2440M}. To address this issue in our study we examine the merger fraction of the AS2COSMOS sources as functions of stellar masses ($M_* > 10^{11} M_{\odot}$ or $M_* < 10^{11} M_{\odot}$) and AGN activity (AGN or non-AGN).

Figure~\ref{figure:merger_fraction} (a) shows the dependence of major merger fractions on stellar masses. We also plot the theoretical prediction using the EAGLE simulation by \citet{2019MNRAS.488.2440M}. We find no significant difference between the massive and less massive subsets of SMGs ($31^{+10}_{-9}$\% and $32\pm6$\%, respectively) and the major merger fractions of the two subsets are consistent with the EAGLE simulation. Figure~\ref{figure:merger_fraction} (b) shows the dependence of major merger fractions on AGN activity. For comparison, we plot the major merger fractions in X-ray selected AGNs \citep{2012ApJ...744..148K} and optical selected AGNs (\citealt{2016ApJ...830..156M} and \citealt{2019ApJ...882..141M}) at $z\sim2$ (see also \citealt{2023OJAp....6E..34V} for the summary of these studies). Notably the major merger fraction in the AS2COSMOS non-AGN sample ($28^{+6}_{-5}$\%) is consistent with those of the general population at $z\sim2$. However, we find that the major merger fraction in the AS2COSMOS AGN sample ($50\pm15$\%) is potentially twice as high as that in the non-AGN sample, which is more significant than the enhancement reported for the general AGN samples at $z\sim2$, although the uncertainties are large. These results suggest that major mergers are not necessarily required for the enhanced star formation in SMGs as predicted by the EAGLE simulation, but may play a key role in triggering AGN activity in bright SMGs. We caution that there still remain large statistical uncertainties and further studies with large samples are needed. For example, dependency of major merger fraction on AGN luminosity cannot be confirmed in this study, but should be investigated in future studies with larger samples. In addition, the selection of AGNs by SED analysis strongly depends on the model assumptions. If we adopt the \citet{2012ApJ...748..142D} criteria for the selection of SED AGN, the major merger fraction in the AGN sample is $43^{+13}_{-12}$\%, which is still almost consistent with the major merger fraction in our AGN sample. The slight difference might originate from the galaxy contamination in the \citet{2012ApJ...748..142D} selection. We also note that the stellar masses of most of the companions are not measured in this study, but it may not affect the main conclusion of this study.

\section{Summary}\label{section:summary}

We performed multi-component SED modeling and X-ray spectral analysis for the sample of bright SMGs from the AS2COSMOS survey. The sample consists of 260 SMGs with $S_{\mathrm{870}\, \mu \mathrm{m}}=0.7\text{--}19.2\,\mathrm{mJy}$, which is effectively complete for SMGs with 850-$\mu$m flux densities $S_{\mathrm{850}\, \mu \mathrm{m}} \geq 6.2\,\mathrm{mJy}$ in the S2COSMOS catalog. Our main results are listed below:
\begin{itemize}
  \item 
  Using an SED fitting methodology, we identified 24 AGN candidates (SED AGNs). Supplemented by 23 X-ray detected AGNs (X-ray AGNs), we construct a sample of 40 AGN candidates, of which seven sources overlap between the two subsets.
  \item 
  The X-ray AGNs have a median of 6 times higher X-ray luminosities than the empirical relation between the X-ray luminosity and far-infrared luminosity for AGN-classified SMGs, assuming Compton-thin absorption. This can be attributed to the selection from the relatively shallow X-ray observations in the COSMOS field. An X-ray stacking analysis shows that the stacked luminosities of X-ray undetected SED AGNs show good agreement with the empirical relation of the AGN-classified SMGs with the assumption of Compton-thin absorption. This indicates that the X-ray undetected SED AGNs are likely to be the same population as the AGN-classified SMGs detected in deeper X-ray studies.
  \item 
  From the dust masses and the sizes of the dust emitting region, the median hydrogen column densities are calculated as $N_{\mathrm{H,dust}}=1.4^{+0.5}_{-0.6}\times 10^{24}\,\mathrm{cm}^{-2}$, $N_{\mathrm{H,dust}}=3.0^{+0.7}_{-1.1}\times 10^{24}\,\mathrm{cm}^{-2}$, and $N_{\mathrm{H,dust}}=2.2^{+0.2}_{-0.1}\times 10^{24}\,\mathrm{cm}^{-2}$ for the X-ray AGNs, SED AGNs, and the rest of AS2COSMOS sources. The SED AGNs have about two times higher column densities than the X-ray AGNs. This may suggest that the SED AGNs are heavily obscured by the host galaxy dust, which makes them difficult to detect with X-ray observations.
  \item 
  Among the seven X-ray detected SED AGNs, six are consistent with the empirical relation of X-ray luminosity and bolometric luminosity of AGNs at $z<3.5$, while one source has about one dex higher X-ray luminosity than the empirical relation. Assuming the empirical relation, the X-ray AGNs are predicted to be Compton-thin AGNs, while the majority of X-ray undetected SED AGNs are nearly Compton-thick AGNs. However, if we consider the X-ray weak cases, the SED AGNs can be Compton-thin AGNs.
  \item 
  In the SFR versus AGN luminosity plane, the SED AGNs are distributed in the AGN-dominant phase, while X-ray AGNs are distributed around the simultaneous evolution lines. Assuming a merger-driven evolutionary scenario, our AGN sample may correspond to the transition phase, where merging has finished but the star formation is not yet quenched. Thus our results suggest that in such phases, AGNs are likely to be heavily obscured by dust or have anomalously suppressed X-ray emissions, which makes them difficult to detect with X-ray observations.
  \item 
  The AGN number fraction in the AS2COSMOS sample is $16^{+3}_{-2}$ per cent for total (SED AGNs and X-ray AGNs) and $9\pm2$ per cent for only X-ray AGNs. These values are lower than the previous estimate with deeper X-ray observation, which can be attributed to the relatively shallow X-ray observations in the COSMOS field and also the selection bias of the SED AGNs.
  \item
  Using visual classification, we identify $47^{+16}_{-15}$\% and $25^{+6}_{-5}$\% of the AGN hosts and galaxies without AGNs as major merger candidates, respectively. The major merger fraction in the AS2COSMOS non-AGN sample is almost consistent with the general population at $z\sim2$. This suggests that major mergers are not necessarily required for the enhanced star formation in SMGs. On the other hand, the major merger fraction in the AS2COSMOS AGN sample is potentially twice as high as that in the AS2COSMOS non-AGN sample. This suggests that major mergers may play a key role in triggering AGN activity in bright SMGs.
\end{itemize}

%% IMPORTANT! The old "\acknowledgment" command has be depreciated. It was
%% not robust enough to handle our new dual anonymous review requirements and
%% thus been replaced with the acknowledgment environment. If you try to
%% compile with \acknowledgment you will get an error print to the screen
%% and in the compiled pdf.
%%
%% Also note that the akcnowlodgment environment does not support long amounts of text. If you have a lot of people and institutions to acknowledge, do not use this command. Instead, create a new \section{Acknowledgments}.
\begin{acknowledgments}

We would like to express our gratitude to Dr. David Rosario for his insightful comment. We also thank Prof. Mikio Kurita for his support of this collaboration. This work was financially supported by JSPS KAKENHI Grant Numbers JP22KJ1990 (R.U.), JP20H01946 (Y.U.), JP20H01953, JP22KK0231, JP23K20240 (H.U.), JP22K21349 (M.K), JP22H01273, JP23K22544 (Y.M.), JP22H04939, JP23K20035, and JP24H00004 (K.K.). C.-C.C. acknowledges support from the National Science and Technology Council of Taiwan (NSTC 111-2112M-001-045MY3), as well as Academia Sinica through the Career Development Award (AS-CDA-112-M02). The co-authors at Durham University acknowledge STFC (ST/X001075/1). Our work utilized data from the S2COSMOS survey (M16AL002) on the JCMT, supplemented by data from S2CLS (MJLSC02) and the JCMT archive. This paper makes use of the following ALMA data: ADS/JAO.ALMA\#2016.1.00463.S. ALMA is a partnership of ESO (representing its member states), NSF (USA), and NINS (Japan), together with NRC (Canada), NSC and ASIAA (Taiwan), and KASI (Republic of Korea), in cooperation with the Republic of Chile. The Joint ALMA Observatory is operated by ESO, AUI/NRAO, and NAOJ. R.U. was supported by the ALMA Japan Research Grant of NAOJ ALMA Project, NAOJ-ALMA-335. The JWST images presented in this study were retrieved from the Dawn JWST Archive (DJA). DJA is an initiative of the Cosmic Dawn Center (DAWN), which is funded by the Danish National Research Foundation under grant DNRF140.

\end{acknowledgments}

%% To help institutions obtain information on the effectiveness of their
%% telescopes the AAS Journals has created a group of keywords for telescope
%% facilities.
%
%% Following the acknowledgments section, use the following syntax and the
%% \facility{} or \facilities{} macros to list the keywords of facilities used
%% in the research for the paper.  Each keyword is check against the master
%% list during copy editing.  Individual instruments can be provided in
%% parentheses, after the keyword, but they are not verified.

\vspace{5mm}
\facilities{CXO, CFHT, Subaru, VISTA, Spitzer, JWST, Herschel, JCMT, ALMA, VLA}

%% Similar to \facility{}, there is the optional \software command to allow
%% authors a place to specify which programs were used during the creation of
%% the manuscript. Authors should list each code and include either a
%% citation or url to the code inside ()s when available.

\software{CIGALE v2022.0 \citep{2019A&A...622A.103B,2020MNRAS.491..740Y,2022ApJ...927..192Y},
          XSPEC \citep{1996ASPC..101...17A},
          CIAO \citep{2006SPIE.6270E..1VF},
          astropy \citep{2013A&A...558A..33A,2018AJ....156..123A}
          }

%% Appendix material should be preceded with a single \appendix command.
%% There should be a \section command for each appendix. Mark appendix
%% subsections with the same markup you use in the main body of the paper.

%% Each Appendix (indicated with \section) will be lettered A, B, C, etc.
%% The equation counter will reset when it encounters the \appendix
%% command and will number appendix equations (A1), (A2), etc. The
%% Figure and Table counter will not reset.

\appendix

\section{MIRI Images of AS2COSMOS Sources}\label{appendix:miri}

\begin{figure*}
  \centering
  \includegraphics[width=\linewidth]{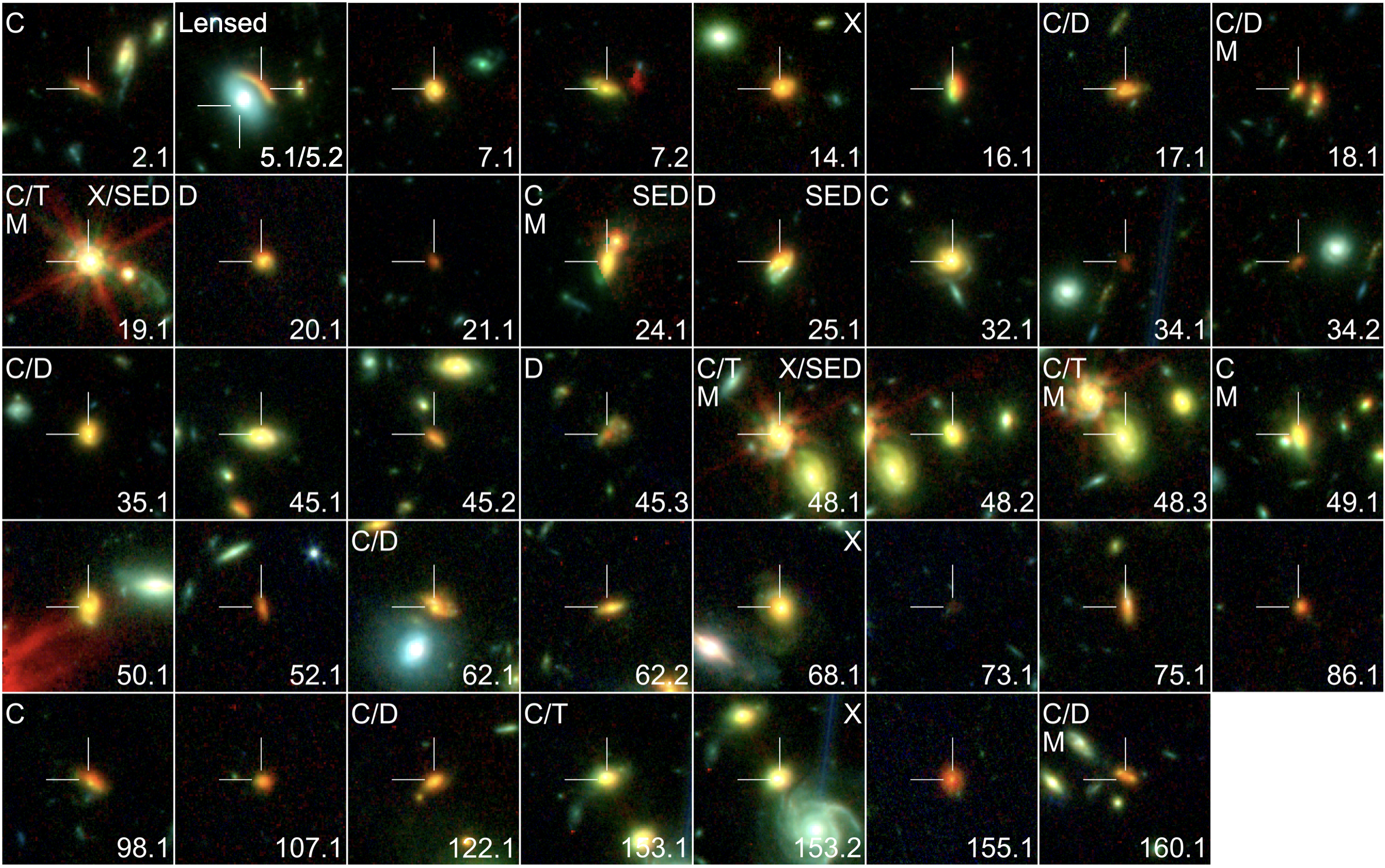}
  \caption{
  8 arcsec $\times$ 8 arcsec JWST images of the 40 AS2COSMOS sources in the coverage of both NIRCam and MIRI imaging. The blue, green, and red colors correspond to the F115W+F150W, F277W+F444W, and F770W filters, respectively. We label the merger candidates, which have tidal features (T), disturbed morphology (D) or possible companions (C) (see Section~\ref{subsection:morphology}). The major merger candidates are indicated by ``M''. The X-ray AGNs and the SED AGNs are indicated by ``X'' and ``SED'' (see Section~\ref{subsubsection:agn_identification} and Section~\ref{subsubsection:x-ray_LFIR}). Note that F150W is not used in the composite image of AS2COS0107.1, because this object is outside the coverage of F150W. For the same reason, F444W is not used in the composite image of AS2COS155.1.}
  \label{figure:JWST_miri}
\end{figure*}

Figure~\ref{figure:JWST_miri} shows the color-composite images of the 40 sources in the coverage of both NIRCam and MIRI. AS2COS0019.1 and AS2COS0048.1 have point-like morphology in the MIRI/F770W band, which is consistent with the results of the SED modeling.

\section{Astrometry Check for X-ray Sources}\label{appendix:x-misid}

\begin{figure}
  \centering
  \includegraphics[width=\linewidth]{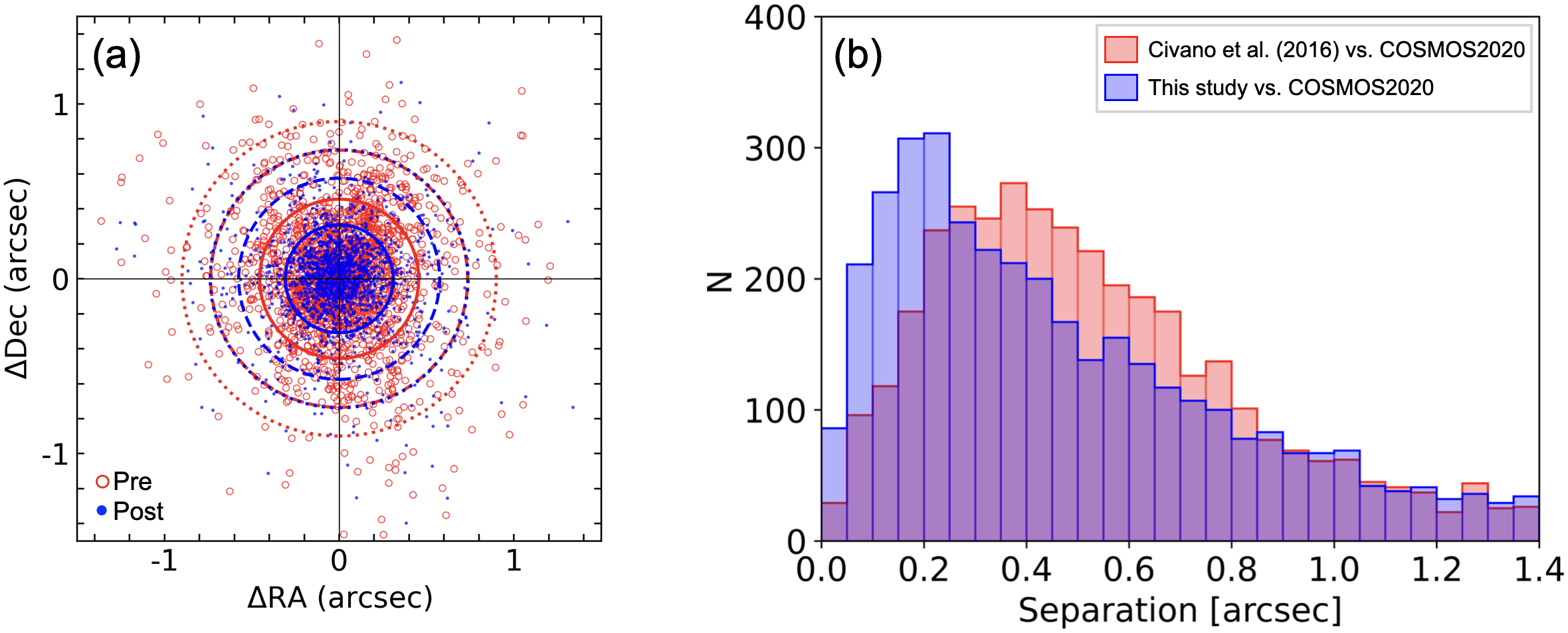}
  \caption{\textit{(a)} 
  Positional offsets between optical sources and X-ray sources used for the astrometry correction. The solid, dotted, and dashed circles encompass 68\%, 90\%, and 95\% of the sources before (red) and after (blue) the correction, respectively. \textit{(b)} Histogram of the separation between the X-ray source positions and the optical to near-infrared source positions in the COSMOS2020 catalog. The red area shows the separation between \citet{2016ApJ...819...62C} and the COSMOS2020 catalog, while the blue area denotes the separation between our X-ray source catalog (Section~\ref{subsubsection:x-ray_extraction}) and the COSMOS2020 catalog.}
  \label{figure:separation}
\end{figure}

Figure~\ref{figure:separation} (a) shows the positional offset between the X-ray source positions and the optical source positions used for the astrometry correction. After the correction, 95\% of the X-ray sources have their optical counterparts within 0.74 arcsec, which is 18\% smaller than the value before the correction (0.90 arcsec). Then, to check the astrometric accuracy, we cross-match the X-ray source positions derived in Section~\ref{subsubsection:x-ray_extraction} to the COSMOS2020 catalog. Figure~\ref{figure:separation} (b) shows the histogram of the separation. For comparison, we plot the separation between the X-ray source positions in \citet{2016ApJ...819...62C} and the optical to near-infrared source positions in the COSMOS2020 catalog. We find that the median separation is about two times smaller in our catalog than \citet{2016ApJ...819...62C}. This suggests that the X-ray source positions in our catalog are better aligned with the COSMOS2020 catalog than \citet{2016ApJ...819...62C}, which is likely because our source positions are calibrated by the COSMOS2020 catalog whereas the CFHT MegaCam catalog \citep{2012A&A...544A.156M} was used as a reference in \citet{2016ApJ...819...62C}.

\begin{figure}
  \centering
  \includegraphics[width=\linewidth]{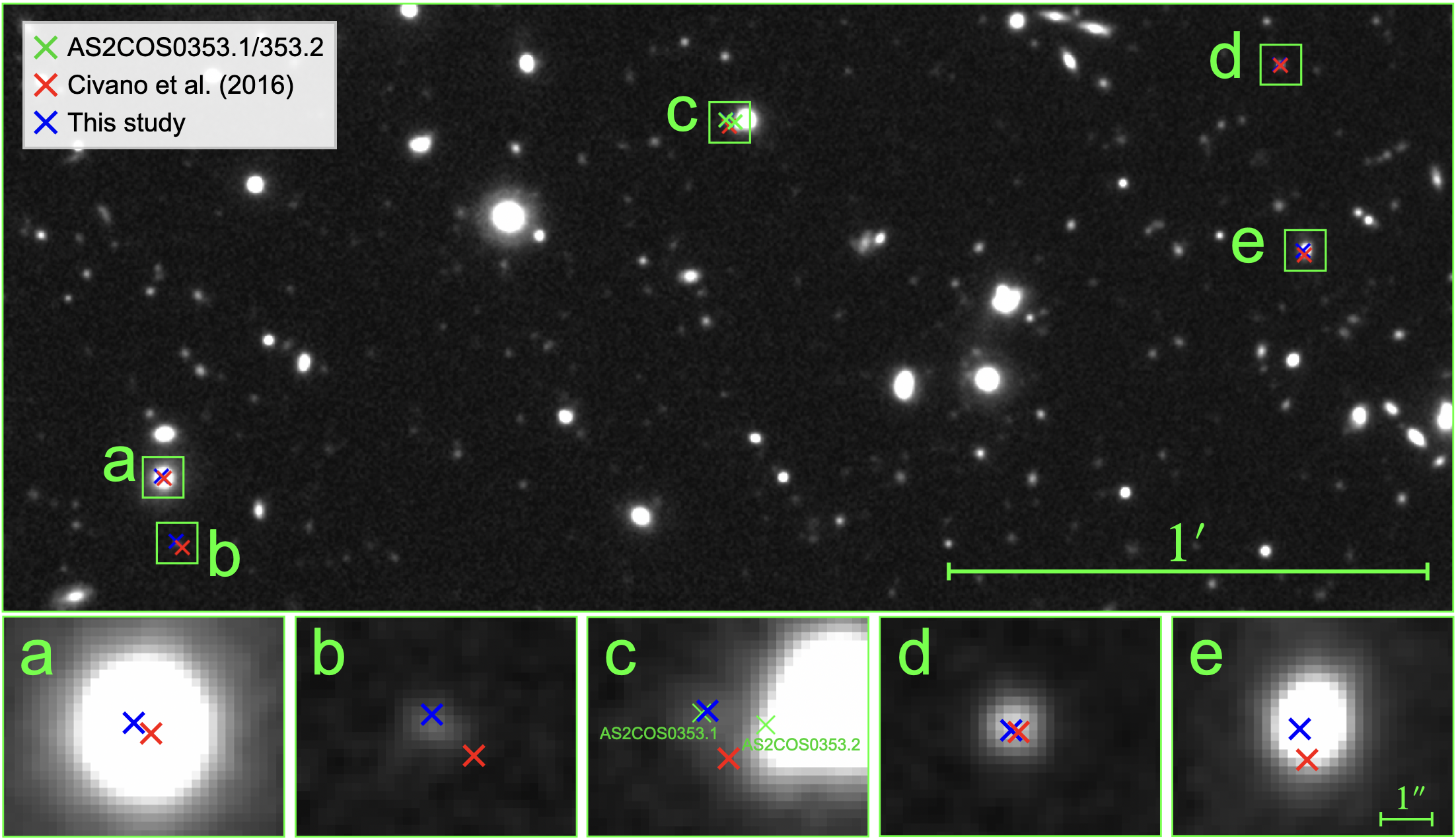}
  \caption{Positions of the X-ray detected sources near AS2COS0353.1 and AS2COS0353.2 plotted over the Ultravista $K_{\mathrm{s}}$-band image. The red points show the X-ray source positions listed in the catalog by \citet{2016ApJ...819...62C}, whereas the blue points show the positions derived in our analysis (Section~\ref{subsubsection:x-ray_extraction}). The green points show the positions of AS2COS0353.1 and AS2COS0353.2. The bottom panels show the zoomed-in images of the X-ray sources.}
  \label{figure:ximage}
\end{figure}

Figure~\ref{figure:ximage} shows the positions of the X-ray detected sources near AS2COS0353.1 and AS2COS0353.2 plotted over the Ultravista $K_{\mathrm{s}}$-band image.\footnote{The $K_{\mathrm{s}}$-band image of the COSMOS field is provided on the COSMOS website (https://cosmos.astro.caltech.edu/page/optical).} The red points show the X-ray source positions listed in the catalog by \citet{2016ApJ...819...62C}, whereas the blue points show the positions derived in our analysis. We notice a systematic offset of about 1 arcsec between the source positions of the two catalogs. As noticeable in boxes (b) and (e) in Figure~\ref{figure:ximage}, the X-ray source positions in our catalog look better aligned with the Ultravista $K_{\mathrm{s}}$-band image than those in \citet{2016ApJ...819...62C}. The X-ray source in the (c) box is first identified as the X-ray counterpart of AS2COS0353.2. However, the corresponding X-ray source in our image looks likely associated with AS2COS0353.1. Therefore, we consider that the X-ray source in the (c) box is associated with AS2COS0353.1.

\section{Details of Iteration of the SED Modeling}\label{appendix:sed_detail}

In this section, we describe the details of the second run in SED modeling. For comparison purposes, we summarise the ``basic'' parameter set used in the first run in Table~\ref{table:modelbasic}. After the first run, we notice some issues in the photometry and the parameter settings. Therefore, we add the following operations and re-perform SED modeling. We confirm that these operations significantly improve the best-fit statistic.

\begin{table*}
  \caption{Basic parameter sets used for the SED modeling. The overall size of the parameter grid is 3,594,240,000.}
  \scriptsize
  \begin{tabularx}{\linewidth}{@{}lll}
    \hline\hline
    Parameter & Symbol & Value\\
    \hline
    \multicolumn3c{SFH (sfhdelayedbq)}\\
    \hline
    E-folding time of the main stellar population & $\tau_{\mathrm{main}}$ [Myr] & 100, 316, 1000, 3162, 10000\\
    Age of the main stellar population & $\mathrm{age_{main}}$ [Myr] & 100, 158, 251, 398, 631, 1000, 1585, 2512, 3981, 6310\\
    Age of the late burst/quench population & $\mathrm{age_{bq}}$ [Myr] & 10\\
    Ratio of the SFR after/before $\mathrm{age_{bq}}$. & $r_{\mathrm{SFR}}$ & 1, 10, 100, 1000\\
    \hline
    \multicolumn3c{SSP (bc03; \citealt{2003MNRAS.344.1000B})}\\
    \hline
    IMF of the stellar model &  &  \cite{2003PASP..115..763C}\\
    Metalicity of the stellar model &  &  0.02\\
    \hline
    \multicolumn3c{Dust attenuation (dustatt\_modified\_starburst; {\citealt{2000ApJ...533..682C,2002ApJS..140..303L,2009A&A...507.1793N}})}\\
    %,2009A&A...507.1793Nを追加
    \hline
    Colour excess of the nebular lines. & $E(B-V)_{\mathrm{lines}}$ & 0.1, 0.2, 0.4, 0.6, 0.8, 1.0, 1.2, 1.4, 1.6, 1.8, 2.4, 2.8, 3.2\\
    Reduction factor to calculate the stellar continuum attenuation & $E(B-V)_{\mathrm{factor}}$ & 0.44\\
    UV bump amplitude &  &  0 (no bump)\\
    Power-law index to modify the attenuation curve & $\delta$ & --0.4, 0.0, 0.4\\
    \hline
    \multicolumn3c{Dust emission (EThemis; \citealt{2024ApJ...965..108U})}\\
    \hline
    Mass fraction of the small hydrocarbon solids & $q_{\mathrm{hac}}$ & 0.01, 0.02, 0.06, 0.10\\
    Minimum radiation field & $U_{\mathrm{min}}$ & 0.0, 0.4, 0.8, 1.2, 1.6, 2.0, 2.4, 3.0\\
    Power-law index of the starlight intensity distribution & $\alpha$ & 2.5, 3.0\\
    Mass fraction of dust illuminated with $U=U_{\mathrm{min}}$ & $1-\gamma$ & 0.9\\
    \hline
    \multicolumn3c{AGN emission (skirtor2016; \citealt{2012MNRAS.420.2756S,2016MNRAS.458.2288S})}\\
    \hline
    Average edge-on optical depth at 9.7 micron & $\tau_{\rm 9.7}$ & 3, 7\\
    Radial gradient of dust density & $p$ & 1.0\\
    Dust density gradient with polar angle & $q$ & 1.0\\
    Half-opening angle of the dust-free cone & $\Delta$\,[\degr] & 40\\
    Ratio of outer to inner radius & $R$ & 20\\
    Inclination & $\theta$\,[\degr] & 30, 60, 80\\
    Fraction of AGN IR luminosity to total IR luminosity\hspace{30pt} & $f_{\rm AGN}$\hspace{40pt} & 0.0, 0.1, 0.2, 0.3, 0.4, 0.5, 0.6, 0.7, 0.8, 0.9\\
    Extinction in polar direction & $E(B-V)$ & 0.0, 0.1\\
    Temperature of polar dust & $T_{\mathrm{pol}}$\,[K] & 100\\
    \hline
    \multicolumn3c{Redshifting}\\
    \hline
    Redshift & $z$ & 0.1--6.0 (step size: 0.1) or fixed at spectroscopic redshifts\\
    \hline
  \end{tabularx}\label{table:modelbasic}
\end{table*}

\subsection{Photometry Optimisation}

In the first run, we notice some inconsistencies between the photometry within individual sources. This can be attributed to the remaining blending effect or a failure in extracting the photometry. For these reasons, we check the images and decide to treat the following photometry as upper limits: AS2COS0001.1 (Spitzer 3.6 $\mu$m), AS2COS0043.1 (Ultravista $YJHK_{s}$), AS2COS0065.2 (Ultravista $YJHK_{s}$), and AS2COS0228.1 (Ultravista $YJHK_{s}$). We also treat the Ultravista $YJHK_{s}$ photometry of AS2COS123.1 as upper limits because we confirm no significant detection in those images. Moreover, we notice systematic offsets between the HSC $grizY$ and Ultravista $YJHK_{s}$ photometry of AS2COS0019.1, AS2COS0086.1, AS2COS0175.1, and AS2COS0203.1. This might be caused by calibration issues. Hence, we replace these photometry with the ones extracted from the COSMOS2020 catalog.

In addition to these operations, we optimize the photometry of some sources by referring to the HSC-SSP DR3 catalog. We replace the optical photometry of AS2COS0063.1, AS2COS0123.1, and AS2COS0228.1 with the ones extracted from the HSC-SSP DR3 catalog \citep{2022PASJ...74..247A}. We also update the Ultravista $YJHK_{s}$ photometry of AS2COS0063.1 with the ones extracted from the Ultravista DR4 catalog, which appear to be less affected by a bright contaminating source. Furthermore, we exclude the Spitzer 4.5 $\mu$m, 5.8 $\mu$m, and 8.0 $\mu$m photometry of AS2COS0228.1 because of the significant contamination by AS2COS0228.2.

\begin{table*}
  \caption{Extended parameter set of the AGN module used for AS2COS175.1.}
  \scriptsize
  \begin{tabularx}{\linewidth}{@{}lll}
    \hline\hline
    Parameter & Symbol & Value\\
    \hline
    \multicolumn3c{AGN emission (skirtor2016; \citealt{2012MNRAS.420.2756S,2016MNRAS.458.2288S})}\\
    \hline
    Average edge-on optical depth at 9.7 micron & $\tau_{\rm 9.7}$ & 3, 5, 7, 9\\
    Radial gradient of dust density & $p$ & 1.0\\
    Dust density gradient with polar angle & $q$ & 1.0\\
    Half-opening angle of the dust-free cone & $\Delta$\,[\degr] & 40\\
    Ratio of outer to inner radius & $R$ & 20\\
    Inclination & $\theta$\,[\degr] & 30, 40, 50, 60, 70, 80\\
    Fraction of AGN IR luminosity to total IR luminosity\hspace{30pt} & $f_{\rm AGN}$\hspace{40pt} & 0.0, 0.1, 0.2, 0.3, 0.4, 0.5, 0.6, 0.7, 0.8, 0.9\\
    Extinction in polar direction & $E(B-V)$ & 0.0, 0.05, 0.1\\
    Temperature of the polar dust & $T_{\mathrm{pol}}$\,[K] & 100\\
    \hline
  \end{tabularx}\label{table:modelAGN}
\end{table*}

\subsection{Misidentified Source}\label{subsection:misidentified}

The potential optical counterpart of AS2COS0072.1 was detected in the DEIMOS 10K spectroscopic survey. They spectroscopically confirmed that the source is at $z=0.802$ with a quality flag of $Q_{f}=4$ (highest quality). However, the submillimeter line scan performed by \citet{2024ApJ...961..226L} showed that AS2COS0072.1 is at $z=3.798$. This inconsistency can be attributed to the misidentification of the counterpart. We thus regard the source at $z=0.802$ as the foreground of AS2COS0072.1 and treat the optical to mid-infrared photometry as upper limits.

We also conclude that the optical to near-infrared counterpart of AS2COS0159.1 in HSC and Ultravista is misidentified. The high-resolution imaging by JWST revealed that there were two sources near the 870-$\mu$m source position. One was identified as a galaxy at $z=0.033$ from the optical spectroscopy by PRIMUS. This source was initially identified as an optical counterpart of AS2COS0159.1, however, we notice that the peak wavelength of the far-infrared SED of AS2COS0159.1 is too long for a galaxy at $z=0.033$.\footnote{The peak of the far-infrared emission of AS2COS0159.1 is ${\sim}350\,\mu$m in the observed frame. This corresponds to 9 K assuming an optically-thin graybody with an emissivity index of 1.8 at $z=0.033$.} Therefore, we consider that the source at $z=0.033$ is foreground of AS2COS00159.1. We confirm that the foreground source was only dominant below the F277W band. We thus decide to treat the optical to near-infrared photometry as upper limits.

\subsection{SED Fitting Parameter Optimization}

With the basic parameter set used in the first run, the mid- to far-infrared SED AS2COS175.1 is not well reproduced. This can be attributed to the oversimplification of the AGN component. Hence, we extend the parameter grids and recalculate the SED of these sources. The extended parameter set is summarized in Table~\ref{table:modelAGN}.

\subsection{Overall Quality of the Final Fits}\label{subsection:chi2}

\begin{figure*}
  \centering
  \includegraphics[width=0.5\linewidth]{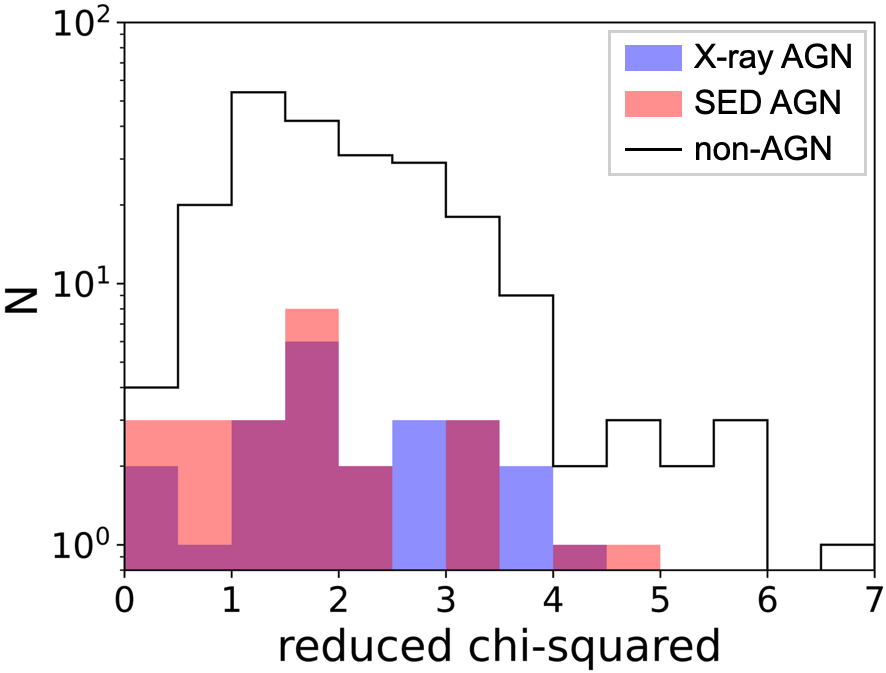}
  \caption{
  Histogram of the reduced $\chi^{2}$ for the final fits. The median and the maximum reduced $\chi^{2}$ are 1.8 and 6.5, respectively.}
  \label{figure:chi2}
\end{figure*}

Following the procedures described above, we finally obtain reasonable fits for all the sources. Figure~\ref{figure:chi2} shows the histogram of the reduced $\chi^{2}$ for the final fits. The median and the maximum reduced $\chi^{2}$ are 1.8 and 6.5, which indicates that the SEDs are well reproduced.

\section{Consistency Check of SED Modeling}\label{appendix:sed_consistency}

\begin{figure*}
  \centering
  \includegraphics[width=\linewidth]{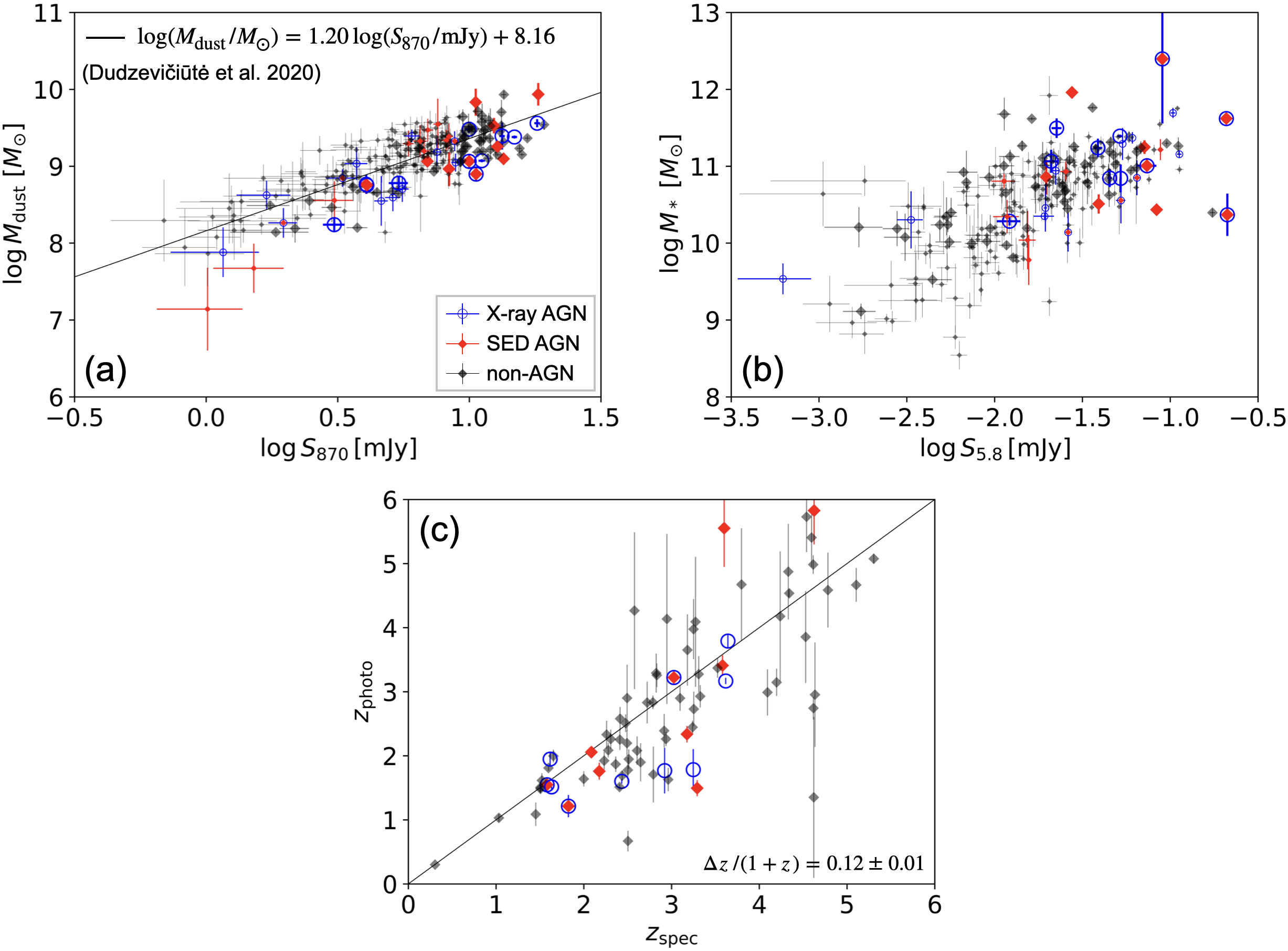}
  \caption{
  Comparison of the physical properties derived by the SED template fitting and the observational properties for the X-ray detected AS2COSMOS sources (X-ray AGN; see Section~\ref{subsubsection:x-ray_LFIR}), SED AGNs, and the rest of the AS2COSMOS sources (non-AGNs). In panel \textit{(a)} and \textit{(b)}, the large symbols show the spec-$z$ sample, while the smaller ones show the photo-$z$ sample. In panel \textit{(b)}, the sample is limited to the sources that are detected in Spitzer 5.8 $\mu$m. In panel \textit{(c)}, only the spec-$z$ sample is plotted.}
  \label{figure:consistency_check}
\end{figure*}

Figure~\ref{figure:consistency_check} compares some observational properties with the physical properties derived by the SED modeling that they are expected to most strongly correlate with. We confirm positive correlations in these plots, supporting the validity of our SED analysis.

\section{Mock Analysis}\label{appendix:sed_degeneracy}

\begin{figure*}
  \centering
  \includegraphics[width=\linewidth]{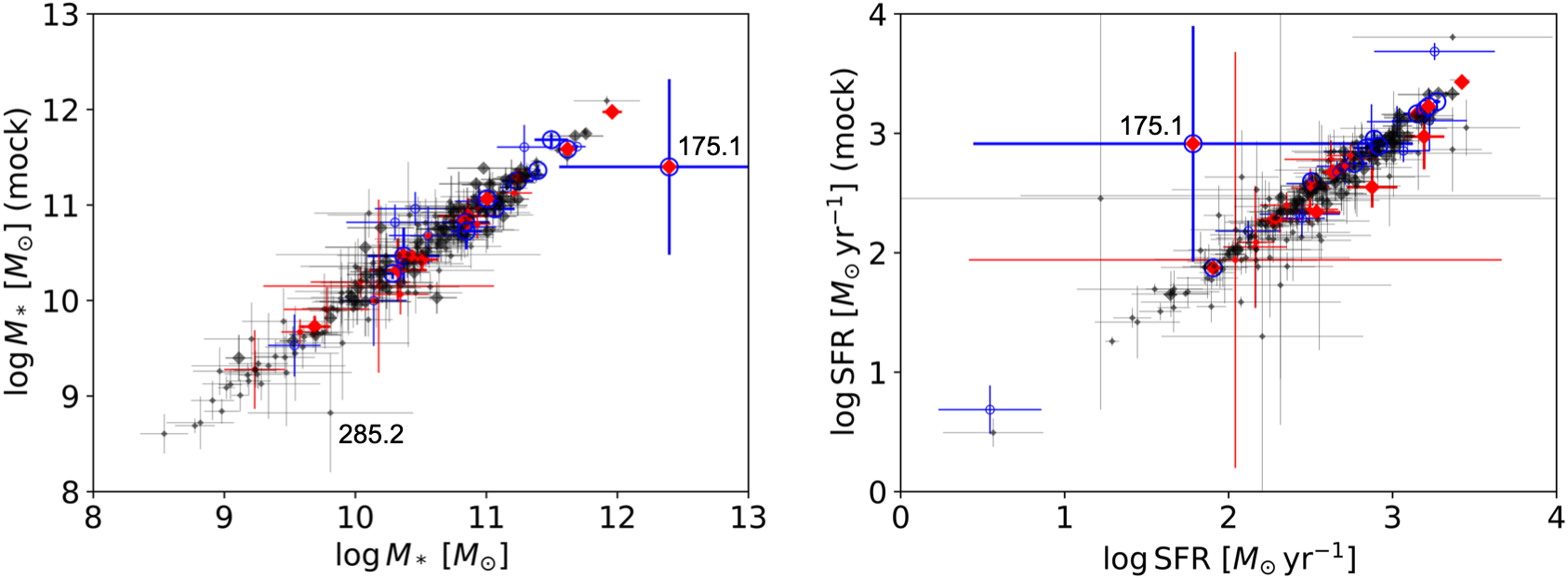}
  \caption{
  Comparison of the stellar mass and SFR derived by the SED analysis with those derived from the mock catalog. Most of the sources align well, but AS2COS0285.2 shows a large discrepancy in stellar mass, and AS2COS0175.1 shows large discrepancies in both stellar mass and SFR. The large symbols show the spec-$z$ sample, while the smaller ones show the photo-$z$ sample.}
  \label{figure:mock}
\end{figure*}

Figure~\ref{figure:mock} compares the stellar masses and SFRs derived by {\sc cigale} with those derived from the mock catalog. We confirm that most of the sources align well, but AS2COS0285.2 shows a large discrepancy in stellar mass, and AS2COS0175.1 shows large discrepancies in both stellar mass and SFR. AS2COS0285.2 is a galaxy without an AGN. This source is not detected in the optical band and has no spectroscopic redshift, which may make it difficult to constrain the stellar component. AS2COS0175.1 is a type-1 SED AGN. According to the SED fitting result, the optical to near-infrared SED of this source is dominated by the AGN emission, which may make the estimations unreliable. We generated probability distribution functions (PDFs) of SFRs and stellar masses for several sources, including the problematic sources mentioned above. We confirmed that most of the PDFs have single-horn shapes, showing that the parameters are well constrained. However, the PDFs of AS2COS0175.1 have multi-horn structures. This indicates that the SFR and stellar mass of AS2COS0175.1 are unreliable.

\section{Background Modeling for Chandra ACIS-I}\label{appendix:background}

Table~\ref{table:modelback} summarizes the best-fit parameters of the Chandra/ACIS-I background modeling (Section~\ref{subsubsection:x-ray_spec_bright}).

\begin{table*}
  \caption{Summary of the background modeling for Chandra ACIS-S ({\sc vfaint} mode).}
  \centering
  \begin{tabularx}{0.5\linewidth}{llll}
    \hline\hline
    Model & Parameter & Value & Unit \\
    \hline
    gaussian  & $E$ & 1.487 & keV \\
    & $\sigma$ & 1 & eV \\
    & norm & $5.3^{+3.9}_{-3.7}$ & $10^{-6}$ photons cm$^{-2}$ s$^{-1}$ \\
    gaussian  & $E$ & 1.557 & keV \\
    & $\sigma$ & 1 & eV \\
    & norm & $7.7^{+3.9}_{-3.7}$ & $10^{-6}$ photons cm$^{-2}$ s$^{-1}$ \\
    gaussian & $E$ & 2.123 & keV \\
    & $\sigma$ & 1 & eV \\
    & norm & $25.5^{+4.7}_{-4.5}$ & $10^{-6}$ photons cm$^{-2}$ s$^{-1}$ \\
    gaussian & $E$ & 2.205 & keV \\
    & $\sigma$ & 1 & eV \\
    & norm & $11.0^{+4.4}_{-4.2}$ & $10^{-6}$ photons cm$^{-2}$ s$^{-1}$ \\
    gaussian & $E$ & 2.410 & keV \\
    & $\sigma$ & 1 & eV \\
    & norm & $4.9^{+3.2}_{-3.1}$ & $10^{-6}$ photons cm$^{-2}$ s$^{-1}$ \\
    gaussian & $E$ & 2.7 & keV \\
    & $\sigma$ & $61^{+39}_{-45}$ & eV \\
    & norm & $8.2^{+4.3}_{-3.7}$ & $10^{-6}$ photons cm$^{-2}$ s$^{-1}$ \\
    gaussian & $E$ & 7.478 & keV \\
    & $\sigma$ & 1 & eV \\
    & norm & $64.3^{+5.5}_{-5.3}$ & $10^{-6}$ photons cm$^{-2}$ s$^{-1}$ \\
    gaussian & $E$ & 9.713 & keV \\
    & $\sigma$ & 1 & eV \\
    & norm & $139.8^{+6.8}_{-6.6}$ & $10^{-6}$ photons cm$^{-2}$ s$^{-1}$ \\
    gaussian & $E$ & 9.628 & keV \\
    & $\sigma$ & 1 & eV \\
    & norm & linked & 1/9 $\times$ norm of Au L$\alpha_1$\\
    gaussian & $E$ & 8.265 & keV \\
    & $\sigma$ & 1 & eV \\
    & norm & $14.5^{+4.2}_{-4.0}$ & $10^{-6}$ photons cm$^{-2}$ s$^{-1}$ \\
    powerlaw & $\Gamma$ & $0.143^{+0.026}_{-0.026}$ & \\
    & norm & $163.2^{+7.4}_{-7.4}$ & $10^{-6}$ \\
    gabs & $E$ & $0.25$ & keV \\
    & $\sigma$ & $153.5$ & eV \\
    & Strength & $0.99^{+2.39}_{-0.66}$ & $10^{4}$ \\
    expdec & $\alpha$ & $1.58^{+0.88}_{-0.49}$ & \\
    & norm & $3.1^{+5.6}_{-0.8}$ & $10^{-4}$ \\
    \hline
  \end{tabularx}\label{table:modelback}
\end{table*}

\section{Figures of Individual Sources}\label{appendix:figure_detail}

\begin{figure*}
  \centering
  \includegraphics[width=\linewidth]{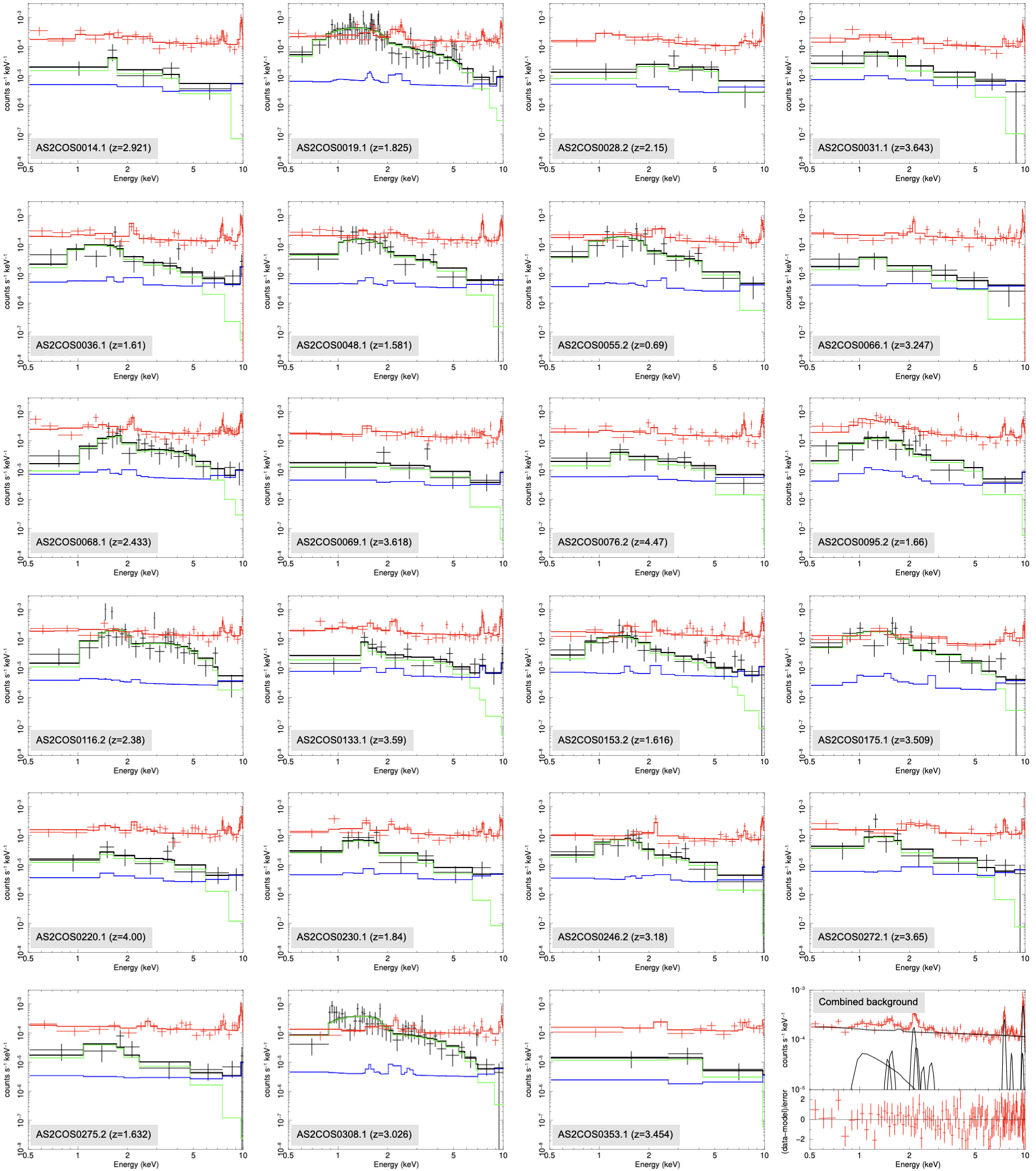}
  \caption{\textit{(AS2COS0014.1--AS2COS0353.1)} The 0.5--10 keV spectra of the 23 X-ray AGNs. The black points and the black solid lines show the observed spectra and the best-fit models. The red points and the red solid lines show the background spectra and the best-fit models. The blue solid lines show the scaled background spectra. The green solid lines show the background-subtracted source spectra. To improve visibility, the source and the background spectra are binned to have at least 1.5$\sigma$ and 3$\sigma$ in each bin, respectively. \textit{(Combined background)} The combined background spectrum. The red points and the red solid line show the background spectra and the best-fit model. The black solid lines show the components of the model. The lower panel shows the residuals. To improve visibility, the combined background spectrum is binned to have at least 7$\sigma$ in each bin. Note that the spectrum binning only affects the presentation of the data and does not affect the spectral analysis results.}
  \label{figure:x-spectrum}
\end{figure*}

\begin{figure*}
  \centering
  \includegraphics[width=\linewidth]{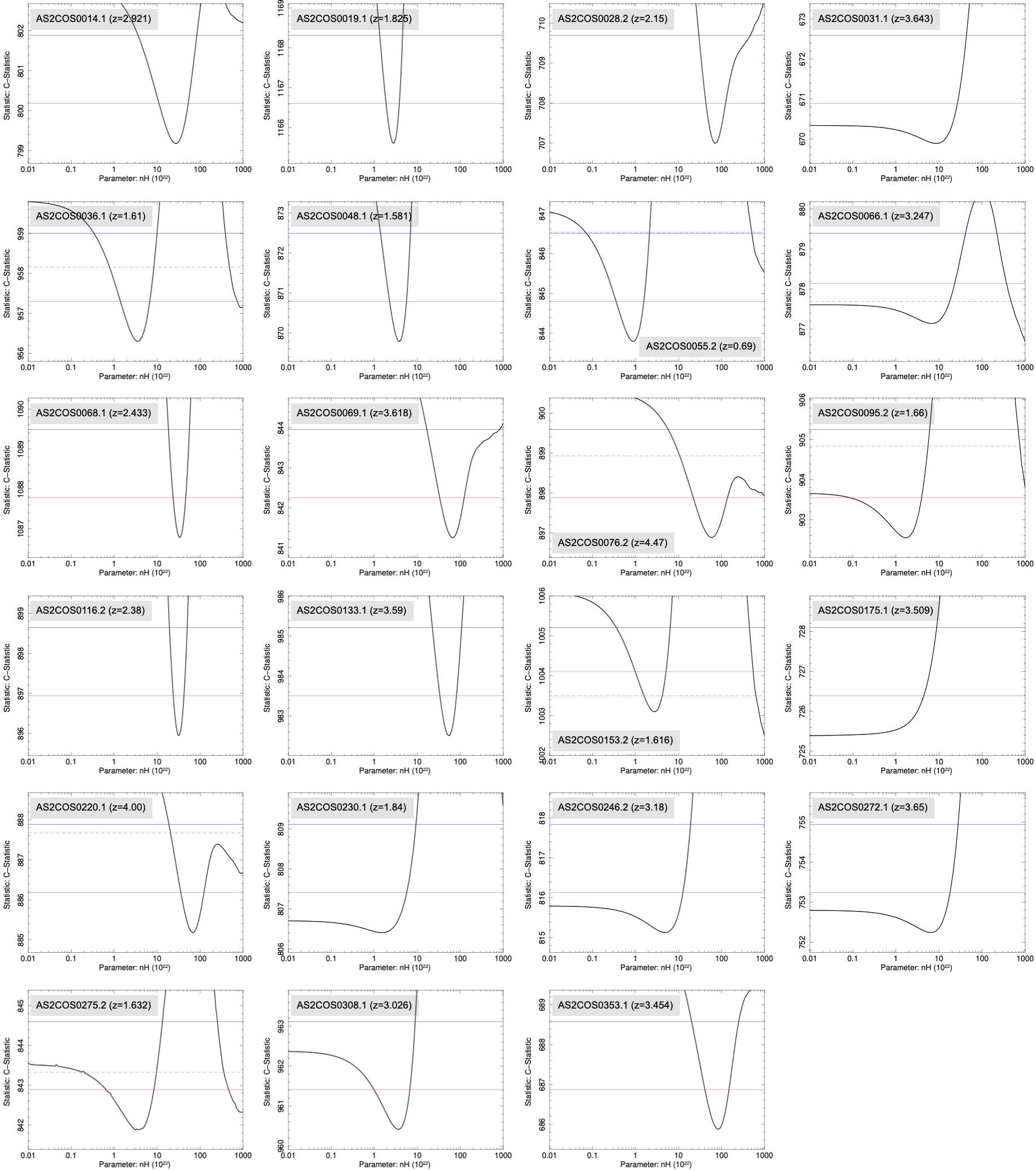}
  \caption{Plots of the goodness of fit as a function of the line-of-sight hydrogen column densities. The vertical axes show the statistical values of the C-statistic. The horizontal axes show the line-of-sight hydrogen column densities in units of $10^{22}\,\mathrm{cm}^{-2}$. The blue lines show the 90 per cent confidence levels compared with the best-fit values. The red solid lines show the 68.27 per cent confidence levels (1$\sigma$) compared with the Compton-thin solutions. The red dashed lines show the 68.27 per cent confidence levels (1$\sigma$) compared with the Compton-thick solutions, if exist.}
  \label{figure:x-contour}
\end{figure*}

\begin{figure*}
  \centering
  \includegraphics[width=\linewidth]{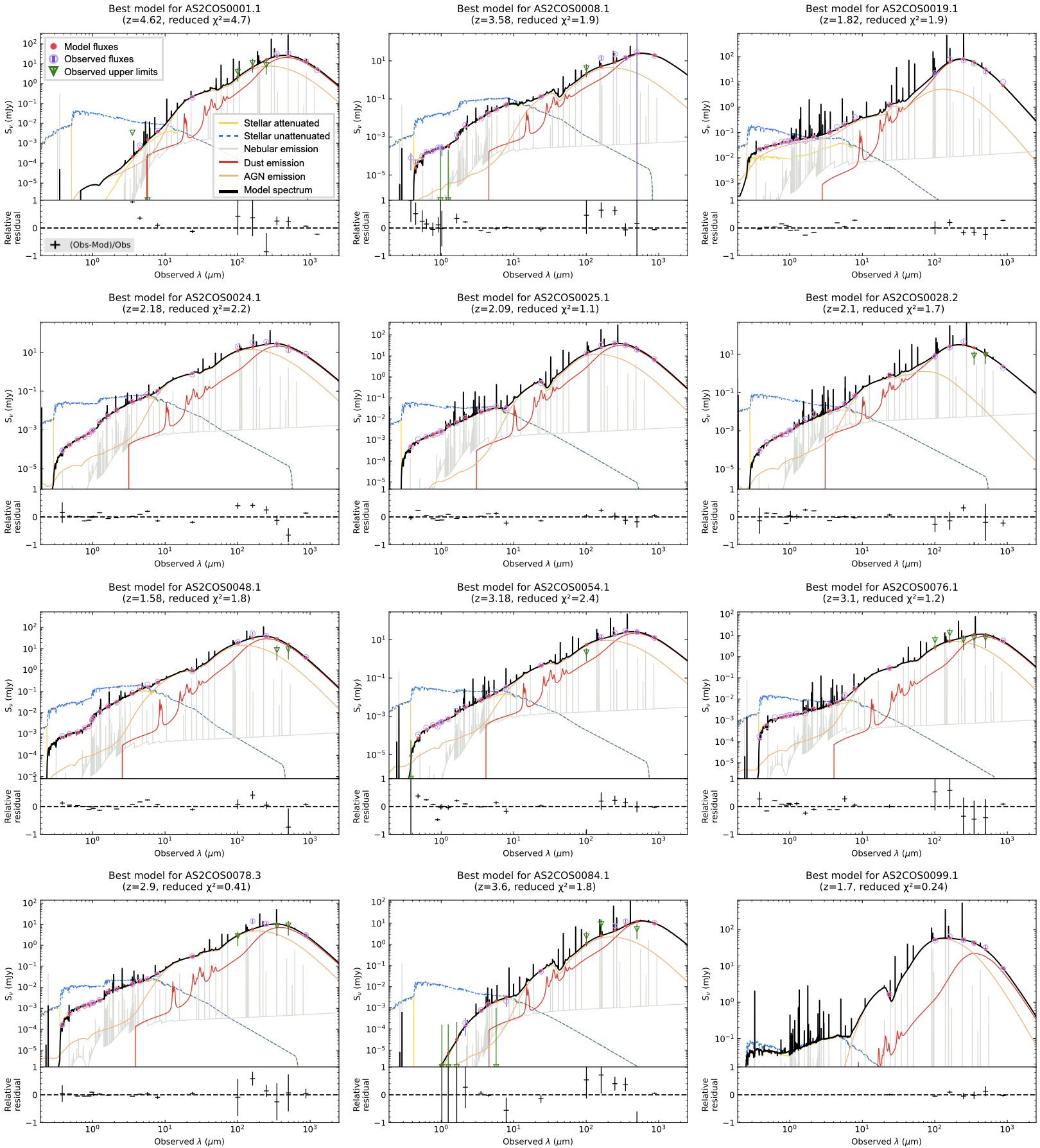}
  \caption{The SEDs and best-fit models of the SED AGNs. The black solid line represents the composite spectrum. The yellow solid line illustrates the stellar emission attenuated by interstellar dust. The blue dashed line depicts the unattenuated stellar emission for a reference purpose. The orange line corresponds to the emission from an AGN. The red line shows the infrared emission from interstellar dust. The gray line denotes the nebulae emission. The observed data points are represented by purple circles, accompanied by 1$\sigma$ error bars. The bottom panel displays the relative residuals.}
  \label{figure:SED}
\end{figure*}

Figure~\ref{figure:x-spectrum} shows the X-ray spectra and best-fit models of the X-ray AGNs. Figure~\ref{figure:x-contour} summarises the plots of the goodness of fit as a function of the line-of-sight hydrogen column densities. Figure~\ref{figure:SED} shows the SEDs and best-fit models of the SED AGNs.

\begin{figure*}
  \centering
  \includegraphics[width=\linewidth]{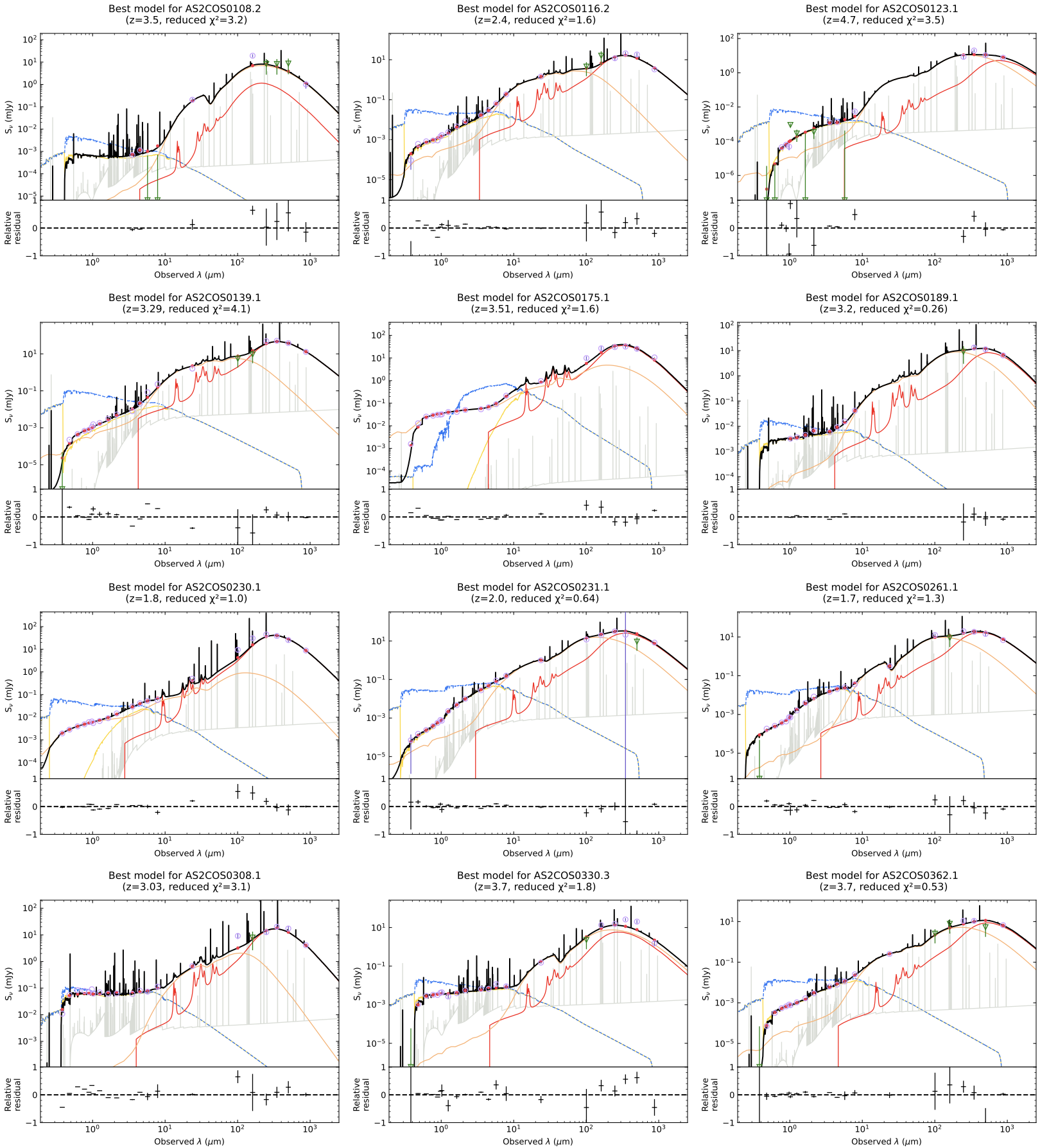}
  {\bf Figure \ref{figure:SED}} (continued)
\end{figure*}

%% For this sample we use BibTeX plus aasjournals.bst to generate the
%% the bibliography. The sample631.bib file was populated from ADS. To
%% get the citations to show in the compiled file do the following:
%%
%% pdflatex sample631.tex
%% bibtext sample631
%% pdflatex sample631.tex
%% pdflatex sample631.tex

\bibliography{as2cosmos}{}
\bibliographystyle{aasjournal}

%% This command is needed to show the entire author+affiliation list when
%% the collaboration and author truncation commands are used.  It has to
%% go at the end of the manuscript.
%\allauthors

%% Include this line if you are using the \added, \replaced, \deleted
%% commands to see a summary list of all changes at the end of the article.
%\listofchanges

\end{document}